\documentclass[a4paper, 12pt, twoside]{report}
\usepackage[a4paper, width=150mm, top=25mm, bottom=25mm]{geometry}
\usepackage{listings}

\usepackage{subcaption}

\usepackage{amsmath}
\usepackage{mathtools}
\usepackage{amssymb}
\usepackage{microtype}
\usepackage{environ}
\usepackage{intcalc}
\usepackage{blkarray}
\newcommand{\NN}{{\mathbb{N}}}

\newcommand{\ZZ}{{\mathbb{Z}}}

\usepackage{tabularx}
\usepackage{graphicx}

\usepackage[colorlinks,pdfpagelabels,pdfstartview = FitH, bookmarksopen = true,bookmarksnumbered = true, linkcolor = black, plainpages = false, hypertexnames = false, citecolor = black]{hyperref}
\usepackage{cleveref}
\usepackage[resetlabels]{multibib}
\newcites{Literature}{Bibliography}
\newcites{Websites}{Web References}

\usepackage[section]{placeins}

\newcommand{\U}{\mathcal{U}}
\renewcommand{\S}{\mathcal{S}}
\newcommand{\SSP}{\ensuremath{\leq_{SSP}}}
\newcommand{\I}{\mathcal{I}}

\usepackage{multicol}
\usepackage{booktabs}

\usepackage{tikz}
\usetikzlibrary{arrows,calc, positioning}

\renewcommand{\phi}{\varphi}
\renewcommand{\theta}{\vartheta}

\usepackage{amsthm}
\theoremstyle{definition}
\newtheorem{definition}{Definition}[section]

\hypersetup{
	pdftoolbar=false,
	pdfmenubar=false,
	colorlinks,
	urlcolor={rwth-magenta},
	linkcolor={rwth-red},
	citecolor={rwth-green}
}

\usepackage{xcolor}
\definecolor{rwth-blue}{cmyk}{1,.5,0,0}\colorlet{rwth-lblue}{rwth-blue!50}\colorlet{rwth-llblue}{rwth-blue!25}
\definecolor{rwth-violet}{cmyk}{.6,.6,0,0}\colorlet{rwth-lviolet}{rwth-violet!50}\colorlet{rwth-llviolet}{rwth-violet!25}
\definecolor{rwth-purple}{cmyk}{.7,1,.35,.15}\colorlet{rwth-lpurple}{rwth-purple!50}\colorlet{rwth-llpurple}{rwth-purple!25}
\definecolor{rwth-carmine}{cmyk}{.25,1,.7,.2}\colorlet{rwth-lcarmine}{rwth-carmine!50}\colorlet{rwth-llcarmine}{rwth-carmine!25}
\definecolor{rwth-red}{cmyk}{.15,1,1,0}\colorlet{rwth-lred}{rwth-red!50}\colorlet{rwth-llred}{rwth-red!25}
\definecolor{rwth-magenta}{cmyk}{0,1,.25,0}\colorlet{rwth-lmagenta}{rwth-magenta!50}\colorlet{rwth-llmagenta}{rwth-magenta!25}
\definecolor{rwth-orange}{cmyk}{0,.4,1,0}\colorlet{rwth-lorange}{rwth-orange!50}\colorlet{rwth-llorange}{rwth-orange!25}
\definecolor{rwth-yellow}{cmyk}{0,0,1,0}\colorlet{rwth-lyellow}{rwth-yellow!50}\colorlet{rwth-llyellow}{rwth-yellow!25}
\definecolor{rwth-grass}{cmyk}{.35,0,1,0}\colorlet{rwth-lgrass}{rwth-grass!50}\colorlet{rwth-llgrass}{rwth-grass!25}
\definecolor{rwth-green}{cmyk}{.7,0,1,0}\colorlet{rwth-lgreen}{rwth-green!50}\colorlet{rwth-llgreen}{rwth-green!25}
\definecolor{rwth-cyan}{cmyk}{1,0,.4,0}\colorlet{rwth-lcyan}{rwth-cyan!50}\colorlet{rwth-llcyan}{rwth-cyan!25}
\definecolor{rwth-teal}{cmyk}{1,.3,.5,.3}\colorlet{rwth-lteal}{rwth-teal!50}\colorlet{rwth-llteal}{rwth-teal!25}
\definecolor{rwth-gold}{cmyk}{.35,.46,.7,.35}
\definecolor{rwth-silver}{cmyk}{.39,.31,.32,.14}

\usepackage{titlesec}
\titleformat*{\subsubsection}{\normalfont}

\usepackage[all]{nowidow}

\usepackage{comment}
\usepackage{todonotes}
\definecolor{lightgreen}{rgb}{0.8,1.0,0.8}

\definecolor{lightgray}{rgb}{.9,.9,.9}
\definecolor{darkgray}{rgb}{.4,.4,.4}
\definecolor{commentgreen}{HTML}{559854}
\definecolor{codeblue}{HTML}{659CD6}
\definecolor{codered}{HTML}{C89278}

\lstdefinelanguage{JavaScript}{
  keywords={break, case, catch, continue, debugger, default, delete, do, else, false, finally, for, function, if, in, instanceof, new, null, return, switch, this, throw, true, try, typeof, var, void, while, with},
  morecomment=[l]{//},
  morecomment=[s]{/*}{*/},
  morestring=[b]',
  morestring=[b]",
  ndkeywords={class, export, boolean, throw, implements, import, this},
  keywordstyle=\color{codeblue}\bfseries,
  ndkeywordstyle=\color{white}\bfseries,
  identifierstyle=\color{white},
  commentstyle=\color{commentgreen}\ttfamily,
  stringstyle=\color{codered}\ttfamily,
  sensitive=true
}

\lstdefinelanguage{HTML5}{
  language=html,
  sensitive=true,	
  alsoletter={<>=-},	
  morecomment=[s]{<!-}{-->},
  tag=[s],
  otherkeywords={
  >,
	<!DOCTYPE,
  </html, <html, <head, <title, </title, <style, </style, <link, </head, <meta, />,
	</body, <body,
	</div, <div, </div>, 
	</p, <p, </p>,
	</script, <script,
  <canvas, /canvas>, <svg, <rect, <animateTransform, </rect>, </svg>, <video, <source, <iframe, </iframe>, </video>, <image, </image>, <header, </header, <article, </article
  },
  ndkeywords={
  =,
  charset=, src=, id=, width=, height=, style=, type=, rel=, href=,
  fill=, attributeName=, begin=, dur=, from=, to=, poster=, controls=, x=, y=, repeatCount=, xlink:href=,
  margin:, padding:, background-image:, border:, top:, left:, position:, width:, height:, margin-top:, margin-bottom:, font-size:, line-height:,
  transform:, -moz-transform:, -webkit-transform:,
  animation:, -webkit-animation:,
  transition:,  transition-duration:, transition-property:, transition-timing-function:,
  }
}

\lstdefinelanguage{CSS}{
  keywords={color,background-image:,margin,padding,font,weight,display,position,top,left,right,bottom,list,style,border,size,white,space,min,width, transition:, transform:, transition-property, transition-duration, transition-timing-function},	
  sensitive=true,
  morecomment=[l]{//},
  morecomment=[s]{/*}{*/},
  morestring=[b]',
  morestring=[b]",
  alsoletter={:},
  alsodigit={-}
}

\def\code#1{\texttt{#1}}

\title{
  Bachelor's Thesis\\
  \textsl{Exploring the Reductions Between SSP-NP-complete Problems and Developing a Compendium Website Displaying the Results}\\[1ex]
}
\date{Summer Semester 2024}
\author{%
  Femke Pfaue\\
  Supervisor: Christoph Grüne\\
  Examiners: Prof. Dr. Peter Rossmanith, Priv.-Doz. Dr. Walter Unger\\
  RWTH Aachen University\\
  Department of Computer Science I1\\
}

\makeatletter
\newcommand{\problemtitle}[1]{\gdef\@problemtitle{#1}}
\newcommand{\probleminstance}[1]{\gdef\@probleminstance{#1}}
\newcommand{\problemuniverse}[1]{\gdef\@problemuniverse{#1}}
\newcommand{\problemquestion}[1]{\gdef\@problemquestion{#1}}
\newcommand{\problemfeassol}[1]{\gdef\@problemfeassol{#1}}
\newcommand{\problemsol}[1]{\gdef\@problemsol{#1}}
\NewEnviron{problem}{  \problemtitle{}\probleminstance{}\problemuniverse{}\problemfeassol{}\problemsol{}
  \BODY
  \par\addvspace{.5\baselineskip}
  \noindent
  \begin{tabularx}{\textwidth}{@{\hspace{\parindent}} l X c}
	\toprule
    \multicolumn{2}{@{\hspace{\parindent}}l}{\textsc{\@problemtitle}} \\
    \textbf{Instance:} & \@probleminstance\\
    \textbf{Universe:} & \@problemuniverse\\
	\textbf{Feasible solution set:} & \@problemfeassol\\
	\textbf{Solution set:} & \@problemsol\\
	\toprule
  \end{tabularx}
  \par\addvspace{.5\baselineskip}
}
\makeatother

\newcommand{\functionG}{
    \paragraph*{Reduction Function}
}
\newcommand{\functionF}{
    \paragraph*{Solution Mapping}
}
\newcommand{\correctness}{
    \paragraph*{Correctness}
}

\newcommand{\clique}{
    \begin{problem}
        \problemtitle{Clique}
        \probleminstance{Graph $G=(V,E)$, number $k\in\NN$}
        \problemuniverse{Vertex set $V=:\U$}
        \problemfeassol{The set of all cliques, thus a set of vertices $C\subseteq V$ such that 
        for all $v,v'\in C$, $v$ is adjacent to $v'$.}
        \problemsol{The set of all cliques of size at least $k$.}
    \end{problem}
}

\newcommand{\IS}{
    \begin{problem}
        \problemtitle{Independent Set}
        \probleminstance{Graph $G=(V,E)$, number $k\in\NN$}
        \problemuniverse{Vertex set $V=:\U$}
        \problemfeassol{The set of all independent sets, thus a set of vertices $V^*\subseteq V$ such that
        for all $v,v'\in V^*$, $v$ is not adjacent to $v'$.}
        \problemsol{The set of all independent sets of size at least $k$.}
    \end{problem}
}

\newcommand{\threeDmatch}{
    \begin{problem}
        \problemtitle{3-Dimensional Matching}
        \probleminstance{Finite Sets $X, Y,Z$, set $U\subseteq X\times Y\times Z$, number $k\in\NN$.}
        \problemuniverse{$\U:=U$.}
        \problemfeassol{The set of all sets $W\subseteq U$ such that for any $(x_i, y_i, z_i),(x_j,y_j,z_j)\in W: x_i\neq x_j,y_i\neq y_j, z_i\neq z_j$.}
        \problemsol{The set of all feasible solutions with $\vert W\vert \geq k$.}
    \end{problem}
}

\newcommand{\threeDExMatch}{
    \begin{problem}
        \problemtitle{3-Dimensional Exact Matching}
        \probleminstance{Finite Sets $X,Y,Z$ such that $\vert X\vert=\vert Y\vert=\vert Z\vert=:q$, set $U\subseteq X\times Y\times Z$.}
        \problemuniverse{$\U:=U$.}
        \problemfeassol{-}
        \problemsol{The set of all sets $W\subseteq U$ such that for any 
        $(x_1, y_1, z_1),(x_2,y_2,z_2)\in W: x_1\neq x_2,y_1\neq y_2, z_1\neq z_2$ and $\vert W\vert = q$.}
    \end{problem}
}

\newcommand{\steiner}{
    \begin{problem}
        \problemtitle{Steiner Tree}
        \probleminstance{Graph $G=(S\cup T, E)$, set of Steiner vertices $S$, set of terminal vertices $T$, edge weights $c:E\rightarrow\NN$, number $k\in\NN$.}
        \problemuniverse{$\U:=E$.}
        \problemfeassol{The set of all sets $A\subseteq E$ such that the $A$ is a tree connecting all terminal vertices from $T$.}
        \problemsol{The set of all feasible solutions $A$ with $\sum_{a'\in A}c(a')\leq k$.}
    \end{problem}
}

\newcommand{\subsum}{
    \begin{problem}
        \problemtitle{Subset Sum}
        \probleminstance{Set of numbers $A=\{a_1,\ldots,a_n\}\subseteq\NN$ and a target value $M\in\NN$.}
        \problemuniverse{$\U:=A$.}
        \problemfeassol{-}
        \problemsol{The set of all sets $A^*\subseteq A$ such that $\sum_{a_i\in A^*}a_i =M$.}
    \end{problem}
}

\newcommand{\setPacking}{
    \begin{problem}
        \problemtitle{Set Packing}
        \probleminstance{A set $\mathfrak{U}$, a family of subsets $\mathfrak{S}\subseteq 2^{\mathfrak{U}}$, and $k\in\NN$}
        \problemuniverse{$\U:=\mathfrak{S}$}
        \problemfeassol{All subfamilies $\mathfrak{C}\subseteq\mathfrak{S}$ such that all sets in $\mathfrak{C}$ are pairwise disjoint.}
        \problemsol{All feasible subfamilies $\mathfrak{C}$ with at least $k$ elements.}
    \end{problem}
}

\newcommand{\sat}{
    \begin{problem}
        \problemtitle{\textsc{Satisfiability}}
        \probleminstance{Literal set $L=\{\ell_1,\ldots,\ell_n\}\cup\{\overline{\ell}_1,\ldots,\overline{\ell}_n\}$, clauses $C\subseteq 2^{\vert L\vert}$.}
        \problemuniverse{$\U := L$.}
        \problemfeassol{-}
        \problemsol{The set of all sets $L^*\subseteq\U$ such that for all $i\in\{1,\ldots,n\}$ we have $\vert L^*\cap\{\ell_i,\overline{\ell}_i\}\vert=1$, and $\vert L^*\cap C_j\vert\geq1$ for all $C_j\in C, j\in \{1,\ldots,\vert C\vert\}$.}
    \end{problem}
}

\newcommand{\IP}{
    \begin{problem}
        \problemtitle{0-1-Integer Programming}
        \probleminstance{A matrix $C\in\ZZ^{n\times m}$ and a vector $b\in\ZZ^{m\times 1}$.}
        \problemuniverse{$\U:=y_1,\ldots,y_n$}
        \problemfeassol{-}
        \problemsol{The set of all subsets $S$ of $y_1,\ldots,y_n$ such that $CX\leq b$ with $X$ being a vector over $\{0,1\}^m$, with $x_i=1$ if and only if $y_i\in S$.}
    \end{problem}
}

\newcommand{\dirHam}{
    \begin{problem}
        \problemtitle{Directed Hamiltonian Cycle}
        \probleminstance{Directed graph $G=(V,A)$.}
        \problemuniverse{Arc set $A=:\U$.}
        \problemfeassol{-}
        \problemsol{The set of all sets $C\subseteq A$ forming a directed Hamiltonian cycle, thus a cycle that includes each vertex exactly once.}
    \end{problem}
}

\newcommand{\undirHam}{
    \begin{problem}
        \problemtitle{Undirected Hamiltonian Cycle}
        \probleminstance{Graph $G=(V,E)$.}
        \problemuniverse{Edge set $\U:=E$.}
        \problemfeassol{-}
        \problemsol{The set of all sets $C\subseteq E$ forming a Hamiltonian cycle, thus a cycle that includes each vertex exactly once.}
    \end{problem}
}

\newcommand{\dirHamPath}{
    \begin{problem}
        \problemtitle{Directed Hamiltonian Path}
        \probleminstance{Directed graph $G=(V,A)$, vertices $s,t\in V$.}
        \problemuniverse{Arc set $\U:=A$.}
        \problemfeassol{-}
        \problemsol{The set of all sets $C\subseteq A$ forming a directed Hamiltonian path from $s$ to $t$, thus a path that includes each vertex exactly once.}
    \end{problem}
}

\newcommand{\undirHamPath}{
    \begin{problem}
        \problemtitle{Undirected Hamiltonian Path}
        \probleminstance{Graph $G=(V,E)$, vertices $s,t\in V$.}
        \problemuniverse{Edge set $\U:=E$.}
        \problemfeassol{-}
        \problemsol{The set of all sets $C\subseteq E$ forming a Hamiltonian path from $s$ to $t$, thus a path that includes each vertex exactly once.}
    \end{problem}
}

\newcommand{\VC}{
    \begin{problem}
        \problemtitle{Vertex Cover}
        \probleminstance{Graph $G=(V,E)$, number $k\in\NN$.}
        \problemuniverse{Vertex set $V=:\U$.}
        \problemfeassol{The set of all vertex covers, thus all sets $V^*\subseteq V$ such that for all $(v,v')\in E$, $v\in V^*$ or $v'\in V^*$, thus, 
        a set of vertices such that all edges are incident to at least one vertex in the set.}
        \problemsol{The set of all vertex covers of size at most $k$.}
    \end{problem}
}

\newcommand{\EC}{
    \begin{problem}
        \problemtitle{Exact Cover}
        \probleminstance{Set $\mathfrak{S}=\{S_1,\ldots, S_m\}$ of sets $S_j$ over the set $U=\{1,\ldots,n\}$.}
        \problemuniverse{$\U=\mathfrak{S}$.}
        \problemfeassol{-}
        \problemsol{Set $T$ of all subsets of $\mathfrak{S}$, $T\subseteq\{S_1,\ldots,S_m\}$ such that all $T_h\in T$ are disjoint and $\bigcup T_h=\bigcup S_j = \{1,\ldots,n\}$.}
    \end{problem}
}

\newcommand{\threesat}{
    \begin{problem}
        \problemtitle{3-SAT}
        \probleminstance{Literal set $L=\{\ell_1,\ldots,\ell_n\}\cup\{\overline{\ell}_1,\ldots,\overline{\ell}_n\}$, 
        clauses $C\subseteq 2^{\vert L\vert}$ with each clause containing at most 3 literals.}
        \problemuniverse{$\U := L$.}
        \problemfeassol{-}
        \problemsol{The set of all sets $L^*\subseteq\U$ such that for all $i\in\{1,\ldots,n\}$ 
        we have $\vert L^*\cap\{\ell_i,\overline{\ell}_i\}\vert=1$, and $\vert L^*\cap C_j\vert\geq1$ for all $C_j\in C, j\in \{1,\ldots,\vert C\vert\}$.}
    \end{problem}
}

\newcommand{\exOne}{
    \begin{problem}
        \problemtitle{Exactly-One-3-SAT}
        \probleminstance{Literal set $L=\{\ell_1,\ldots,\ell_n\}\cup\{\overline{\ell}_1,\ldots,\overline{\ell}_n\}$, 
        clauses $C\subseteq 2^{\vert L\vert}$ with each clause containing at most 3 literals.}
        \problemuniverse{$\U := L$.}
        \problemfeassol{-}
        \problemsol{The set of all sets $L^*\subseteq\U$ such that for all $i\in\{1,\ldots,n\}$ 
        we have $\vert L^*\cap\{\ell_i,\overline{\ell}_i\}\vert=1$, and $\vert L^*\cap C_j\vert=1$ 
        for all $C_j\in C, j\in \{1,\ldots,\vert C\vert\}$.}
    \end{problem}
}

\newcommand{\partition}{
    \begin{problem}
        \problemtitle{Partition}
        \probleminstance{Set of numbers $A=\{a_1,\ldots,a_n\}\in\NN$ with weights $w(a_i)\in\NN$.}
        \problemuniverse{$\U := A$.}
        \problemfeassol{-}
        \problemsol{The set of all sets $A^*\subseteq \U$ such that $a_n\in A^*$ 
        and $\sum_{a_i\in A^*}a_i=\sum_{a_j\notin A^*} a_j$.}
    \end{problem}
}

\newcommand{\TSP}{
    \begin{problem}
        \problemtitle{Traveling Salesperson Problem}
        \probleminstance{Complete graph $G=(V,E)$, edge weight function $w:E\rightarrow\ZZ$, number $k\in\NN$.}
        \problemuniverse{Edge set $E=:\U$.}
        \problemfeassol{The set of all TSP tours $T\subseteq E$, thus the set of all Hamiltonian cycles of the graph.}
        \problemsol{The set of all feasible $T$ with $w(T)\leq k$.}
    \end{problem} 
}

\newcommand{\combiAuction}{
    \begin{problem}
        \problemtitle{Combinatorial Auction}
        \probleminstance{Numbers $n,k\in\NN$, list of pairs $P=\{(S_i,x_i)\}_{i=1}^m$, where $S_i\subseteq\{1,\ldots,n\}, x_i\in\NN$.}
        \problemuniverse{The set of sets $\{S_1,\ldots, S_m\}=:\U$.}
        \problemfeassol{The set of all disjoint sets $S^*\subseteq\{S_1,\ldots, S_m\}$.}
        \problemsol{The set of all feasible solutions $S^*$ with $\sum_{j=1}^l x_{ij}\geq k$.}
    \end{problem}
}

\newcommand{\sequencing}{
    \begin{problem}
        \problemtitle{Sequencing}
        \probleminstance{The number of jobs $p$, an execution time vector$(T_1,\ldots,T_p)\in\ZZ^p$, 
        a deadline vector $(D_1,\ldots, D_p)\in\ZZ^p$, a penalty vector $(P_1,\ldots,P_p)\in\ZZ^p$ 
        and a number $k\in\NN$.}
        \problemuniverse{$\U:=\{1,\ldots,p\}$.}
        \problemfeassol{All permutations $\pi$ of $\{1,\ldots,p\}$.}
        \problemsol{All permutations $\pi$ of $\{1,\ldots,p\}$ such that 
        $(\sum_{j=1}^p(\text{ if } T_{\pi(1)}+\cdots+T_{\pi(j)}>D_{\pi(j)} \text{ then } P_{\pi(j)} \text{ else } 0))\leq k$.}
    \end{problem}
}

\newcommand{\graphColoring}{
    \begin{problem}
        \problemtitle{\textsc{Graph Coloring}}
        \probleminstance{Graph $G=(V,E)$, number $k\in\NN$.}
        \problemuniverse{$\U:=V\times \{1,\ldots, k\}$.}
        \problemfeassol{-}
        \problemsol{The set of all $A=\{(v,i)\}\subset V\times \{1,\ldots,k\}$ such that 
        for each $(v,i),(w,j)\in A$ it holds that $v\neq w$ and $i\neq j$ if $(v,w)\in E$.}
    \end{problem}
}

\tikzset{
  dot/.style={circle, fill, inner sep=1pt, label=#1},
  solutionnode/.style={
    circle, fill, inner sep=1pt, label=#1, fill=rwth-blue
    }}
\newcommand{\VCexample}{
    \textsc{Vertex Cover}\\ instance:\\
    \begin{tikzpicture}[on grid, auto, node distance=2cm]
        \node[dot=left:$a$, color=rwth-blue] (a) {};
        \node[dot=right:$b$] (b) [right=of a] {};
        \node[dot=left:$c$, color=rwth-blue] (c) [below=of a] {};
        \node[dot=right:$d$] (d) [right=of c] {};

        \path[]
        (a) edge node {1} (b)
        (a) edge node {2} (c)
        (c) edge node {3} (d);
    \end{tikzpicture}
    ~\\ $k=2$
}

\newcommand{\ECexample}{
    \textsc{Exact Cover} instance:\\
    Elements: $\{a,b,c,d,e\}$\\
    Set of subsets $\mathfrak{S}$:
    \begin{align*}
        \{ \fcolorbox{rwth-blue}{white}{$S_1$}&=\{a,b,c\},\\
        S_2&=\{a,c,d\},\\
        \fcolorbox{rwth-blue}{white}{$S_3$}&=\{d,e\},\\
        S_4&=\{a,d\}\}
    \end{align*}
}

\newcommand{\SATexample}{
    \textsc{SAT} instance:\\
    $\phi = (x_1\vee x_2 \vee \overline{x}_3) \wedge (\overline{x}_1\vee x_2)\wedge(\overline{x}_1\vee \overline{x}_4)$\\
    Example solution:
    \begin{align*}
        \fcolorbox{rwth-blue}{white}{$x_1$}&=1\\
        \fcolorbox{rwth-blue}{white}{$x_2$}&=1\\
        x_3&=0\\
        x_4&=0
    \end{align*}
}

\newcommand{\threeSATexample}{
    \textsc{3-SAT} instance:\\
    $\phi = (\fcolorbox{rwth-blue}{white}{$x_1$}\vee \fcolorbox{rwth-blue}{white}{$x_2$} \vee \overline{x}_3) \wedge (\overline{x}_1\vee \fcolorbox{rwth-blue}{white}{$x_3$}\vee x_4)$\\
    Example solution:
    \begin{align*}
        x_1&=1\\
        x_2&=1\\
        x_3&=1\\
        x_4&=0
    \end{align*}
}

\newcommand{\threeDimexample}{
    \textsc{3-Dimensional Matching} instance:\\
    \begin{align*}
        X=&\{
            x_1(1), \overline{x}_1(1), x_1(2), \overline{x}_1(2), \\
            &x_2(1), \overline{x}_2(1), x_2(2), \overline{x}_2(2), \\
            &x_3(1), \overline{x}_3(1), x_3(2), \overline{x}_3(2), \\
            &x_4(1), \overline{x}_4(1), x_4(2), \overline{x}_4(2)
        \}\\
        Y=&\{
            a_1(1), a_1(2), a_2(1), a_2(2), \\
            &a_3(1), a_3(2), a_4(1), a_4(2)
        \}\\
        \cup&\{
            s_1(1), s_1(2)
        \}\\
        Z=&\{
            b_1(1), b_1(2), b_2(1), b_2(2), \\
            &b_3(1), b_3(2), b_4(1), b_4(2)
        \}\\
        \cup&\{
            s_2(1), s_2(2)
        \}\\
        k=&2(4+1)=10
    \end{align*}
}

\newcommand{\threeDimPic}{
    \resizebox{\textwidth}{!}{
    \begin{tikzpicture}
        \literalGadget{1}{2}{0}{4}{90}
        \literalGadget{2}{2}{5}{0}{0}
        \literalGadget{3}{2}{0}{-4}{180}
        \literalGadget{4}{2}{-5}{0}{0}
        
        \node at (1,1) (s_11) {$s_1(1)$};
        \node at (-1,1) (s_12) {$s_1(2)$};
        \node at (1,-1) (s_21) {$s_2(1)$};
        \node at (-1,-1) (s_22) {$s_2(2)$};
    
        \draw[draw, fill=rwth-red, fill opacity=0.1] \hedgem{x11}{s_11}{s_21}{6mm};
        \draw[draw, fill=rwth-red, fill opacity=0.1] \hedgem{x21}{s_21}{s_11}{6mm};
        \draw[draw, fill=rwth-red, fill opacity=0.1] \hedgem{notx31}{s_11}{s_21}{6mm};
        \draw[draw, fill=rwth-red, fill opacity=0.1] \hedgem{notx12}{s_12}{s_22}{6mm};
        \draw[draw, fill=rwth-red, fill opacity=0.1] \hedgem{x32}{s_12}{s_22}{6mm};
        \draw[draw, fill=rwth-red, fill opacity=0.1] \hedgem{x42}{s_12}{s_22}{6mm};
    \end{tikzpicture}
    }
}

\newcommand{\literalGadget}[5]{
    \begin{scope}[shift={(#3,#4)}, rotate around={#5:(0,0)}]
        \foreach \a in {1,...,#2}{
            
            \node (x#1\a) at (\a*360/#2: 2cm) {$x_{#1}(\a)$};
            \node (notx#1\a) at (\a*360/#2+180/#2: 2cm) {$\overline{x}_{#1}(\a)$};
            \node (a#1\a) at (\a*360/#2+90/#2: 1cm) {$a_{#1}(\a)$};
            \node (b#1\a) at (\a*360/#2+180/#2+90/#2: 1cm) {$b_{#1}(\a)$};
        }
    \end{scope}
    \foreach \a in {1,...,#2}{
        \pgfmathsetmacro{\prev}{int(\intcalcMod{\a-2}{#2}+1)}
        \draw[draw, fill=rwth-blue, fill opacity=0.1] \hedgem{notx#1\a}{a#1\a}{b#1\a}{6mm};
        \draw[draw] \hedgem{x#1\a}{b#1\prev}{a#1\a}{6mm};
    }
}

\newcommand{\literalGadgetSolution}[7]{
    \begin{scope}[shift={(#3,#4)}, rotate around={#5:(0,0)}]
        \foreach \a in {1,...,#2}{
            
            \node (x#1\a) at (\a*360/#2: 2cm) {$x_{#1}(\a)$};
            \node (notx#1\a) at (\a*360/#2+180/#2: 2cm) {$\overline{x}_{#1}(\a)$};
            \node (a#1\a) at (\a*360/#2+90/#2: 1cm) {$a_{#1}(\a)$};
            \node (b#1\a) at (\a*360/#2+180/#2+90/#2: 1cm) {$b_{#1}(\a)$};
        }
    \end{scope}
    \foreach \a in {1,...,#2}{
        \pgfmathsetmacro{\prev}{int(\intcalcMod{\a-2}{#2}+1)}
        \ifthenelse{\equal{#6}{1}}{
            \draw[draw] \hedgem{notx#1\a}{a#1\a}{b#1\a}{6mm};
            \draw[draw] \hedgem{x#1\a}{b#1\prev}{a#1\a}{6mm};
        }{
            \draw[draw, fill=rwth-blue, fill opacity=0.1] \hedgem{notx#1\a}{a#1\a}{b#1\a}{6mm};
            \draw[draw] \hedgem{x#1\a}{b#1\prev}{a#1\a}{6mm};
        }
        
    }
}

\def\rotateclockwise#1{
  \newdimen\xrw
  \pgfextractx{\xrw}{#1}
  \newdimen\yrw
  \pgfextracty{\yrw}{#1}
  \pgfpoint{\yrw}{-\xrw}
}

\def\rotatecounterclockwise#1{
  \newdimen\xrcw
  \pgfextractx{\xrcw}{#1}
  \newdimen\yrcw
  \pgfextracty{\yrcw}{#1}
  \pgfpoint{-\yrcw}{\xrcw}
}

\def\outsidespacerpgfclockwise#1#2#3{
  \pgfpointscale{#3}{
    \rotateclockwise{
      \pgfpointnormalised{
        \pgfpointdiff{#1}{#2}}}}
}

\def\outsidespacerpgfcounterclockwise#1#2#3{
  \pgfpointscale{#3}{
    \rotatecounterclockwise{
      \pgfpointnormalised{
        \pgfpointdiff{#1}{#2}}}}
}

\def\outsidepgfclockwise#1#2#3{
  \pgfpointadd{#2}{\outsidespacerpgfclockwise{#1}{#2}{#3}}
}

\def\outsidepgfcounterclockwise#1#2#3{
  \pgfpointadd{#2}{\outsidespacerpgfcounterclockwise{#1}{#2}{#3}}
}

\def\outside#1#2#3{
  ($ (#2) ! #3 ! -90 : (#1) $)
}

\def\cornerpgf#1#2#3#4{
  \pgfextra{
    \pgfmathanglebetweenpoints{#2}{\outsidepgfcounterclockwise{#1}{#2}{#4}}
    \let\anglea\pgfmathresult
    \let\startangle\pgfmathresult

    \pgfmathanglebetweenpoints{#2}{\outsidepgfclockwise{#3}{#2}{#4}}
    \pgfmathparse{\pgfmathresult - \anglea}
    \pgfmathroundto{\pgfmathresult}
    \let\arcangle\pgfmathresult
    \ifthenelse{180=\arcangle \or 180<\arcangle}{
      \pgfmathparse{-360 + \arcangle}}{
      \pgfmathparse{\arcangle}}
    \let\deltaangle\pgfmathresult

    \newdimen\x
    \pgfextractx{\x}{\outsidepgfcounterclockwise{#1}{#2}{#4}}
    \newdimen\y
    \pgfextracty{\y}{\outsidepgfcounterclockwise{#1}{#2}{#4}}
  }
  -- (\x,\y) arc [start angle=\startangle, delta angle=\deltaangle, radius=#4]
}

\def\corner#1#2#3#4{
  \cornerpgf{\pgfpointanchor{#1}{center}}{\pgfpointanchor{#2}{center}}{\pgfpointanchor{#3}{center}}{#4}
}

\def\hedgem#1#2#3#4{
  
  \outside{#1}{#2}{#4}
  \pgfextra{
    \def\hgnodea{#1}
    \def\hgnodeb{#2}
  }
  foreach \c in {#3} {
    \corner{\hgnodea}{\hgnodeb}{\c}{#4}
    \pgfextra{
      \global\let\hgnodea\hgnodeb
      \global\let\hgnodeb\c
    }
  }
  \corner{\hgnodea}{\hgnodeb}{#1}{#4}
  \corner{\hgnodeb}{#1}{#2}{#4}
  -- cycle
}

\begin{document}
\maketitle
\begin{abstract}
  SSP reductions are a type of polynomial reductions that also preserve the solutions of the instances.
  This means that there is a mapping from every solution in the original instance to a solution in the 
  instance which is reduced to, such that from every solution to the instance that is reduced to, 
  one can directly deduce a solution to the original instance. 
  SSP reductions can be used to show SSP-NP completeness of a problem, which is interesting because it has
  been proven that the network interdiction variant of an SSP-NP-complete problem is $\Sigma_2^p$-complete,
  as well as the min-max regret optimization variant.
  These variants are min-max optimization problems, which are optimization problems in which the objective
  is to minimize the maximum outcome. An example that highlights the interest of this topic is a power network
  that supplies a large amount of people with electricity, and therefore is prone to be attacked by terrorists.
  Lets say the government wants to minimize the damage that can be achieved by the terrorists, e.g. by protecting 
  the electricity poles which were found to be most important for distributing power.
  Most theoretical computer scientists assume min-max optimization problems to generally be $\Sigma_2^p$-complete,
  and with the help of the SSP-framework, this was shown for the two specific subareas of network interdiction
  and min-max regret robust optimization.

  This paper is devoted to the exploration of SSP reductions and the structured 
  collection of SSP-NP-complete problems as well as SSP reductions. 
  Additionally, a compendium website is developed, in which the SSP-NP-complete problems as well as SSP reductions
  are displayed in a graph, so that it becomes easy to understand the relations between the problems.
  The website is also extendable to include problems from other complexity classes and reductions that are
  not SSP reductions, such as gap-preserving reductions, fixed parameter tractable reductions or fine-grained reductions.
  Overall, 19 new SSP reductions are found, resulting in eight
  new SSP-NP-completeness proofs, and the compendium website is developed to enable 
  theoretical computer scientists to easily access the results and provide a knowledge base for future research.
\end{abstract}

\section*{Acknowledgements}
I would like to thank my supervisor Christoph Grüne for his amazing support and guidance throughout the project.
He never hesitated to help me with any questions or proofreading my work, and I am very grateful for that.
I also want to thank my parents and Emil Houben for providing me feedback and knowledge.

\tableofcontents
\chapter{Introduction}\label{chap:intro}
\section{Motivation}
A current topic in the theoretical computer science community is the study of robust and bilevel optimization.
Both areas are part of the larger area of min-max optimization, but with different focuses.
Robust optimization aims at finding solutions that are robust against uncertainties in the input data,
and bilevel optimization contains one optimization problem within another one, thus, there are two, possibly conflicting,
decision makers that have an hierarchical relationship, and each wants to optimize their problem.
An example for a bilevel optimization problem is the question of which
electricity poles are most vulnerable to an attack, because they are the most important ones for some power network.
In this scenario, one wants to minimize the possible damage an attack can result in by protecting the most vital 
poles, and the attacker wants to maximize the damage with the constraints that some poles are protected and thus 
cannot be attacked.
The overall objective in this case is to minimize the maximum damage that can be achieved by the attacker.
This idea can also be viewed as a game between an attacker and a defender, where the defender tries to minimize 
the maximal solution.
First, the defender can delete a given number of elements from the given instance, and then the attacker tries
to maximize the objective with respect to the modifications the defender made.
An example for a min-max optimization problem is the min-max variant of the well-known clique problem, 
where the goal is to find the largest clique in a given graph.
In the min-max variant of clique with a graph $G$ and a number $k\in\NN$ given, the attacker wants to find a clique
as large as possible, thus \textit{maximizing} the objective, and the defender wants to \textit{minimize} 
the largest clique possible. The defender can delete up to $k$ vertices from the graph, and then the attacker 
looks for the largest clique in the modified graph.

Min-max optimization is interesting because it has many practical applications~\citeLiterature{Sinha2017}, 
as for example in traffic management, where the number of people that can be transported should be maximized,
and the individual people try to minimize their personal travel time in parallel,
or in defense mechanisms, which the power network attack problem described above is an example of. 
Min-max optimization can also be used for modeling for many economical questions, 
in which some revenue is to be maximized, lets say the pricing
of a product must be calculated, and the company wants to maximize the revenue, whilst the customers
try to minimize their costs, et cetera. Generally, real-world problems that include a hierarchical relationship
between two decision levels can be modeled as bilevel optimization problems~\citeLiterature{colson2007overview}.
Thus, the complexity of these problems is of interest to many different fields.

It is quite likely that min-max optimization variants of NP-complete problems are 
$\Sigma_2^p$-complete, which means that they are at least as hard to solve as problems in the complexity class NP,
and it is assumed that they are strictly harder to solve than NP-complete problems.
$\Sigma_2^p$ is a level in the polynomial hierarchy, in which P and NP are contained as well.
The polynomial hierarchy consists of a complexity class $\Sigma_k^p$ for every $k\in\NN$, with $\Sigma_0^p$ being
better known as $P$, and $\Sigma_1^p$ as $NP$. 
It is widely believed that each level of the polynomial hierarchy is strictly contained 
in the next higher level, but the question whether the polynomial hierarchy collapses or not is still open.
Thus, if the polynomial hierarchy does not collapse to a level below $\Sigma_2^p$, these $\Sigma_2^p$-complete 
problems are even harder to solve than NP-complete problems.

However, although P and NP have been studied a lot, and there are not many known complete problems for the 
complexity class $\Sigma_2^p$. 
This is where the concept of SSP-NP-completeness comes into play, as with this concept,
Grüne and Wulf~\citeLiterature{grune2023large} were able to find 70 new $\Sigma_2^p$ and $\Sigma_3^p$-complete problems
at once, and these results are going to be extended in this paper.
For some SSP-NP-complete problem, the network interdiction variant as well as the 
min-max regret robust optimization variant of the problem are known to be $\Sigma_2^p$-complete,
and the two-stage adjustable optimization variant is known to be $\Sigma_3^p$-complete, 
which is shown in~\citeLiterature{grune2023large}.

\section{Related Work}
Subset Search Problems (SSP) were introduced by Grüne and Wulf in 2023~\citeLiterature{grune2023large}.
They introduce SSP problems and reductions and prove that for SSP-NP-complete problems, 
the network interdiction variants as well as the min-max regret robust optimization variants are $\Sigma_2^p$-complete,
and the two-stage adjustable optimization variants even $\Sigma_3^p$-complete.

Background knowledge on the complexity class NP can be found in Garey and Johnson's guide on NP-completeness
and Arora and Barak's book on computational complexity~\citeLiterature{garey1979computers,arora2009computational},
and on the polynomial hierarchy in the paper by Stockmeyer from 1976, where he introduced the polynomial hierarchy~\citeLiterature{Stockmeyer1976hierarchy}.

There are many compendiums on problems that are NP-, or P-complete,
such as the book on P-completeness by Greenlaw, Hoover and Ruzzo written in 1995~\citeLiterature{greenlaw1995limits},
which can be recommended due to its entertaining writing style, and of course contains many P-completeness results.
The book on NP-completeness by Garey and Johnson~\citeLiterature{garey1979computers} is a compendium on NP-complete 
problems, was written in 1979 and is still relevant today,
and Karp's paper on the computational complexity of combinatorial problems from 1975~\citeLiterature{karp1975}
contains 21 NP-complete problems.
Furthermore, the article by Schaefer and Umans~\citeLiterature{schaefer2002completeness} from 2008 
contains a list of problems that are complete for 
different classes within the polynomial hierarchy, as well as additional hardness and approximation results.

In the field of min-max optimization and the completeness of these problems,
the paper on complexity of robust multi-stage problems by Goerigk, Lendl and Wulf from 2022~\citeLiterature{goerigk2023complexity}
proved the $\Sigma_3^p$-completeness of some robust two-stage adjustable and robust recoverable problems,
and Coco, Santos and Noronha showed that the min-max regret Maximum Benefit Set Covering Problem is $\Sigma_2^p$-hard~\citeLiterature{Coco2022computational}.
Deineko and Woeginger proved a natural robust optimization variant of the knapsack problem to be complete for the second level of the polynomial hierarchy
in~\citeLiterature{Deineko2010pinpointing}.
Johannes introduced reductions between $\Sigma_2^p$-complete problems to find new $\Sigma_2^p$-complete problems
in~\citeLiterature{johannes2011new} in 2011, thus using a similar approach to Grüne and Wulf.

In the field of bilevel optimization, the thesis by Dempe and Zemkoho on bilevel optimization~\citeLiterature{dempe2020bilevel}
is a good starting point that introduces bilevel optimization and its applications and different variants,
as well as the shorter overview of bilevel optimization by Colson, Marcotte and Savard~\citeLiterature{colson2007overview}.
The book~\citeLiterature{bard2013practical} by Bard is on algorithms for bilevel optimization problems,
and also provides a lot of background knowledge. In 1995, Migdalas wrote an article on bilevel optimization concentrated
on traffic planning~\citeLiterature{migdalas1995bilevel}, which is a very intuitive and practical example for 
bilevel optimization.

On robust optimization, Ben-Tal, El Ghaoui and Nemirovski wrote a book in 2009~\citeLiterature{ben-Tal2009robust},
and part three of the book handles multi-stage robust optimization.
Gabrel, Murat and Thiele wrote a compendium on the research results in the field of robust optimization between
2007 and 2014~\citeLiterature{gabrel2014recent}, and Gorissen, Yanıkoğlu and den Hertog wrote about how to apply
robust optimization in practice~\citeLiterature{Gorissen2015practical}.\\

There are also already some compendium websites online,
however, none of the websites we could find uses the approach of visualizing the problems and reductions in a graph.
The website on NP optimization problems by Kann and Crescenzi~\citeWebsites{npWebsite}, which focuses on
approximability results for NP optimization problems, as well as the Complexity Zoo~\citeWebsites{complexityZoo},
that contains a large collection of complexity classes and what is known about them,
are both organized as a collaborative website where users can contribute content, but in difference to the 
website developed in this work, the entries are displayed on different pages and the connection between them
are not graphically showed.
House of Graphs~\citeWebsites{houseOfGraphs} contains nice navigation tools, where one can select many
parameters and gets graphs fulfilling these, and it aims at collecting graphs that are interesting for researchers.
One last website, which is not related to this work by its content, but is interesting for anyone interested
in programming is The Arcane Algorithm Archive~\citeWebsites{algorithmArchive}, 
which has the goal to provide a guide for important and often used algorithms for all languages,
and is also organized as a collaborative platform.

\section{Contributions}
The objective of this paper is to find more problems and reductions to which the SSP concept can be applied,
and to develop a website on which the results are displayed in a structured way.
In total, eight new SSP-NP-complete problems are found, 
and 19 new SSP reductions are presented.
The reductions were found by examining many already existing polynomial reductions between NP-complete problems for 
the SSP property, and modifying them to achieve the SSP property, either only slightly by adding the solution mapping
function, or some reductions needed to be adapted more drastically to achieve the SSP property.
The reductions that are examined and presented in this paper are the reductions from Karp's 
famous paper on the computational complexity of combinatorial problems~\citeLiterature{karp1975},
as well as the six reductions presented by Garey and Johnson in their book on NP-completeness~\citeLiterature{garey1979computers},
and some reductions from the book on computational complexity by Arora and Barak~\citeLiterature{arora2009computational}.
Most of the reductions also needed some modifications to fit into the SSP framework,
and for all reductions, the function mapping the solutions of the original instances to the solutions
of the instances they are reduced to was added.
The problems that were shown to be SSP-NP-complete by this work the first time are \textsc{0-1-Integer Programming},
\textsc{Exact Cover}, \textsc{Set Packing}, \textsc{Undirected Hamiltonian Path}, \textsc{Exactly-One-3SAT},
\textsc{combinatorial Auction}, \textsc{3-Dimensional Matching} and \textsc{3-Dimensional Exact Matching}.\\

Additionally, a compendium website was developed and filled with the current known results on SSP-NP-completeness.
The website displays the problems and reductions between them in a graph, such that the relations between
the problems are easily visible and structured. In the graph, the nodes represent the problems, and the directed edges
correspond to the reductions between them. Further information on the problems and reductions are shown
by clicking on the corresponding graph element, and the website is designed to be extendable for other complexity
classes or types of reductions as well, such as fine-grained reductions, gap-preserving reductions,
or fixed parameter tractable reductions. 
This way, the use case extends further and the website can be used as a general knowledge base for reductions.
The vision is to have multiple graphs for different purposes, and at the current time,
the website contains the main structure and one graph for SSP reductions.
It is designed to be usable by researchers who want to contribute to or learn about SSP reductions,
and everyone can submit new reductions or make requests to change entries, for example if errors are detected.

\section{Structure}
This work is structured into two parts, the theoretical part and the documentation of the website.

\Cref{chap:theoretical} contains the theoretical background knowledge,
as well as the results of the examination of the reductions for the SSP property, both the successful ones
as well as reductions that did not appear to be easily transformable into SSP reductions.

In the preliminaries (\cref{sec:preliminaries}), Subset Search Problems and the main concepts around
reductions are explained, and in~\cref{subsec:reductions}, 19 SSP reductions are presented,
which are graphically displayed in a tree in~\cref{fig:reductiontree}.
The reductions are structured to separate the reduction function $g$ from the mapping function $f$,
and the correctness of the reductions are proven in an own paragraph.

Next, in~\cref{subsec:noreductions}, reductions tat were examined for the SSP property
but could not be transformed into SSP reductions are presented, and for understanding the concept
of SSP reductions, it can be helpful to have a look at the reduction from \textsc{Clique} to \textsc{Independent Set} in~\cref{subsubsec:clique-is}.
\Cref{subsec:excludedProblems} explains why some problems do not need to be considered in the SSP context,
and shows this for the \textsc{Graph Coloring} problem.\\

In \cref{chap:documentation}, the compendium website is documented.
First, in~\cref{sec:concept}, an overview of the ideas behind and the concept of the website is given,
and after that in~\cref{sec:structure}, the structure and architecture of the website are explained
and accessibility concerns are addressed in~\cref{sec:accessibility}.
Then,~\cref{sec:libraries} gives an overview of the external libraries used in the project,
why they were chosen and what their purpose in the context of the website is.

Lastly, some crucial functions from the code are presented and explained in~\cref{sec:code},
and the work is concluded in~\cref{chap:conclusion}.

\chapter{Theoretical Overview}\label{chap:theoretical}
\section{Preliminaries}
\label{sec:preliminaries}
In this work, we denote the power set of some set $S$ as $2^S$.\\

For understanding the concept of SSP-NP-completeness, we first need to introduce some basic concepts.
A language is a set $L\subseteq{\{0,1\}}^*$.
\begin{definition}[Subset Search Problem (SSP)]
    A Subset Search Problem, short SSP, is a tuple $\Pi=(\I, \U, \S)$, such that
    \begin{itemize}
        \item $\I\subseteq{\{0,1\}}^*$ is a language,
        \item for each instance $I\in \I$, there is some set $\U(I)$, the universe of $I$,
        \item and for each instance $I\in \I$, there is a (potentially empty) set
        $\S(I)\subseteq 2^{\U(I)}$, 
        which is called the solution set associated to the instance $I$. 
    \end{itemize}
\end{definition}
\begin{definition}[Polynomial Reduction]
    A polynomial reduction is a mapping $g:{\{0,1\}}^*\rightarrow{\{0,1\}}^*$ computable in polynomial time 
    from one language $L$ to another language $L'$ 
    such that for all $x\in {\{0,1\}}^*:\: x\in L \Leftrightarrow g(x)\in L'$. 
    The language $L$ is polynomially reducible to $L'$, denoted as $L\leq_p L'$, if there is a polynomial reduction from $L$ to $L'$.
\end{definition}

A function $g$ is computable in polynomial time if on input $x$, it only takes $O(\vert x\vert^c)$ 
steps on a Turing machine to compute the function $g$, with $c$ being some constant.\\
SSP reductions, denoted as $L\SSP{}L'$, are polynomial reductions with an additional property.
This property is that any solution of $L'$ directly translates into a solution for $L$, via the mapping function $f$.
This is formalized as follows: 
\begin{definition}[SSP-Reduction]
    Let $\Pi=(\I,\U,\S)$, $\Pi'=(\I',\U',\S')$ be two SSP-problems. We say there is an SSP reduction from $\Pi$ to $\Pi'$ ($\Pi\SSP\Pi'$) if
    \begin{itemize}
        \item $\Pi\leq_p\Pi'$ (i.e. $\S(I)\neq\emptyset$ if and only if $\S'(g(I))\neq\emptyset$, 
        with $g$ being the reduction function) and
        \item for every instance $I\in\I$ there exist a function $f_I$ such that
        \begin{itemize}
            \item $f_I$ is computable in polynomial time in $\vert I\vert$ and
            \item for all instances $I\in \I$, the function $f_I: \U(I)\rightarrow\U'(g(I))$ is an 
            injective function
            mapping from the universe of the instance $I$ to the universe of the instance $g(I)$ such that 
            \[\{f_I(S)\mid S\in\S(I)\}=\{S'\cap f_I(\U(I))\mid S'\in\S'(g(I))\}\]
        \end{itemize} 
    \end{itemize}
\end{definition}
This means that the image of a solution $S$ of $I$ under the function $f_I$ is exactly a solution of $g(I)$
intersected with the image of the function $f_I$, thus, the function $f_I$ maps a solution of $I$ to a solution
of $g(I)$. Intuitively, this property can be described as the solution-preserving
property of the reduction.\\

SSP reductions are transitive, thus, if $\Pi\SSP\Pi'$ and $\Pi'\SSP\Pi''$, then $\Pi\SSP\Pi''$.\\

Decision problems are classified into so-called complexity classes, 
which consist of problems that can be solved within some time or space bounds. 
From here on, we only concern ourselves with the classes NP and SSP-NP.
NP is the class of languages $L$ that are accepted by a non-deterministic Turing machine (TM) within polynomial time. 
This is equivalent to saying $L$ is verifiable in polynomial time, thus:\\
\begin{definition}[NP, NP-hardness, NP-completeness]
    A language $L\subseteq\{0,1\}^*$ is in NP if there exists a polynomial $p:\NN\rightarrow\NN$ and a polynomial-time
    TM $V$, which is called a verifier for $L$, such that for every $x\in\{0,1\}^*$: 
    \[x\in L\Leftrightarrow\exists y \in{\{0,1\}}^{p(\vert x\vert)}, \text{ such that } V\text{ accepts }y\#x.\]
    A language $L$ is NP-hard if and only if every problem $L'\in$ NP is polynomially reducible to $L$,
    and $L$ is NP-complete if it is NP-hard and in NP.
\end{definition}

The terms for SSP problems are defined analogously:
A problem lies in SSP-NP if it is an SSP-problem, that is also verifiable in polynomial time, hence, 
intuitively, it is the class NP regarding SSP-problems.\\
SSP-NP-hardness is defined as NP-hardness, but all problems $L'\in$ SSP-NP must be SSP-reducible to $L$, 
and the definition for SSP-NP-completeness is analogous.\\

The following conventions are used:
Undirected graphs are always denoted as $G=(V,E)$, directed graphs as $G=(V,A)$, and weights for edges by $c$. 
Both directed and undirected edges are denoted as $(v,w)$, where $v$ and $w$ are the vertices the edge is incident to.
Directed edges are also called arcs.
The neighborhood of a vertex $v\in V$, denoted as $N[v]$, is the set of all vertices adjacent to $v$, 
thus the set of all $v'\in V$ such that $(v,v')\in E$.
For problems with a threshold, we always use $k$ for the threshold.
Instances denoted with an apostrophe, e.g., $G'$, are the instance that is reduced to, 
and instances with an asterisk, e.g., $G^*$, are a solution to the instance $G$.

\section{Examining Reductions For the SSP Property}\label{sec:main}
This section is devoted to the examination of already existing reductions for the SSP property, and the
modification to those reductions to achieve the SSP property.
First, in~\cref{subsec:reductions}, reductions that proved to have the SSP property are listed and explained in detail, 
and afterwards, reductions that are no SSP reductions are composed in~\cref{subsec:noreductions} and it is 
argued why these reductions can not be easily transformed into SSP reductions.\\
However, some problems are hard to formulate as SSP problems intuitively, and quickly loose their meaning when being translated
into a problem on a higher complexity class. As one of the main focuses of SSP-NP-completeness proofs is, that the
min-max variants of the SSP-NP-complete problems are automatically $\Sigma_2^p$-complete, and the min-max-min variants
are even $\Sigma_3^p$-complete~\citeLiterature{grune2023large}, these problems are explicitly excluded in this paper. 
More detail on this will be given in~\cref{subsec:excludedProblems}.
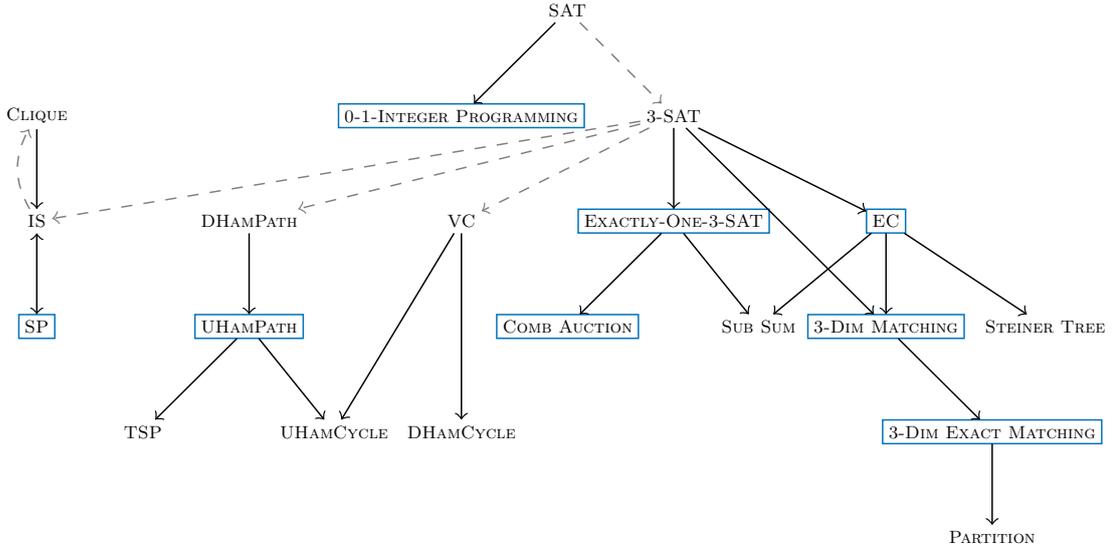
\begin{figure}
    \centering
    \resizebox{\textwidth}{!}{
        \begin{tikzpicture}[every node/.style={scale=0.4}, on grid]
            \node[] (sat) {\textsc{SAT}};
            \node[draw=rwth-blue] (01int) [below left=of sat] {\textsc{0-1-Integer Programming}};
            \node[] (3sat) [below right=of sat] {\textsc{3-SAT}};
            \node[] (vc) [below left=1 and 2 of 3sat] {\textsc{VC}};
            \node[] (dirHamCyc) [below=2 of vc] {\textsc{DHamCycle}};
            \node[] (dirHamPath) [left=2 of vc] {\textsc{DHamPath}};
            \node[draw=rwth-blue] (undirHamPath) [below=of dirHamPath] {\textsc{UHamPath}};
            \node[] (tsp) [below left=of undirHamPath] {\textsc{TSP}};
            \node[] (undirHamCyc) [below right=1 and 0.8 of undirHamPath] {\textsc{UHamCycle}};

            \node[] (is) [left=2 of dirHamPath] {\textsc{IS}};
            \node[] (clique) [above=of is] {\textsc{Clique}};
            \node[draw=rwth-blue] (sp) [below=of is] {\textsc{SP}};
            \node[draw=rwth-blue] (ex13sat) [below=of 3sat] {\textsc{Exactly-One-3-SAT}};
            \node[draw=rwth-blue] (combAuct) [below left=of ex13sat] {\textsc{Comb Auction}};
            \node[] (subsum) [below right=1 and 0.8 of ex13sat] {\textsc{Sub Sum}};
            \node[draw=rwth-blue] (ec) [right=2 of ex13sat] {\textsc{EC}};
            \node[draw=rwth-blue] (3dimMatch) [below=of ec] {\textsc{3-Dim Matching}};
            \node[draw=rwth-blue] (ex3dimMatch) [below right=of 3dimMatch] {\textsc{3-Dim Exact Matching}};
            \node[] (partition) [below=of ex3dimMatch] {\textsc{Partition}};
            \node[] (steiner) [below right=1 and 1.5 of ec] {\textsc{Steiner Tree}};

            \path[<->]
                (is) edge node {} (sp);
            \path[->]
                (clique) edge node {} (is)
                (dirHamPath) edge node {} (undirHamPath)
                (undirHamPath) edge node {} (tsp)
                (undirHamPath) edge node {} (undirHamCyc)
                (vc) edge node {} (undirHamCyc)
                (vc) edge node {} (dirHamCyc)
                (sat) edge node {} (01int)
                (3sat) edge node {} (ex13sat)
                (3sat) edge node {} (3dimMatch)
                (ex13sat) edge node {} (combAuct)
                (ex13sat) edge node {} (subsum)
                (ec) edge node {} (3dimMatch)
                (ec) edge node {} (subsum)
                (ec) edge node {} (steiner)
                (3dimMatch) edge node {} (ex3dimMatch)
                (ex3dimMatch) edge node {} (partition)
                (3sat) edge node {} (ec);

            \path[->, color=gray, style=dashed]
                (sat) edge node {} (3sat)
                (3sat) edge node {} (vc)
                (3sat) edge node {} (is)
                (is) edge [bend left] node {} (clique)
                (3sat) edge node {} (dirHamPath);

        \end{tikzpicture}
    }
    \caption{The SSP reductions found in this paper. The gray dashed lines are reductions by Grüne and Wulf,
    that help to see how the problems are all SSP-NP-complete.}
    \label{fig:reductiontree}
\end{figure}

\subsection{SSP-Reductions}
\label{subsec:reductions}
In this section, newly found SSP reductions based on Karp's paper~\citeLiterature{karp1975},
Arora and Barak~\citeLiterature{arora2009computational}, and Gary and Johnson~\citeLiterature{garey1979computers} are presented,
using the SSP framework by Grüne and Wulf~\citeLiterature{grune2023large}.\\
\Cref{fig:reductiontree} shows all the reductions that are presented in this section.
The first reductions are based on the original NP-completeness proofs by Karp~\citeLiterature{karp1975}.
The reductions that did not show to be SSP reductions or easily modified 
to be one are not included in this section and can be found in \cref{subsec:noreductions}.
Some reductions are also examined in~\citeLiterature{grune2023large} and thus not presented in this paper.
For each reduction, an example is given, and where possible, a solution of the example is provided in the form
of blue boxes or blue marked elements.

\subsubsection{\textsc{Independent Set} \SSP{} \textsc{Set Packing}}
This reduction is based on \textsc{Clique} $\leq_p$ \textsc{Set Packing} by Karp~\citeLiterature{karp1975}, 
but is changed to reducing from \textsc{Independent Set}, as this simplifies the process by avoiding one step.

\IS
\setPacking

\functionG
Let $G=(V,E)$ be the graph of the \textsc{Independent Set} instance and $k$ the threshold.
The reduction $g:\{0,1\}^*\rightarrow\{0,1\}^*$ transforms this instance into a \textsc{Set Packing} 
instance $(\mathfrak{U}', \mathfrak{S}', k')$ as follows:
We transform each vertex $v_i\in V$ to two elements $s_i\in\mathfrak{U}'$ and $i\in\mathfrak{U}'$. 
For each vertex $v_i$, a subset $S_i\in \mathfrak{S}'$ is created that contains the index $i$ of the vertex $v_i$, $s_i$,
and the elements corresponding to the vertices $v_j$ adjacent to $v_i$, thus, $S_i=\{i, s_i\}\cup \{s_j\mid v_j\in N[v_i]\}$. 
We include $i$ in the set to ensure that each vertex is transformed into a different set.
Thus, an edge $(v_i, v_j)\in E$ implies that $S_i\cap S_j\neq\emptyset$, as then $s_i$ appears in $S_j$,
and two sets are disjoint if there is no edge $(v_i, v_j)\in E$.
Finally, we set $k'$ to $k$. The idea is shown in~\cref{fig:is-setpacking}.\\
The reduction is computable in polynomial time, as we only need to go
through the adjacency matrix $\vert V\vert$ times.

\functionF
$f_{(G,k)}(v_i)=S_i=\{i, s_i\}\cup\{s_j\mid v_j\in N[v_i]\}\in\mathfrak{S}'$
as $S_i$ appears in a solution of 
set packing if and only if $v_i$ appears in an independent set.

\correctness
If there is a subfamily $\mathfrak{C}\subseteq\mathfrak{S}'$ with $\vert\mathfrak{C}\vert\geq k'$
such that all sets in $\mathfrak{C}$ are pairwise disjoint, the corresponding vertices are pairwise not adjacent to each other
and form an independent set of size at least $k$.
If there is an independent set of size at least $k$ in $G$, all of the other vertices
do not appear in the corresponding sets, and therefore they form a set packing with 
a size greater or equal to $k'$. 
The embedding functions $f_I$ map to correct elements, as $S_i$ and $S_j$ are pairwise disjoint if and only if $v_i, v_j$ are not adjacent to each other.
Thus, we can see that the following holds:
\begin{align*}
    &\{f_{(G,k)}(S) \mid S\in \S(\textsc{Independent Set})\}\\
    =\:&\{f_{(G,k)}(\mathfrak{C}) \mid\mathfrak{C}\subseteq V, \vert \mathfrak{C}\vert\geq k, \text{ such that } \forall v_i, v_j\in \mathfrak{C}: (v_i,v_j)\notin E\}\\
    =\:&\{\mathfrak{S} \mid \mathfrak{S}=\{S_{i_1},\ldots, S_{i_{\ell}}\}, \ell\geq k \text{ such that } \forall i_j,i_{j'}\in\mathfrak{S}: S_{i_j}\text{ and }S_{i_{j'}}\text{ are pairwise}\\ &\text{disjoint}\}\\
    =\:&\{S'\mid S'\in \S(\textsc{Set Packing})\}\\
    =\:&\{S'\cap f_{(G,k)}(V)\mid S'\in \S(\textsc{Set Packing})\}
\end{align*}

\subsubsection{\textsc{Set Packing} \SSP{} \textsc{Independent Set}}
The other way around, the reduction works almost analogously.
Thus, all subsets $S\in\mathfrak{S}$ are viewed as vertices, and two vertices are adjacent to each other if their sets are not disjoint.
For example, if two sets $S$ and $S'$ both contain the element $e$, then the edge $(v_S, v_{S'})$ is in $E$.

\functionG
The reduction $g:\{0,1\}^*\rightarrow\{0,1\}^*$ from the \textsc{Set Packing} instance $(\mathfrak{U}, \mathfrak{S}, k)$
to the \textsc{Independent Set} instance $(G', k')$ is defined as follows:
For each set $S\in\mathfrak{S}$, we create a vertex $v_S\in V'$. The vertex $v_S$ gets connected to all vertices $v_{S'}$ 
such that $S\cap S'\neq\emptyset$, thus $(v_S, v_{S'})\in E$ if and only if $S\cap S'\neq\emptyset$.
We set $k'=k$.

\functionF
We define $f_{(\mathfrak{U}, \mathfrak{S}, k)}(S)=v_S$. 

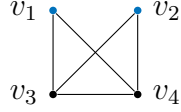
\begin{figure}
    \begin{minipage}[t]{0.49\textwidth}
        \textsc{Set Packing} instance:\\
        Universe $\mathfrak{U}=\{1,2,3\}$\\
        Subsets $\mathfrak{S} =\\
            \{\fcolorbox{rwth-blue}{white}{$\{1,2\}$},
            \fcolorbox{rwth-blue}{white}{$\{3\}$},
            \{2,3\},
            \{1,2,3\}\}$\\
        Number $k=3$
    \end{minipage}\hfill
    \begin{minipage}[t]{0.49\textwidth}
        \textsc{Independent Set} instance:\\
        \begin{tikzpicture}
            \node[solutionnode=left:$v_1$, fill=rwth-blue] (v_1) {};
            \node[solutionnode=right:$v_2$, fill=rwth-blue] (v_2) [right=of v_1] {};
            \node[dot=left:$v_3$] (v_3) [below=of v_1] {};
            \node[dot=right:$v_4$] (v_4) [right=of v_3] {};
    
            \path[]
            (v_1) edge node {} (v_3)
            (v_1) edge node {} (v_4)
            (v_2) edge node {} (v_3)
            (v_2) edge node {} (v_4)
            (v_3) edge node {} (v_4);
        \end{tikzpicture}
        ~\\$k'=3$
    \end{minipage}\hfill
    \caption{Set Packing instance and the corresponding graph $G'$.}
    \label{fig:setpacking-is}
\end{figure}

\correctness
The argument is analogous to the reduction \textsc{Independent Set} \SSP{} \textsc{Set Packing} before:
The running time is $O(n^2)$, as the subset needs to be compared to each other and then, a graph with $\vert\mathfrak{S}\vert$ vertices is created.
If there is a \textsc{Set Packing} instance of size at least $k$, the corresponding vertices are pairwise not connected, 
as there are no intersections between the sets. Thus, the vertices form an independent set of size at least $k$. 
If there is an \textsc{Independent Set} instance of size greater or equal to $k$, 
the corresponding sets form a \textsc{Set Packing} instance with the same argument. 
An example is shown in~\cref{fig:setpacking-is}.

The embedding functions $f_I$ are correct, as $v_i, v_j$ are not adjacent to each other if and only if $S_i, S_j$ are pairwise disjoint.
Thus, we can see that the following holds:
\begin{align*}
    &\{f_{(G,k)}(S) \mid S\in \S(\textsc{Set Packing})\}\\
    =\:&\{f_{(G,k)}(\mathfrak{C}) \mid 
    \mathfrak{S}=\{S_{i_1},\ldots, S_{i_{\ell}}\}, \ell\geq k \text{ such that } 
    \forall i_j,i_{j'}\in\mathfrak{S}: S_{i_j}\text{ and }S_{i_{j'}}\text{ are}\\
    &\text{pairwise disjoint}\}\\
    =\:&\{\mathfrak{S} \mid 
    \mathfrak{C}\subseteq V, \vert \mathfrak{C}\vert\geq k, \text{ such that } 
    \forall v_i, v_j\in \mathfrak{C}: (v_i,v_j)\notin E \}\\
    =\:&\{S'\mid S'\in \S(\textsc{Independent Set})\}\\
    =\:&\{S'\cap f_{(G,k)}(V)\mid S'\in \S(\textsc{Independent Set})\}
\end{align*}

\begin{figure}
    \begin{minipage}[t]{0.49\textwidth}
        \textsc{Independent Set} instance:\\
        \begin{tikzpicture}
            \node[solutionnode=left:$v_1$] (v_1) at (0,0) {};
            \node[solutionnode=left:$v_2$] (v_2) at (0,-2) {};
            \node[solutionnode=left:$v_3$] (v_3) at (2,-1) {};
            \node[dot=right:$v_4$] (v_4) [right=of v_3] {};
        
            \path[]
            (v_1) edge node {} (v_4)
            (v_2) edge node {} (v_4);
        \end{tikzpicture}
        ~\\$k=3$
    \end{minipage}\hfill
    \begin{minipage}[t]{0.49\textwidth}
        \textsc{Set Packing} instance:\\
        Universe $\mathfrak{U}'=\{1,2,3,4\}$\\
        Subsets $\mathfrak{S}' = \{\fcolorbox{rwth-blue}{white}{$\{1, v_1,v_4\}$},
        \fcolorbox{rwth-blue}{white}{$\{2,v_2,v_4\}$},\\
        \fcolorbox{rwth-blue}{white}{$\{3,v_3\}$}, 
        \{4, v_4, v_1, v_2\}  \}$\\
        Number $k'=k=3$
    \end{minipage}\hfill
    \caption{Graph $G$ with $k=3$ and the resulting Set Packing instance.}
    \label{fig:is-setpacking}
\end{figure}

\subsubsection{\textsc{SAT} \SSP{} \textsc{0-1-Integer Programming}}
\sat
The definition of the problem \textsc{0-1-Integer Programming} differs slightly from the definition Karp used in his paper~\citeLiterature{karp1975},
as this is the more common definition in literature:
\IP
The reduction from \textsc{SATISFIABILITY}, short \textsc{SAT}, to \textsc{0-1-Integer Programming} presented in 
Karp's paper~\citeLiterature{karp1975} cannot be transformed into a SSP reduction, which is explained in~\cref{subsec:noreductions}.
To still add \textsc{0-1-Integer Programming} to the known SSP-NP-complete problems,
another reduction is presented here.
In this reduction, we map a \textsc{SAT} instance $(L,C)$ to a \textsc{0-1-Integer Programming} instance $(C',b')$.
The matrix $C'$ has a row for each clause and additionally rows that ensure that for each variable,
either the positive or the negative literal is set to true, but never both.
The columns correspond to the literals of the \textsc{SAT} instance, and each clause row contains
$-1$ for each literal appearing in the corresponding clause, and zeros for the other literals.
The solution vector $X'$ corresponds to an assignment of the literals, and in the product $C'X'$,
for each row corresponding to a clause, the entry must be at most $-1$. This is because then, at least one literal in the
corresponding clause is satisfied, as each satisfied literal adds $-1$. 
It must be ensured that a variable can only be set \textit{either} to true \textit{or} false,
thus, the additional rows are added to the matrix $C'$, for each $x_i\in\phi$ a row containing a one at the
columns for $x_i$ and $\overline{x}_i$.
Then, in the product $C'X'$, the value in that row must be at most one, so that not
both $x_i$ and $\overline{x}_i$ can be set to true, and the same is done to ensure that for each variable $x_i$,
at least one literal $x_i$ or $\overline{x}_i$ is set to true.

\functionG
The function $g$ from the \textsc{SAT} instance $(L,C)$ to the \textsc{0-1-Integer Programming} instance $(C',b')$ is defined as follows:
\begin{align*}
    &\text{ for } i\leq \vert C\vert\text{ (clauses)}:\\
    c'_{ij}&=\begin{cases}
        -1 & \text{if } l_j\in C_i \\
        0 & \text{if } l_j\notin C_i
    \end{cases}\\
    b'_i&= -1\\
    \vspace{1em}\\
    &\text{ for } \vert C\vert<i\leq \vert C\vert+\vert L\vert \text{ (at most one of $x_i$ and $\overline{x}_i$ is true)}:\\
    c'_{ij}&=\begin{cases}
        1 & \text{if } l_j=x_i \text{ or } l_j=\overline{x}_i\\
        0 & \text{otherwise}
    \end{cases}\\
    b'_i&= 1\\
    \vspace{1em}\\
    &\text{ for } \vert C\vert + \vert L \vert < i \text{ (at least one of $x_i$ or $\overline{x}_i$ is true)}:\\
    c'_{ij}&=\begin{cases}
        -1 & \text{if } l_j=x_i \text{ or } l_j=\overline{x}_i\\
        0 & \text{otherwise}
    \end{cases}\\
    b'_i&= -1
\end{align*}

\functionF
The functions $f_I$ are defined as $f_{(L,C)}(\ell_i)=x'_i$.

\correctness
Let $(L,C)$ be a YES-instance of \textsc{SAT} and $L^*\subseteq L$ a solution.
Then, for each clause $C_i$, at least one literal is satisfied, and for each $i$, either $x_i$ or $\overline{x}_i$ is set to true.
We set the solution variables $Y'$ to $y'_i=f_{(L,C)}(\ell_i)$, which means that in the vector $X'$, $x_i=1$
if and only if $\ell_i\in L^*$.
Let $C_j$ be a satisfied clause. Then, the row $C'_j$ contains at least one $-1$ at some $c'_{ji}$
such that $X'_i=1$, and thus $x_j\leq -1\leq b'_j$.
For the additional rows, we know that, as $L^*$ is a satisfying assignment, for each $i$,
exactly one of the literals $x_i, \overline{x}_i$ is set to true.
Thus, for the rows that enforce that at least one of the literals is false,
the entry in the product $C'X'$ is at most one, and for the rows that enforce 
that at least one of the literals is true, the entry is at most $-1$.
Thus, the mapping function $f_{(L,C)}$ maps to a solution of the \textsc{0-1-Integer Programming} instance.\\

Let us assume that $(C',b')$ is a YES-instance of \textsc{0-1-Integer Programming}.
Then, $C'X'\leq b'$. Thus, for each clause row, the entry in the product $C'X'$ is at most $-1$,
which directly means that at the corresponding literal satisfies the clause.
The extra rows ensuring that at most one literal $x_i,\overline{x}_i$ is set to true
multiplied with the solution vector $X'$ are at most one, which means that for each row $\vert C\vert+i$,
at most one of the literals $\ell_{2i}, \ell_{2i+1}$ is set to true, and the same argumentation holds 
for the next rows. Thus, the solution $X'$ corresponds to a satisfying assignment of the \textsc{SAT} instance.\\

We have shown that the mapping function works correctly and the reduction is a correct SSP reduction:
\begin{align*}
    & \{f_{(L,C)}(S) \mid S\in\S(\textsc{SAT})\}\\
    =&\{f_{(L,C)}(L^*)\mid L^*\subset L \text{ such that }\vert L^*\cap\{\ell_i,\overline{\ell}_i\}\vert=1 \text{ and }\vert L^*\cap C_j\vert\geq1\text{ for all }C_j\in C\} \\
    =&\{ x'_i \mid \ell_i\in L^* \}\\
    =&\{X'\in\{0,1\}^n\mid C'X'\leq b' \}\\
    =&\{S'\mid S'\in \S(\textsc{0-1-Integer Programming})\}\\
    =&\{S'\cap f_{(L,C)}(L)\mid S'\in \S(\textsc{0-1-Integer Programming})\}
\end{align*}

\begin{figure}
    \centering
    \vbox{
    \begin{minipage}[t]{0.24\textwidth}
        \SATexample
    \end{minipage}\hfill
    \begin{minipage}[t]{0.74\textwidth}
        \textsc{0-1-Integer Programming} instance:\\
        \begin{align*}
            C=&\begin{blockarray}{ccccccccc}
                & x_1 & \overline{x}_1 & x_2 &\overline{x}_2 & x_3 & \overline{x}_3 & x_4 & \overline{x}_4 \\
                \begin{block}{c(cccccccc)}
                C_1 & -1 & 0 & -1 & 0 & 0 & -1 &  0 & 0\\
                C_2 & 0 & -1 & -1 & 0 &  0 & 0 &  0 & 0\\
                C_3 & 0 & -1 & 0 &  0 &  0 & 0 & 0 & -1\\
                x_1\text{ false } & 1 & 1 & 0 & 0 & 0 & 0 & 0 & 0\\
                x_2\text{ false }  & 0 & 0 & 1 & 1 & 0 & 0 & 0 & 0\\
                x_3\text{ false }  & 0 & 0 & 0 & 0 & 1 & 1 & 0 & 0\\
                x_4\text{ false }  & 0 & 0 & 0 & 0 & 0 & 0 & 1 & 1\\
                x_1\text{ true } & -1 & -1 & 0 & 0 & 0 & 0 & 0 & 0\\
                x_2\text{ true } & 0 & 0 & -1 & -1 & 0 & 0 & 0 & 0\\
                x_3\text{ true } & 0 & 0 & 0 & 0 & -1 & -1 & 0 & 0\\
                x_4\text{ true } & 0 & 0 & 0 & 0 & 0 & 0 & -1 & -1\\
                \end{block}
            \end{blockarray}\\
            b=&\begin{blockarray}{ccccccccccc}
                C_1 & C_2 & C_3 & x_1 & x_2 & x_3 & x_4 & \overline{x}_1 & \overline{x}_2 & \overline{x}_3 & \overline{x}_4\\
                \begin{block}{(ccccccccccc)}
                    -1 & -1 & -1 & 1 & 1 & 1 & 1 & -1 & -1 & -1 & -1\\
                \end{block}
            \end{blockarray}\\
            &\text{Example solution:}\\
            X'=&\begin{blockarray}{cccccccc}
                \fcolorbox{rwth-blue}{white}{$x'_1$} & \overline{x'}_1 & \fcolorbox{rwth-blue}{white}{$x'_2$} & \overline{x'}_2 & x'_3 & \fcolorbox{rwth-blue}{white}{$\overline{x'}_3$} & x'_4 & \fcolorbox{rwth-blue}{white}{$\overline{x'}_4$}\\
                \begin{block}{(cccccccc)}
                    1 & 0 & 1 & 0 & 0 & 1 & 0 & 1\\
                \end{block}
            \end{blockarray}
        \end{align*}
    \end{minipage}\hfill
    }
\end{figure}

\subsubsection{\textsc{Vertex Cover} \SSP{} \textsc{Directed Hamiltonian Cycle}}\label{subsubsec:vc-dirHamCyc}
\VC
\dirHam
\label{VC-HC}

In~\citeLiterature{karp1975}, Karp introduces a reduction from \textsc{Vertex Cover} to \textsc{Directed Hamiltonian Cycle}, which is explained more precisely in~\citeLiterature{aho1974design}. 
The following reduction is based on Karp's reduction, with some minor changes to fulfill the SSP property.\\
The idea behind this reduction is that each edge, which needs to be covered by the vertex cover, is transformed into a gadget.
Each gadget can be traversed in three different ways, depending on which of the vertices incident to the edge 
is part of the vertex cover, or even both.
The gadgets are then connected in a way that the next gadget can only be reached through an edge that connects
a vertex in the gadget that corresponds to some $v\in V$ with a vertex in the next gadget
that also corresponds to the same $v\in V$.\\
Additionally, there are $k$ extra vertices $c_1,\ldots, c_k$ that are each connected to the gadgets. 
This way, each path between the $c$ vertices corresponds to one vertex in the vertex cover.
Because a vertex cover can also have a smaller size than $k$, the $c$ form a clique, 
that acts as a garbage collector, so that they can always be visited.

\begin{figure}
    \begin{minipage}[t]{0.19\textwidth}
        \VCexample
    \end{minipage}\hfill
    \begin{minipage}[t]{0.79\textwidth}
        \textsc{Directed Hamiltonian Cycle} instance:\\
        \resizebox{\textwidth}{!}{
        \begin{tikzpicture}[on grid, auto, node distance=2cm and 4cm]
            \node[] (b10) [] {$\langle b, 1, 0\rangle$}; 
            \node[] (b11) [right=of b10] {$\langle b, 1, 1\rangle$}; 
            \node[] (a10) [below=of b10] {$\langle a, 1, 0\rangle$}; 
            \node[] (a11) [right=of a10] {$\langle a, 1, 1\rangle$}; 
            \node[] (a20) [below=of a10] {$\langle a, 2, 0\rangle$}; 
            \node[] (a21) [right=of a20] {$\langle a, 2, 1\rangle$};
            \node[] (c20) [below=of a20] {$\langle c, 2, 0\rangle$}; 
            \node[] (c21) [right=of c20] {$\langle c, 2, 1\rangle$}; 
            \node[] (c30) [below=of c20] {$\langle c, 3, 0\rangle$}; 
            \node[] (c31) [right=of c30] {$\langle c, 3, 1\rangle$};
            \node[] (d30) [below=of c30] {$\langle d, 3, 0\rangle$}; 
            \node[] (d31) [right=of d30] {$\langle d, 3, 1\rangle$};
            \node[dot=right:$a'$] (a')  [below right=1cm and 4cm of a11] {};
            \node[dot=right:$b'$] (b')  [right=of b11] {};
            \node[dot=right:$c'$] (c')  [below right=1cm and 4cm of c21] {};
            \node[dot=right:$d'$] (d')  [right=of d31] {};
            \node[dot=right:$c_1$] (c_1) [right=of a'] {};
            \node[dot=right:$c_2$] (c_2) [right=of c'] {};
    
        \foreach \x/\y in {b10/b11,
                            a10/a11,
                            a20/a21,
                            c20/c21,
                            c30/c31,
                            d30/d31}
        \path[->, rwth-blue] (\x) edge node {} (\y);
    
        \foreach \x/\y in {b10/a10,
                            b11/a11,
                            a20/c20,
                            a21/c21,
                            c30/d30,
                            c31/d31}
        \path[->, rwth-magenta, dotted] 
            (\x) edge [bend right] node {} (\y)
            (\y) edge [bend right] node {} (\x);
    
        \path[->, rwth-orange, densely dashed]
            (a11) edge node {} (a20)
            (c21) edge node {} (c30);
    
        \foreach \x in {b11, a21, c31, d31}
        \foreach \y in {b', a', c', d'}
        \foreach \z in {c_1, c_2}
        \path[->]
            (\x) edge [rwth-cyan, dashdotted] node {} (\z)
            (\z) edge [rwth-green, loosely dashed] node {} (\y)
            (a') edge [rwth-green, loosely dashed] node {} (a10)
            (b') edge [rwth-green, bend right, loosely dashed] node {} (b10)
            (c') edge [rwth-green, loosely dashed] node {} (c20)
            (d') edge [rwth-green, bend left, loosely dashed] node {} (d30);
    
        \foreach \a in {b', c', d'}
        \foreach \b in {a', c', d'}
        \foreach \c in {a', b', d'}
        \foreach \d in {a', b', c'}
        \path[->, loosely dashdotted]
            (a') edge [bend right] node {} (\a)
            (b') edge [bend right] node {} (\b)
            (c') edge [bend right] node {} (\c)
            (d') edge [bend right] node {} (\d);
    
        \path[->, loosely dashdotted]
            (c_1) edge [bend right] node {} (c_2)
            (c_2) edge [bend right] node {} (c_1);
        \end{tikzpicture}
        }
    \end{minipage}\hfill
    \caption{A \textsc{Vertex Cover} instance $(G,k)$ and the corresponding directed graph constructed as described in~\cref{VC-HC}}
    \label{fig:example_hamilton}
\end{figure}
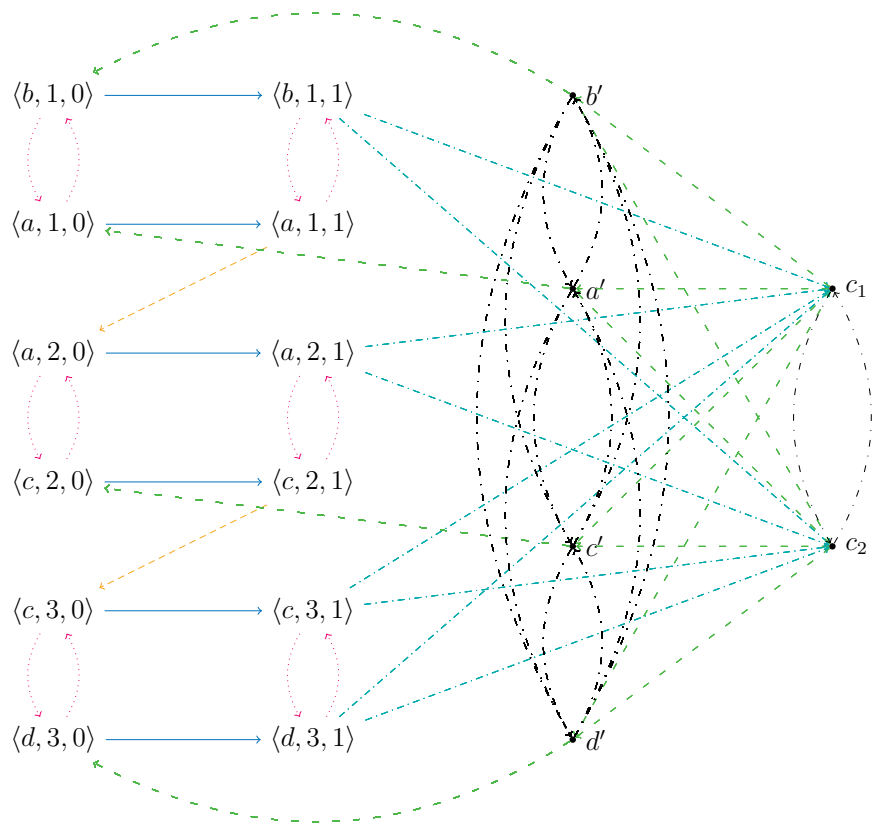

\functionG
The reduction function $g$ takes the input $(G, k)$, thus an undirected graph and a number $k\in\NN$ and 
maps it to a directed graph $G'=(V',A')$. 
We assume without loss of generality that the edges are numbered, thus $E=\{e_1,\ldots,e_n\},\:n:=\vert E\vert$. 
For each edge $e_i\in E$, there are four vertices in $V'$. 
To denote vertices that contain multiple parts, we use the notation $\langle a, b, c\rangle$.\\
$V'$ contains one additional vertex for each $v\in V$ and $k$ extra vertices: 
\[V'=\{c_1,\ldots,c_k\}\cup\{\langle v,i,a\rangle\mid v\in V,\: e_i\in E,\: e_i=(v,v^*),\: a\in\{0,1\}\}\cup\{ v'\mid v\in V \}\]
Then, the arcs are defined as follows:
\begin{align}
    A'= &\{\color{rwth-blue}(\langle v,i,0\rangle,\langle v,i,1\rangle)\color{black}\mid \langle v,i,0\rangle\in V'\}\label{eq:1}\tag{1}\\
    \cup\: & \{\color{rwth-magenta}(\langle v,i,a\rangle, \langle v^*,i,a\rangle)\color{black}\mid 
    e_i\in E,\: e_i=(v,v^*),\: a\in\{a,1\}\}\label{eq:2}\tag{2}\\
    \cup\: & \{\color{rwth-orange}(\langle v,i,1\rangle, \langle v,j,0\rangle)\color{black}\mid 
    (v,v^*)\in E,\: \nexists h,\: i<h<j\label{eq:3}\tag{3} 
    \text{ s.t. } (v,h)\in E\}\notag{}\\
    \cup\: & \{\color{rwth-cyan}(\langle v,i,1\rangle, c_b)\color{black}\mid 1\leq b\leq k, \nexists h>i\label{eq:4}\tag{4}
    \text{ s.t. } (v,h)\in E\}\notag{}\\
    \cup\: & \{\color{rwth-green}(c_b, v'), (v', \langle v,i,0\rangle)\color{black}\mid 1\leq b\leq k, v\in V, \nexists h<i \label{eq:5}\tag{5}
    \text{ s.t. } (v,h)\in E\}\notag{}\\
    \cup\: & \{(v_i',v_j')\mid 1\leq i,j\leq \vert V\vert, i\neq j \}\label{eq:6}\tag{6}\\
    \cup\: & \{(c_i, c_j)\mid 1\leq i,j\leq k, i\neq j\}\label{eq:7}\tag{7}
\end{align}
The dark blue solid arcs from \cref{eq:1} connect the two vertices corresponding to a vertex-edge pair from $G$, 
going from 0 to 1. The pink dotted arcs (\cref{eq:2}) connect the two ends $v, v^*$ of $i$ in $G$ in both directions, 
for the third component being 0 and 1 each. Thus, for each edge from $G$, the four vertices in $G'$ form a gadget like shown in~\cref{fig:hamilton}.\\
The in- and outgoing arcs can either be orange and dashed(\cref{eq:3}), cyan and dashdotted(\cref{eq:4}) or green and loosely dashed(\cref{eq:5}).
Orange (dashed) arcs connect gadgets with each other that both represent 
edges adjacent to the same vertex $v$, and following each other directly in the order that was assigned to them in the beginning. 
Cyan (dashdotted) and blue (solid) arcs connect the vertices $c_1,\ldots,c_k$ to the gadgets as follows: 
for each $v$, the node $\langle v,i,1\rangle$ with $i$ being the edge with the \textit{highest} number 
adjacent to $v$ has a cyan(dashdotted) arc to each $c_1,\ldots,c_k$, and the other way around, 
for each $v$, the node $\langle v,i,0\rangle$ with $i$ being the edge with the \textit{lowest} number 
adjacent to $v$ has an incoming green green (loosely dashed) arc from $v'$, and $v'$ again has an incoming green (loosely dashed) arc 
from each $c_1,\ldots,c_k$. Thus, each $\langle v,i,0\rangle$ has either one incoming orange (dashed) arc, 
or one incoming green (loosely dashed) arc, and each $\langle v,i,1\rangle$ has either one outgoing orange (dashed) arc, 
or $k$ outgoing cyan (dashdotted) arcs. Furthermore, the vertices $v_1,\ldots,v_{\vert V\vert}$ and $c_1,\ldots, c_k$ form
a clique each, which is done in \cref{eq:6,eq:7}. This is to ensure that a path corresponding to a vertex cover can also have less than $k$ vertices, as this way,
the vertices $c_1,\ldots, c_k$ can always be visited.\\
The running time is polynomially bounded, as there are $O(\vert E\vert+k+\vert V\vert)$ vertices in $G'$. 
In \cref{fig:example_hamilton}, an example instance of \textsc{Vertex Cover} $(G, k)$ and the corresponding DIR-HAM-CYCLE instance $G'$ are shown.

\functionF
We define the functions $f_{(G,k)}(v)$ as follows: each $v\in V$ is mapped to the arc $(v', \langle v,i,0\rangle)$ 
such that $e_i$ is the smallest edge incident to $v$.

\correctness
Let $C=(a_1,\ldots,a_m)$ be an Hamiltonian Cycle on $G'$ with $m=\vert V'\vert$. 
\begin{enumerate}
    \item Each path from some $c_i$ to some $c_j$ corresponds to one vertex $v\in V$ that is part of the vertex cover and contains only edges that are incident to this $v$, as explained in~\citeLiterature{aho1974design}.
    \item As the vertices $v_1',\ldots v_k'$ form a clique, they can always be visited in any order.
    \item If a gadget is first entered through some vertex $\langle v,i,0\rangle$, it must be left through $\langle v,i,1\rangle$. Either, all 4 vertices are visited in between, or the other two are left out and must be visited earlier or later.
    \item\begin{itemize}
        \item If there is no orange edge leaving $\langle v,i,1\rangle$, the next edge will be a cyan edge, thus completing this sub-path.
        \item Otherwise, one specific other gadget can be entered, which again is entered through some vertex containing $v$ (a gadget representing an edge adjacent to $v$), thus the same repeats.
    \end{itemize}
    \item Let us assume the first gadget is entered trough $\langle v^*,i,0\rangle$, 
    where $v^*$ is not part of a valid vertex cover. 
    The first case is not possible, as in that case, the path would cover only one edge, 
    thus it would not matter which endpoint is chosen for the vertex cover. In the second case, 
    the next gadget will be entered through $\langle v^*,i',0\rangle$, thus represent an edge incident to $v^*$. 
    As there can only be one edge between two vertices in an undirected graph, this directly implies that only $v^*$ can cover both edges, 
    again leading to a contradiction.
    \item As each path only contains gadgets representing edges adjacent to the same $v\in V$, all the gadgets must be entered (and left) through some vertex containing this $v$.
    \item Thus, the green arc going into the first gadget on the part must be the arc $(v', \langle v,i,0\rangle)$. 
\end{enumerate}
With this argumentation we have shown that
\begin{align*}
    &\{f_{(G,k)}(S)\mid S\in\S(\textsc{Vertex Cover})\}\\
    =& \{ f_{(G,k)}(C) \mid C\subseteq V, \vert C\vert\leq k\text{ such that } \forall(v,w)\in E: v\in C \text{ or } w\in C\} \\
    =& \{E \mid E=\{(v',\langle v,i,0\rangle) \mid v\in C\}\}\\
    =& \{ E \mid E = \{(v',\langle v,i,0\rangle)\mid e_i \text{ is edge with the smallest } i \text{ incident to }v\text{ in }G, \\
    & (v',\langle v,i,0\rangle)\text{ is part of a solution in }G'\}  \}\\
    =& \{S'\cap f_{(G,k)}(V)\mid S'\in \S(\textsc{Dir Hamiltonian Cycle})\}
\end{align*}
and therefore, it is indeed an SSP reduction.

\begin{figure}[hb!]
    \centering
    \begin{tikzpicture}[shorten >=1pt,node distance=3cm,on grid,auto] 
       \node[] (u1)   {$\langle u,i, 0\rangle$}; 
       \node[] (u2) [right=of u1] {$\langle u,i, 1\rangle$}; 
       \node[] (v1) [below=of u1] {$\langle v,i, 0\rangle$}; 
       \node[] (v2) [right=of v1] {$\langle v,i, 1\rangle$};
       \node[] (u1') [left=of u1]  {}; 
       \node[] (u2') [right=of u2] {}; 
       \node[] (v1') [left=of v1] {}; 
       \node[] (v2') [right=of v2] {};
       
        \path[->] 
        (u1) edge [bend left, color=rwth-magenta] node {} (v1)
        (v1) edge [bend left, color=rwth-magenta] node {} (u1)
        (u2) edge [bend left, color=rwth-magenta] node {} (v2)
        (v2) edge [bend left, color=rwth-magenta] node {} (u2)
        
        (u1) edge [color=rwth-blue] node {} (u2)
        (v1) edge [color=rwth-blue] node {} (v2)
        
        (u1') edge [] node {} (u1)
        (v1') edge [] node {} (v1)
        (u2) edge [] node {} (u2')
        (v2) edge [] node {} (v2');
    \end{tikzpicture}
    \caption{A gadget in $G'$ corresponding to one edge from $G$.}
    \label{fig:hamilton}
\end{figure}
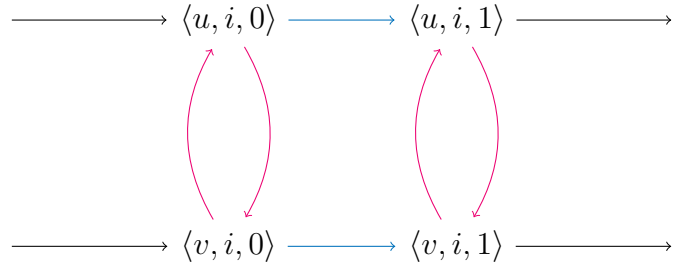

\FloatBarrier

\subsubsection{\textsc{Vertex Cover} \SSP{} \textsc{Undirected Hamiltonian Cycle}}
\VC
\undirHam

This reduction originates from \citeLiterature[p.~56-60]{garey1979computers} by Garey and Johnson.
It is similar to the reduction from \textsc{Vertex Cover} to \textsc{Directed Hamiltonian Cycle} presented in~\cref{subsubsec:vc-dirHamCyc},
but the gadget corresponding to the edges are more complicated, as we do not have directed edges but need to make
sure they are traversed in a certain order.

\begin{figure}
    \centering
    \textsc{Vertex Cover} instance:\\
    \begin{tikzpicture}[on grid, auto, node distance=2cm]
        \node[dot=left:$a$, color=rwth-blue] (a) {};
        \node[dot=right:$b$] (b) [right=of a] {};
        \node[dot=left:$c$, color=rwth-blue] (c) [below=of a] {};
        \node[dot=right:$d$] (d) [right=of c] {};

        \path[]
        (a) edge node {1} (b)
        (a) edge node {2} (c)
        (c) edge node {3} (d);
    \end{tikzpicture}
    ~\\ $k=2$\\
    \textsc{Undirected Hamiltonian Cycle} instance:
    \resizebox{\textwidth}{!}{
    \begin{tikzpicture}
        \node[] (b11) {$\langle b,1,1\rangle$};
        \node[] (a11) [right=of b11] {$\langle a,1,1\rangle$};

        \node[] (a21) [right=of a11]{$\langle a,2,1\rangle$};
        \node[] (c21) [right=of a21] {$\langle c,2,1\rangle$};

        \node[] (c31) [right=of c21]{$\langle c,3,1\rangle$};
        \node[] (d31) [right=of c31] {$\langle d,3,1\rangle$};

        \node[] (a') [above right=1 and 0.5 of a11] {$a'$};
        \node[] (b') [above=of b11] {$b'$};
        \node[] (c') [above right=1 and 0.5 of c21] {$c'$};
        \node[] (d') [above=of d31] {$d'$};
            
        \foreach \x[evaluate=\x as \y using int(\x-1)] in {2,...,6} {
            \foreach \z/\position in {a/1, b/1, a/2, c/2, c/3, d/3} {
                \node (\z\position\x) [below=of \z\position\y] {$\langle \z,\position,\x\rangle$};
                \ifthenelse{\equal{\y}{3}}{
                    \ifthenelse{\equal{\z}{a}}{
                        \path
                        (\z\position\y) edge node {} (\z\position\x);
                    }{
                        \ifthenelse{\equal{\z}{c}}{
                            \path
                            (\z\position\y) edge node {} (\z\position\x); 
                        }{
                            \path[color=rwth-blue]
                    (\z\position\y) edge node {} (\z\position\x);
                        }
                    }
                }{
                    \path[color=rwth-blue]
                    (\z\position\y) edge node {} (\z\position\x);
                }  
            }
        }
        \node[] (a1) [right=of d32] {$a_1$};
        \node[] (a2) [right=of d34] {$a_2$};

        \path 
            (a11) edge node {} (b13)
            (b11) edge [color=rwth-blue] node {} (a13)
            (a23) edge node {} (c21)
            (c23) edge node {} (a21)
            (c31) edge node {} (d33)
            (d31) edge [color=rwth-blue] node {} (c33)
            
            (a14) edge [color=rwth-blue] node {} (b16)
            (b14) edge node {} (a16)
            (a26) edge node {} (c24)
            (c26) edge node {} (a24)
            (c34) edge [color=rwth-blue] node {} (d36)
            (d34) edge node {} (c36);
             
            \node[] (a61') [below=of a16]{};
            \node[] (a62') [below=of a26]{};
            \node[] (b6') [below=of b16]{};
            \node[] (c26') [below=of c26]{};
            \node[] (c36') [below=of c36]{};
            \node[] (d6') [below=of d36]{};
            \path[dashed, color=gray]
                (a') edge [solid, color=rwth-blue] node {} (a11)
                (b') edge node {} (b11)
                (c') edge [solid, color=rwth-blue] node {} (c21)
                (d') edge node {} (d31)

                (a62') edge [solid, color=rwth-blue] node {} (a26)
                (b6') edge node {} (b16)
                (c36') edge [solid, color=rwth-blue] node {} (c36)
                (d6') edge node {} (d36);
        
        \path
            (a16) edge [color=rwth-blue] node {} (a21)
            (c26) edge [color=rwth-blue] node {} (c31);

        \node[] (a1above) [above=of a1]{};
        \node[] (a1below) [below=of a1]{};
        \node[] (a2above) [above=of a2]{};
        \node[] (a2below) [below=of a2]{};

        \path[]
            (a1) edge [bend left, color=rwth-blue] node {} (a2);
        \path[dashed, color=rwth-blue]
            (a1above) edge node {} (a1)
            (a1below) edge node {} (a1)
            (a2above) edge node {} (a2)
            (a2below) edge node {} (a2);
        
        \path[dashed, color=gray, bend left]
            (b') edge node {} (a')
            (a') edge node {} (c')
            (a') edge node {} (d')
            (b') edge node {} (c')
            (b') edge node {} (d')
            (c') edge node {} (d');

        \node[] (a'above) [above=2 of a']{};
        \node[] (b'above) [above=2 of b']{};
        \node[] (c'above) [above=2 of c']{};
        \node[] (d'above) [above=2 of d']{};
        \path[dashed, color=gray]
            (a'above) edge [solid, color=rwth-blue] node {} (a')
            (b'above) edge node {} (b')
            (c'above) edge [solid, color=rwth-blue] node {} (c')
            (d'above) edge node {} (d');
        
    \end{tikzpicture}
    }
    \caption{Example for the reduction from \textsc{Vertex Cover} to HAMILTONIAN CYCLE. 
    The edges going out of the gadgets at the bottom are each connected to both $a_1$ and $a_2$.}
    \label{fig:VC-HamCyc-example}
\end{figure}
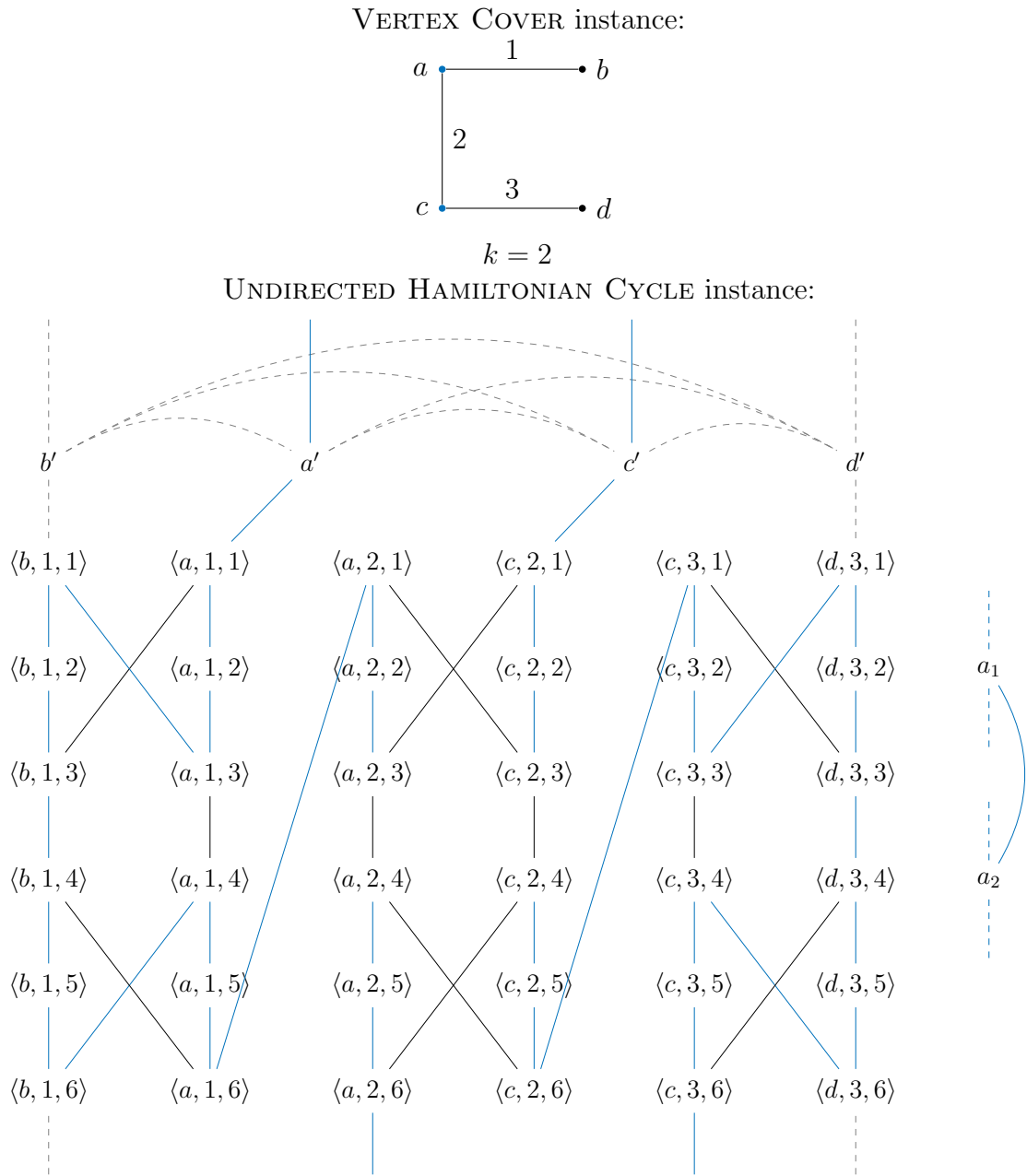

In this reduction, each edge is represented by a gadget that is composed of vertices corresponding to the 
endpoints of the edge and extra vertices that ensure that each gadget can only be traversed in three 
different ways in order to cover all vertices in the gadget, as can be seen in~\cref{fig:VC-HamCyc-example}. 
That is, if a gadget for the edge $(v,w)$ is entered through a vertex corresponding
to $v$, it must be left through a vertex corresponding to $v$ (and equally for $w$). 
Additionally, each gadget has vertices which are connected to vertices outside of the gadget, 
corresponding either to $v$ or $w$. The gadgets are connected in a way that two gadgets with the same
endpoint $v$ are connected through adding an edge between the vertices corresponding to $v$ in the gadgets, 
thus, two gadgets are connected if and only if they are adjacent in the original graph. 
Additionally, for each vertex $v\in V$, a vertex $v'\in V$ is introduced, which is connected to the gadget
corresponding to the edge with the lowest index incident to $v$.
Finally, $k$ extra vertices $a_1,\ldots, a_k$ are introduced, that are connected to all $v'$ and the lower end of the gadgets.
These $k$ extra vertices basically model the $k$ vertices in the vertex cover,
as this way, $k$ paths can be formed, and each path can only contain gadgets corresponding to edges incident to the same vertex.

\functionG
To construct the graph $G'$ from the \textsc{Vertex Cover} instance $(G,k)$, the following steps are performed:
First, we introduce the vertices $V'$ as follows, under the assumption $k\leq\vert V\vert$. 
We use $\langle a,b,c \rangle$ to denote a vertex that is composed of the different parts $a, b$ and $c$.
\begin{align*}
    V'=&\{a_i\mid 1\leq i\leq k\}\\
    &\cup\{\langle u,e,i\rangle, \langle v,e,i\rangle \mid 1\leq i\leq 6, (u,v)\in E \}\\
    &\cup\{v'\mid v\in V\}
\end{align*}
Thus, we have $k$ vertices $a_i$, which will later be used to select the $k$ vertices of the vertex cover,
and six vertices for each edge-vertex pair $(u,e)$ and $(v,e)$ for each edge $e=(u,v)\in E$. 
These vertices will form a gadget that ensures that at least one endpoint $u$ or $v$ is part of the vertex cover.
There are six vertices for each edge-vertex combination to later ensure that the gadget can only be traversed in three ways.
The vertices $v'$ are used to connect the gadgets to the $a_i$-vertices and to be able to define the mapping
functions $f_I$.

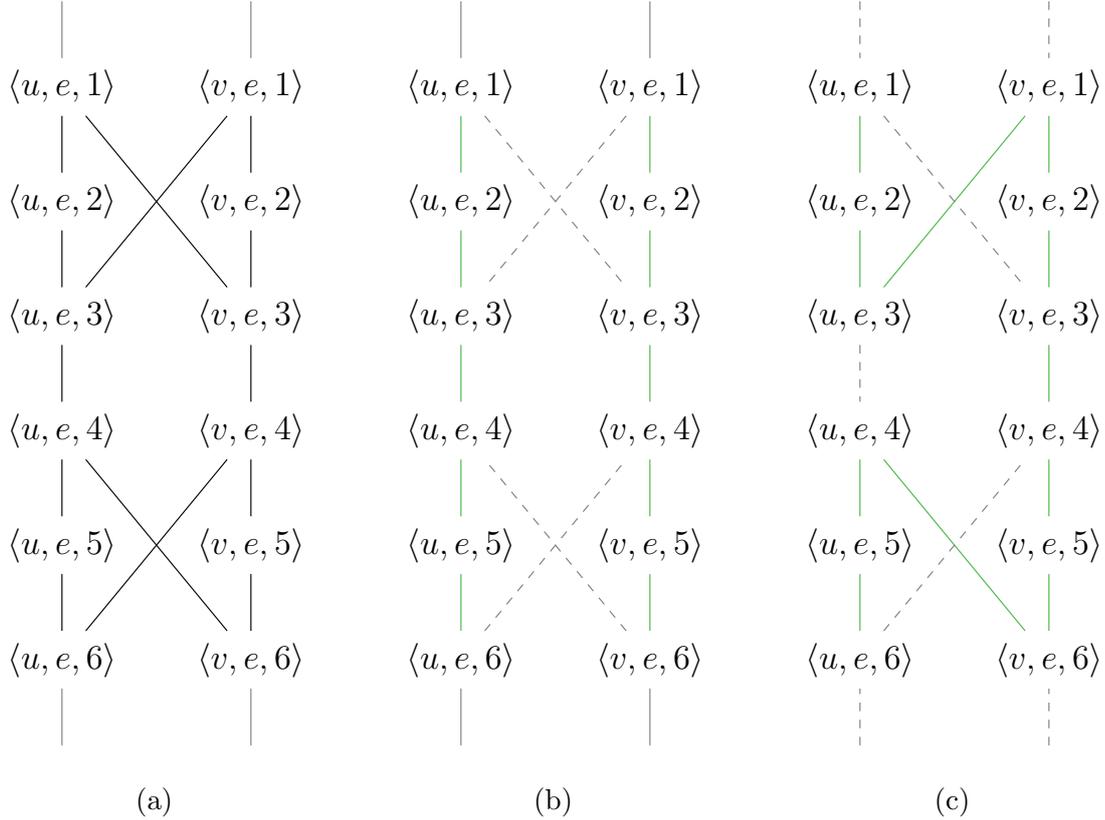
\begin{figure}[h]
    \centering
    \begin{subfigure}{0.3\textwidth}
        \resizebox{\textwidth}{!}{
        \begin{tikzpicture}[node distance=0.7cm]
            \node[] (u1) {$\langle u,e,1\rangle$};
            \node[] (v1) [right=of u1] {$\langle v,e,1\rangle$};
             
            \foreach \x[evaluate=\x as \y using int(\x-1)] in {2,...,6} {
                \node[] (u\x) [below=of u\y] {$\langle u,e,\x\rangle$};
                \node[] (v\x) [below=of v\y] {$\langle v,e,\x\rangle$};
                \path
                    (u\y) edge node {} (u\x)
                    (v\y) edge node {} (v\x);
            }
            \path 
                (u1) edge node {} (v3)
                (v1) edge node {} (u3)
                (u6) edge node {} (v4)
                (v6) edge node {} (u4);
            \node[] (u1') [above=of u1]{};
            \node[] (v1') [above=of v1]{};
            \node[] (u6') [below=of u6]{};
            \node[] (v6') [below=of v6]{};
            \path[color=gray]
                (u1') edge node {} (u1)
                (v1') edge node {} (v1)
                (u6') edge node {} (u6)
                (v6') edge node {} (v6);
        \end{tikzpicture}
        }
        \caption{}
    \end{subfigure}\hfill
    \begin{subfigure}{0.3\textwidth}
        \resizebox{\textwidth}{!}{
        \begin{tikzpicture}[node distance=0.7cm]
            \node[] (u1) {$\langle u,e,1\rangle$};
            \node[] (v1) [right=of u1] {$\langle v,e,1\rangle$};
             
            \foreach \x[evaluate=\x as \y using int(\x-1)] in {2,...,6} {
                \node[] (u\x) [below=of u\y] {$\langle u,e,\x\rangle$};
                \node[] (v\x) [below=of v\y] {$\langle v,e,\x\rangle$};
                \path[color=rwth-green]
                    (u\y) edge node {} (u\x)
                    (v\y) edge node {} (v\x);
            }
            \path[dashed, color=gray]
                (u1) edge node {} (v3)
                (v1) edge node {} (u3)
                (u6) edge node {} (v4)
                (v6) edge node {} (u4);
            \node[] (u1') [above=of u1]{};
            \node[] (v1') [above=of v1]{};
            \node[] (u6') [below=of u6]{};
            \node[] (v6') [below=of v6]{};
            \path[color=gray]
                (u1') edge node {} (u1)
                (v1') edge node {} (v1)
                (u6') edge node {} (u6)
                (v6') edge node {} (v6);
        \end{tikzpicture}
        }
        \caption{}
        \label{fig:hamCyc-gadget:b}
    \end{subfigure}\hfill
    \begin{subfigure}{0.3\textwidth}
        \resizebox{\textwidth}{!}{
        \begin{tikzpicture}[node distance=0.7cm]
            \node[] (u1) {$\langle u,e,1\rangle$};
            \node[] (v1) [right=of u1] {$\langle v,e,1\rangle$};
             
            \foreach \x[evaluate=\x as \y using int(\x-1)] in {2,...,6} {
                \node[] (u\x) [below=of u\y] {$\langle u,e,\x\rangle$};
                \node[] (v\x) [below=of v\y] {$\langle v,e,\x\rangle$};
            }
            \path[color=rwth-green]
                (u1) edge [solid] node {} (u2)
                (u2) edge [solid] node {} (u3)
                (u4) edge [solid] node {} (u5)
                (u5) edge [solid] node {} (u6)
                (v1) edge [solid] node {} (v2)
                (v2) edge [solid] node {} (v3)
                (v3) edge [solid] node {} (v4)
                (v4) edge [solid] node {} (v5)
                (v5) edge [solid] node {} (v6)
                (v1) edge [solid] node {} (u3)
                (v6) edge [solid] node {} (u4);
            \path[dashed, color=gray]
                (u3) edge node {} (u4)
                (u1) edge node {} (v3)
                (u6) edge node {} (v4);
            \node[] (u1') [above=of u1]{};
            \node[] (v1') [above=of v1]{};
            \node[] (u6') [below=of u6]{};
            \node[] (v6') [below=of v6]{};
            \path[dashed, color=gray]
                (u1') edge node {} (u1)
                (v1') edge node {} (v1)
                (u6') edge node {} (u6)
                (v6') edge node {} (v6);
        \end{tikzpicture}
        }
        \caption{}
        \label{fig:hamCyc-gadget:c}
    \end{subfigure}\hfill
    \caption{Gadget for an edge $(u,v)$ in the reduction from \textsc{Vertex Cover} to HAMILTONIAN CYCLE. 
    The outgoing edges are potential edges connecting the gadget to the rest of the graph
    and the solid green edges in the second and third gadget show the possible paths through the gadget, 
    where the last path can also be mirrored.}
    \label{fig:hamCyc-gadget}
\end{figure}

Each gadget has the following edges: let $e=(u,v)\in E$. Then, the edges of that gadget are defined as follows:
\begin{align*}
    E'_e=&\{(\langle u,e,i\rangle,\langle u,e,i+1\rangle), (\langle v,e,i\rangle,\langle v,e,i+1\rangle)\mid 1\leq i\leq 5 \}\\
    \cup&\{(\langle u,e,3\rangle,\langle v,e,1\rangle), (\langle v,e,3\rangle,\langle u,e,1\rangle)\}\\
    \cup&\{(\langle u,e,6\rangle,\langle v,e,4\rangle), (\langle v,e,6\rangle,\langle u,e,4\rangle)\}
\end{align*}
Thus, a gadget for an edge $(u,v)$ looks as shown in~\cref{fig:hamCyc-gadget}.

As one can see in~\cref{fig:hamCyc-gadget}, each gadget can be traversed in three ways: either starting in 
$\langle u,e,1\rangle$ and ending in $\langle u,e,6\rangle$, going through all vertices in between, 
or starting in $\langle v,e,1\rangle$ and ending in $\langle v,e,6\rangle$ the same way but mirrored,
or going through the $e$-vertices and $v$-vertices separately.

Next, the gadgets need to be connected. This is done using the $v'$- and $a_i$-vertices as well as edges 
connecting gadgets to each other.
Let us pick one vertex $v$ and its incident edges $(u,v)\in E$. Let us assume the edges $(u,v)$ are arbitrarily ordered,
thus we have the edges $e_{v(1)},\ldots,e_{v(d)}$ incident to $v$, with $d$ being the degree of $v$.
We can connect all the gadgets corresponding to these edges incident to $v$ as follows:
\[E'_v = \{(\langle v, e_{v(i)}, 6\rangle, \langle v, e_{v(i+1)}, 1\rangle)\mid 1\leq i<d\}\]
Thus, we obtain a path connecting all gadgets corresponding to edges incident to $v$.

Then, we connect the $a_i$-vertices to the gadgets and the $v'$-vertices as follows:
\begin{align*}
    E''=
    &\{
        (a_i,v') \mid 1\leq i\leq k
    \}\\
    &\cup \{
        (v', \langle v, e_{v(1)},1\rangle) \mid v\in V
    \} \\
    &\cup \{
        (\langle v,e_{v(d)},6\rangle, a_i) \mid 1\leq i\leq k, v\in V
    \}\\
\end{align*}
This means that each $a_i$ vertex is connected to each $v'$ vertex, and each $v'$ vertex is connected
to the vertex corresponding to the edge with the lowest index incident to the corresponding vertex $v\in V$.
Additionally, the vertex corresponding to the edge with the \textit{highest} index incident to $v$ is connected
to each $a_i$ vertex. 

We connect the $a_i$-vertices and the $v'$ vertices to each form a clique:
\[
    A'=\{(a_i,a_j)\mid 1\leq i<j\leq k\} \cup \{ (v', w')\mid v,w\in V, v\neq w \}
\]

Finally, we have all edges by setting 
\[ E'=\bigcup_{e\in E}E'_e\bigcup_{v\in V}E'_v\cup E''\cup A'\]
An example for this reduction is shown in~\cref{fig:VC-HamCyc-example}.

\functionF
The mapping functions $f_I: V\rightarrow E, f_{(G,k)}(v) = (v', \langle v, e_{v(1)}, 1\rangle)$ maps each
vertex to the edge going from $v'$ to the first vertex corresponding to $v$ in the first gadget of $v$, 
as this edge is used in the cycle if and only if $v$ is part of the vertex cover.

\correctness
First, we can see that the reduction can be performed in polynomial time, 
as the number of vertices and edges in $G'$ is polynomial in the size of $G$.

Let $V^*\subseteq V$ be a vertex cover of $G$ of size at most $k$.
Then, each edge is incident to at least one $v^*\in V^*$.
This means that each gadget in the \textsc{Hamiltonian Cycle} instance $(G')$ can be traversed like shown in
\cref{fig:hamCyc-gadget:b} if $\{u,v\}\cap V^*=\{u,v\}$, and like shown in~\cref{fig:hamCyc-gadget:c}, 
starting in $\langle v,e,1\rangle$ if $\{u,v\}\cap V^*=\{v\}$ and starting in $\langle u,e,1\rangle$
if $\{u,v\}\cap V^*=\{u\}$. 
As one of these three possibilities must always hold, since $V^*$ is a vertex cover, one ot the three traversions 
is always possible. For each $v^*\in V^*$, the edges connecting the gadgets, thus 
$(\langle v^*,e_{v(i)},6\rangle, \langle v^*, e_{v(i+1)}, 1\rangle)$ are selected.
Additionally, $(a_i, v^{*'})$, $(v^{*'}, \langle v^*, e_{v(1)}, 1\rangle)$ and 
$(\langle v^{*'}, e_{v(d)}, 6\rangle, a_{i+1})$ are selected for each $v^*\in V^*$.
This forms a path from some $a_i$ to $a_{i+1}$, containing each vertex that contains $v^*$, and no vertex more than once.
As $\vert V^*\vert$ is at most $k$, there are at most $k$ of these paths, thus, they can all be connected through
the $a_i$ vertices and the $v^*$ vertices.
Note that the vertices $v'_1,\ldots,v'_{\vert V\vert}$ and $a_1,\ldots,a_k$ each form a clique, and thus can always be visited.
If $\vert V^*\vert$ is less then $k$, the remaining $a_i$ can be visited by going through 
$a_{k-\vert V^*\vert},\ldots,a_k$ at the end of the cycle.
Furthermore, the $v'$ with $v\notin V^*$ can also be visited, as they form a clique with each other, thus,
after visiting the left over $a_i$ vertices, all $v'$ with $v\notin V^*$ can be visited.
Then, the cycle is closed by going from the last $v'$ to $a_1$, when the cycle starts in $a_1$.
Thus, for each $v^*\in V^*$, the mapping $f_I(v^*)=(v^{*'},\langle v,e_{v(1)},1\rangle)$ is included
in the path corresponding to $v^*$ and all paths combined form a Hamiltonian cycle in $G'$.\\

The other way around, let $v'_1,\ldots,v'_{\vert V\vert}$ be a Hamiltonian cycle in $G'$.
Lets consider a subpath going from some $a_j$ to some $a_l$ without any $a$-vertices in between.
The path either is empty or starts with an edge $(a_j, v')$. Then, either the path continues with $(v', v_1')$,
which can be neglected, or with $(v', \langle v, e_{v(1)}, 1\rangle)$.
As a gadget is left through the vertex $(\langle v,e_{v(i)},6\rangle)$ if and only if it is entered through 
the vertex $(\langle v,e_{v(i)},1\rangle)$,
and, by construction, the outgoing edge must lead to either $a_l$ or to $(\langle v,e_{v(i+1)},1\rangle)$,
only gadgets corresponding to edges incident to $v\in V$ in the original graph are traversed in this subpath.
Thus, the subpath corresponds to a vertex in the vertex cover. 

Furthermore, the path includes the edge $(v', \langle v, e_{v(1)}, 1\rangle)$, because otherwise, the gadgets
cannot be entered.\\
As there are no more than $k$ paths and we assume $v'_1,\ldots,v'_{\vert V\vert}$ to be a Hamiltonian cycle, thus,
every gadget must have been traversed, hence, each gadget is part of a path corresponding to some $v\in V$,
these $v$ must form a vertex cover.\\
Note that in the whole correctness proof, the paths can always be traversed in the reversed way without
changing the correctness.
Thus, the vertices corresponding to the paths form a vertex cover of size at most $k$, the reduction is correct and

\begin{align*}
    & \{f_{(G,k)}(S) \mid S\in\S(\textsc{Vertex Cover})\}\\
    =& \{f_{(G,k)}(V^*) \mid V^*\subseteq V, \vert V^*\vert\leq k\text{ such that } \forall(v,w)\in E: v\in V^* \text{ or } w\in V^* \} \\
    =& \{ (v', \langle v, e_{v(1)}, 1\rangle) \mid v \text{ is part of the vertex cover of }G \}\\
    =& \{ (v', \langle v, e_{v(1)}, 1\rangle) \mid (v', \langle v, e_{v(1)}, 1\rangle) \text{ is part of the Hamiltonian cycle}\}\\
    =& \{S'\cap f_{(G,k)}(V)\mid S'\in \S(\textsc{Undir Hamiltonian cycle})\}
\end{align*}

\subsubsection{\textsc{Exact Cover} \SSP{} \textsc{Steiner Tree}}
\EC
\steiner

In this reduction, first mentioned by Karp~\citeLiterature{karp1975}, each set $S_j$ from the 
\textsc{Exact Cover} instance $(U, \mathfrak{S})$ is represented by a Steiner vertex $S'_j$,
and each element $u_i\in U$ is represented by a terminal vertex $u'_i$. 
Additionally, a root vertex $v_0$ is introduced, 
that is connected to all Steiner vertices $S'_j$, and the Steiner vertices are 
connected to the terminal vertices $u'_i$ depending on whether $u_i$ appears in the set $S_j$ or not.
The weights of the edges and $k'$ are set in a way that in order to connect the terminal vertices $u'_i$ to the root $v_0$,
one must take the edge from $v_0$ to a Steiner vertex $S'_j$ that is connected to $u'_i$.

\functionG
Formally, the reduction function $g$ from $(U, \mathfrak{S})$ to \\$(G', S', T', c', k')$ is defined as follows
with $m:=\vert \mathfrak{S}\vert$ and $n:=\vert U\vert$:
\begin{align}
    S'& = \{S'_1,\ldots,S'_m\}\label{steiner-1}\\
    T'& = \{v_0\}\cup\{u'_1,\ldots, u'_n\}\label{steiner-2}\\
    E'& = \{(v_0, S'_j)\mid j\in \{1,\ldots, m\}\}\cup \{(S'_j, u_i)\mid j\in \{1,\ldots, m\}, u_i\in S_j \}\label{steiner-3}\\
    c'((v_0, S'_j))& = \vert S_j\vert\label{steiner-4}\\
    c'((S'_j, u_i))& = n+1\label{steiner-5}\\
    k'& = n^2+2n
\end{align}
The weight of an edge between $v_0$ and $S'_j$ is the number of elements in $S_j$~\cref{steiner-4},
and the weight of the edges between $S'_j$ and $u'_i$ is set to $n+1$~\cref{steiner-5}.
Then, $k'$ is set to $n+n(n+1)=n^2+2n$. This enforces that it is only feasible
to take the edges connecting the root to the Steiner vertices that correspond to the exact cover, 
and to the terminal vertices that correspond to the elements in the sets, because by choosing the vertices 
connecting the $u'_i$ to the tree, the total weight is $n(n+1)=n^2+n$, thus, the weight that is left is $n$, 
so it is not feasible to add any edge between an $S'_j$ and a $u'_i$, and $v_0$ can still be connected, 
as all edges from $v_0$ to the $S'_j$ together have weight $n$.
Thus, in the end, all components are connected by $v_0$ and all terminal vertices are part of the tree.

In~\cref{steiner-1}, the Steiner vertices are defined from the sets of \textsc{Exact Cover}, then, in~\cref{steiner-2}, 
the terminal vertices, the root vertex $v_0$ and the $u'_i$ corresponding to elements from the \textsc{Exact Cover} instance, are defined.
The edges are defined in~\cref{steiner-3}, where edges are added between $v_0$ and the Steiner vertices as well
as between $S'_j$ and the $u'_i$ that appear in $S_j$. The weights are set in~\cref{steiner-4} and~\cref{steiner-5}.

\functionF
We define the functions $f_{(U, \mathfrak{S})}(S_j) = (v_0, S'_j)$.

\correctness
The reduction runs in polynomial time, since we obtain a number of vertices linear in the input size.
Let $(U, \mathfrak{S})$ be a YES-instance of the \textsc{Exact Cover} problem and $(G', S', T', c', k')$ be the corresponding instance of the \textsc{Steiner Tree} problem.
Let $\mathfrak{S}^*\subseteq\mathfrak{S}$ be a solution to the instance. Then, the tree defined by the edges $\{(v_0, S'_j)\mid S_j\in\mathfrak{S}^*\}\cup\{(S'_j, u'_i)\mid u_i\in S_j\}$
is a solution to the \textsc{Steiner Tree} instance. All terminal vertices are part of the tree, as the exact cover contains all elements $u_i$ in its sets $S_j\in\mathfrak{S}^*$
and $v_0$ is connected to all $S_j\in\mathfrak{S}^*$, thus, the edges form a tree, because only edges incident to $S_j\in\mathfrak{S}^*$ are part of the edges chosen,
and all $S'_j$ are connected to $v_0$. The total weight of the tree is $n^2+2n$, as the weight of the edges connecting $v_0$ to the $S'_j$ is $n$,
and all $n$ edges are connected to exactly one $S'_j$, which leads to a weight of $n+n(n+1)=n^2+2n$.
Thus, the solution is given exactly by the function $f_{(U, \mathfrak{S})}$.\\

To prove the other direction, let us assume there is a feasible solution $A^*$ for the \textsc{Steiner Tree} instance. 
Then, $A^*$ is a tree connecting all terminal vertices from $T$ with a weight of at most $k'=n^2+2n$. 
Because the weight of the edges adjacent to any $u'_i$ is $n+1$ and there are $n$ different $u'_i$, 
each $u'_i$ is connected to at most one $S'_j$, since otherwise, the weight of the tree would exceed $k'$.
After connecting the $u'_i$ to the Steiner vertices, either, all $u'_i$ are connected to the same $S'_j$, 
or there are multiple disconnected components.
In the first case, the edge $(v_0,S'_j)$ must be part of the solution to connect the last terminal vertex, 
which leads to a total weight of $k'$, completing the tree.
Otherwise, the components must be connected. 
At this point, the weight of the tree is already $n^2+n$, thus, only a weight of $n$ can be added to not exceed $k'$.
Thus, the components cannot be connected via some $u'_i$, as the weight of an extra edge connecting $u'_i$ to some 
$S'_j$ is $n+1$. Hence, all components must be connected to $v_0$ via the Steiner vertices.
As we assume $A^*$ to be a solution, and the weights of the edges connecting $v_0$ to $S'_j$ is exactly the number of terminal vertices that are adjacent
to $S'_j$, for each $S'_j$ in the solution, all edges incident to $S'_j$ must be in the solution, as otherwise,
the weight would extend $k'$. Thus, the sets $S_j$ corresponding to the Steiner vertices $S'_j$ of which 
$(v_0, S'_j)$ is in the solution of the \textsc{Steiner Tree} instance form a solution to the \textsc{Exact Cover} instance. 
Again, $f_{(U, \mathfrak{S})}$ directly gives us the corresponding solution.\\
Thus, 
\begin{align*}
    & \{f_{(U, \mathfrak{S})}(S) \mid S\in\S(\textsc{Exact Cover})\}\\
   =& \{f_{(U, \mathfrak{S})}(T) \mid T\subseteq\mathfrak{S}, T_i, T_j\text{ are disjoint }\forall T_i, T_j\in T, \bigcup T_i = \bigcup \mathfrak{S} \} \\
   =& \{E \mid E=\{ (v_0, s'_j) \mid S_j\in T \}, (v_0,S'_j)\text{ is part of a solution to the}\\
   &\text{\textsc{Steiner Tree} instance}\}\\
   =& \{S'\cap f_{(U, \mathfrak{S})}(\mathfrak{S})\mid S'\in \S(\textsc{Steiner Tree})\}
\end{align*}

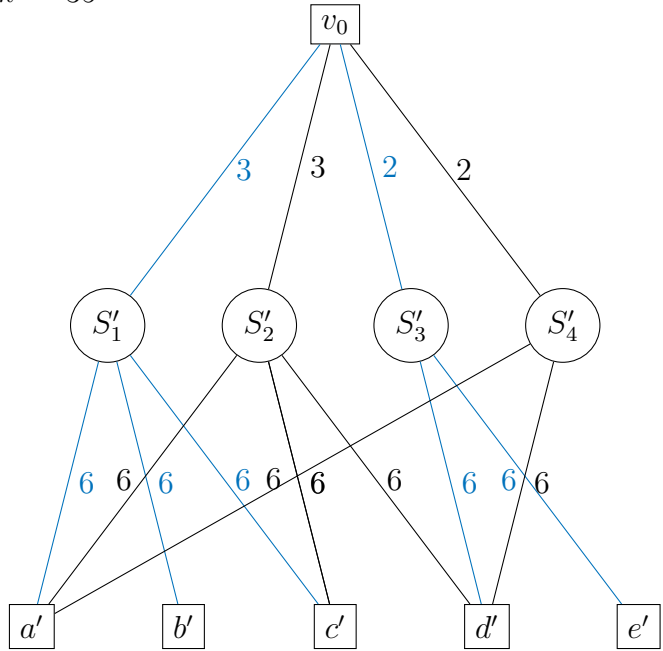
\begin{figure}
    \centering
    \begin{minipage}[t]{0.39\textwidth}
        \ECexample
    \end{minipage}\hfill
    \begin{minipage}[t]{0.59\textwidth}
        \textsc{Steiner Tree} instance:\\
        $T' = \{v_0, a',b',c',d',e'\}\\
        S' = \{S'_1,S'_2,S'_3,S'_4\}\\
        k'=35$\\
        \resizebox{\textwidth}{!}{
        \begin{tikzpicture}
            \node[draw, rectangle] (v0) at (4,8) {$v_0$};
            \node[draw, circle] (S1) at (1,4) {$S'_1$};
            \node[draw, circle] (S2) at (3,4) {$S'_2$};
            \node[draw, circle] (S3) at (5,4) {$S'_3$};
            \node[draw, circle] (S4) at (7,4) {$S'_4$};
            \node[draw, rectangle] (a) at (0,0) {$a'$};
            \node[draw, rectangle] (b) at (2,0) {$b'$};
            \node[draw, rectangle] (c) at (4,0) {$c'$};
            \node[draw, rectangle] (d) at (6,0) {$d'$};
            \node[draw, rectangle] (e) at (8,0) {$e'$};

            \path[]
                (v0) edge [color=rwth-blue] node[right] {3} (S1)
                (v0) edge node[right] {3} (S2)
                (v0) edge [color=rwth-blue] node[right] {2} (S3)
                (v0) edge node[right] {2} (S4)
                (S1) edge [color=rwth-blue] node[right] {6} (a)
                (S1) edge [color=rwth-blue] node[right] {6} (b)
                (S1) edge [color=rwth-blue] node[right] {6} (c)
                (S2) edge node[right] {6} (c)
                (S2) edge node[right] {6} (c)
                (S2) edge node[right] {6} (d)
                (S2) edge node[left] {6} (a)
                (S3) edge [color=rwth-blue] node[right] {6} (d)
                (S3) edge [color=rwth-blue] node[left] {6} (e)
                (S4) edge node[left] {6} (a)
                (S4) edge node[right] {6} (d);
        \end{tikzpicture}
        }
    \end{minipage}\hfill
    \caption{Example of a reduction from \textsc{Exact Cover} to \textsc{Steiner Tree}.}
    \label{fig:exact-cover-steiner}
\end{figure}

\subsubsection{\textsc{Exact Cover} \SSP{} \textsc{3-Dimensional Matching}}
\EC
Karp defines the problem \textsc{3-Dimensional Matching} slightly different~\citeLiterature{karp1975}, 
but as the following definition used is the more common definition
for the problem, and Karp's definition is a special case thereof, we use this definition:
\threeDmatch

This reduction is a bit more complicated and not as intuitive as the previous ones. It is based on Karp~\citeLiterature{karp1975}.

\functionG
Without loss of generality, we assume $\vert S_j\vert\geq 2$ for all $j\in\{1,\ldots,m\}$. 
The dimensions $X',Y',Z'$ in the \textsc{3-Dimensional Matching} instance $(X',Y',Z',U',k')$ are composed of
pairs that correspond to the elements from $U$ of the \textsc{Exact Cover} instance $(U, \mathfrak{S})$, 
thus the elements are of the form $(i,j)$, such that $u_i\in S_j$ and  $X':=Y':=Z':=\{(i,j)\mid u_i\in S_j\}$.
Next, we define an arbitrary injective function $\alpha: U\rightarrow X'$, 
thus mapping each element $u_i\in U$ to some element in $X'$.
Additionally, $\pi: X'\rightarrow X'$ is a permutation that maps each pair $(i,j)$ to the pair with the same $j$ and 
the next higher $i'$ such that $u_i'\in S_j$, thus 
\[\pi((i,j))=(i',j) \mid u_{i'}\in S_j,\: \nexists h, i<h<i'\text{ s.t. }u_h\in S_j.\]

Next up we need to create the tuples $(x,y,z)\in X'\times Y'\times Z'$. 
The tuples are divided into two sets, $U_1$ and $U_2$.
$U_1$ contains the tuples that contain the elements which $\alpha$ maps to in the first dimension,
and these are the sets that correspond to the elements of the \textsc{Exact Cover} instance.
The set $U_2$ is the garbage collection set:

\begin{align*}
    U_1&:=\{(\alpha(u_i),(i,j),(i,j))\mid(i,j)\in X'\}\\
    U_2&:=\{(x, (i,j), \pi((i,j)))\mid x\in X',\: x\neq\alpha(u_{i'})\text{ for any } i'\text{ and any } (i,j)\in X'\}\\
    U'&:=U_1\cup U_2
\end{align*}
In $U_1$, the second and third dimension are the same in each tuples, with the first dimension containing an element
from $\alpha$. As we need to select as many tuples from $U_1$ as we had elements in the original instance, 
this means that for each element to which an element from the original instance is mapped, one
of the tuples from $U_1$ must be chosen.

In the garbage collection set, the remaining elements that are not part of $\alpha$ appear in the first dimension,
paired with any element $y\in Y'$ and the permutation of $y$ in the third dimension, thus $\pi(y)\in Z'$.
This way, since we again need to select one tuple for each element from the first dimension,
the tuples that are selected have to be from different $S_j$ than the tuples selected in $U_1$, as otherwise, the
second or third dimension would overlap at some point. 

Thus, $U_1$ contains only tuples with the first element being part of the image of $\alpha$, 
and $U_2$ contains tuples with the first element not being part of the image of $\alpha$.
Finally, we set $k'=\vert X'\vert$.

An example of the reduction can be seen in~\cref{fig:exact-cover-3Dmatch}, however, as the reduction is quite complicated,
an extra explanation is needed:\\
In $U_1$ we can see, that the first component is always an $\alpha(x_i)$. In $U_2$, all possible first components that
are not in the image of $\alpha$ are matched with all possible second components, and their permutation in the third component.
The dots in~\cref{eq:U2.2,eq:U2.4} indicate that the tuples are generated in the same way as the tuples in~\cref{eq:U2.1,eq:U2.3,eq:U2.5}.
Now, one can select the tuples~\cref{eq:U1.1,eq:U1.2,eq:U1.3,eq:U1.4,eq:U1.5}, 
which correspond to the elements $x_1,\ldots, x_5$ in $S_1$ and $S_2$.
Then, $n=5$ tuples have been selected. We need to select two more tuples from $U_2$, one with $(2,3)$ in the first
dimension and one with $(4,3)$ in the first dimension. These tuples must have elements in the second and third dimension
that correspond to a set $S_j$, in this case $S_2$, that is not part of the selections from $U_1$, 
because of the permutation $\pi$.

\functionF 
We define the functions $f_I:\mathfrak{S}\rightarrow U'$ by
$f_{(U, \mathfrak{S})}(S_j) = (\alpha(u_i), (i,j), (i,j))$ for the smallest $i$ such that $u_i\in S_j$.

\correctness
First, one can see that $\vert X'\vert = \sum_{j\in\{1,\ldots,m\}}\vert S_j\vert$.
A maximum of $n$ tuples $(\alpha(u_i),(i,j),(i,j))$ can be chosen from $U_1$, as there are exactly $n$ different $\alpha(u_i)$.\\
Assume, $W^*$ is a solution to the \textsc{3-Dimensional Matching} instance. Then, in each dimension, each tuple $(i,j)\in X'$ is covered exactly once.
If a tuple $(\alpha(u_i),(i,j),(i,j))\in W^*$, the tuples $(\alpha(u_{i'}),(i',j),(i',j))$ must also be contained in $W^*$ for all $i'\neq i, u_{i'}\in S_j$,
as otherwise, at least one pair $(i',j)$ can not be chosen as an element in the second dimension, because of the nature of the permutation $\pi$,
which would lead to $\pi(i',j)=(i,j)$ for that $i'$, and thus, the third dimension of the two tuples being the same.
We conclude: if one tuple from $U_1$ containing $j$ is chosen, no tuple in $U_2\cap W^*$ can contain $j$, thus, every pair $(i',j)$ must be
contained in some tuple in $U_1\cap W^*$. Especially, the tuple $(\alpha(u_i),(i,j),(i,j))$ with $i$ being the smallest index of $u_i\in S_j$ must be in $W^*$,
which indicates that $S_j$ is part of the solution to the \textsc{Exact Cover} instance.\\
Furthermore, if $(\alpha(u_i),(i,j),(i,j))\in W^*$, no other tuple containing $i$ can be in $U_1\cap W^*$, 
as the first dimension would be the same in that case.
Thus, $U_1\cap W^*$ contains $\alpha(u_i)$ for all $i$, and only $j$ so that the $S_j$ are pairwise disjoint.
The maximum of tuples $(x, (i,j), \pi((i,j)))\in U_2$ that can be part of the solution is $\vert X'\vert - n$, 
which is the number of different $x$, thus for $W^*$ to be a solution, $\vert W^*\cap U_2\vert = \vert X'\vert - n$ must hold.
The remaining $(i,j)$ that are not part of the second and third dimension of any tuple in $U_1\cap W^*$ can be combined with the $x$ randomly, 
as we have already argued that the permutation can not result in a conflict.\\

Assume, $\mathfrak{S}^*\subseteq\mathfrak{S}$ is a solution of the \textsc{Exact Cover} instance.
Then, one can derive the set $W^*$ to the solution easily by applying $f$ to all $S_j\in \mathfrak{S}^*$ and combine the remaining
$(i,j)$ with any $x$ to compose the remaining tuples in $U_2\cap W^*$, which is a solution to the \textsc{3-Dimensional Matching} instance
by the same argument as above.\\
We thus have shown that the reduction is a correct SSP reduction:
\begin{align*}
    & \{f_{(U, \mathfrak{S})}(S) \mid S\in\S(\textsc{Exact Cover})\}\\
   =& \{f_{(U, \mathfrak{S})}(\mathfrak{S}^*) \mid T\subseteq\mathfrak{S}, T_i, T_j\text{ are disjoint }\forall T_i, T_j\in T, \bigcup T_i = \bigcup \mathfrak{S} \} \\
   =& \{ W \mid W :\{(\alpha(u_i), (i,j), (i,j)) \mid i\text{ is the smallest number}\\
   & \text{ such that } u_i\in S_j\text{ and } S_j\in T\} \text{ and }(\alpha(u_i), (i,j), (i,j))\text{ is part of a solution to the }\\
   &\text{\textsc{3-Dimensional Matching} instance}\}\\
   =& \{S'\cap f_{(U, \mathfrak{S})}(\mathfrak{S})\mid S'\in \S(\textsc{3-Dimensional Matching})\}
\end{align*}

The reduction runs in polynomial time, as at most $3(nm)$ elements are generated, 
and is also a reduction from \textsc{Exact Cover} to \textsc{3-Dimensional Exact Matching}, 
as $k'$ is set to the size of the dimensions, which all have the same size.
\begin{figure}
    \centering
    \begin{minipage}[t]{0.49\textwidth}
        \textsc{Exact Cover} instance:\\
        Elements: $\{a,b,c,d,e\}$\\
        Set of subsets $\mathfrak{S}$:
        \begin{align*}
            \{ \fcolorbox{rwth-blue}{white}{$S_1$}&=\{x_1,x_2,x_4\},\\
            \fcolorbox{rwth-blue}{white}{$S_2$}&=\{x_3,x_5\},\\
            S_3&=\{x_2,x_4\}\}
        \end{align*}
    \end{minipage}\hfill
    \begin{minipage}[t]{0.49\textwidth}
        \textsc{3-Dimensional Matching} instance:
        $X'=Y'=Z'=\\
        \{(1,1),(2,1),(4,1),
        (3,2),(5,2),(2,3),(4,3)\}$
        \begin{align*}
            \alpha(x_1)&=(1,1)\\
            \alpha(x_2)&=(2,1)\\
            \alpha(x_3)&=(4,1)\\
            \alpha(x_4)&=(3,2)\\
            \alpha(x_5)&=(5,2)
        \end{align*}
        \begin{align}
            U_1&=\{\notag{}\\
            &\fcolorbox{rwth-blue}{white}{$((1,1),(1,1),(1,1))$},\label{eq:U1.1}\\
            &\fcolorbox{rwth-blue}{white}{$((2,1),(2,1),(2,1))$},\label{eq:U1.2}\\
            &\fcolorbox{rwth-blue}{white}{$((3,2),(4,1),(4,1))$},\label{eq:U1.3}\\
            &\fcolorbox{rwth-blue}{white}{$((4,1),(3,2),(3,2))$},\label{eq:U1.4}\\
            &\fcolorbox{rwth-blue}{white}{$((5,2),(5,2),(5,2))$},\label{eq:U1.5}\\
            &((2,1),(2,3),(2,3)),\label{eq:U1.6}\\
            &((3,2),(4,3),(4,3))\}\label{eq:U1.7}\\
            U_2&=\{\notag{}\\
            &((2,3),(1,1),(2,1)),\label{eq:U2.1}\\
            &((2,3),(2,1),(4,1)),\notag{}\\
            &((2,3),(4,1),(1,1)),\notag{}\\
            &\ldots,\label{eq:U2.2}\\
            &\fcolorbox{rwth-blue}{white}{$((2,3),(2,3),(4,3))$},\label{eq:U2.3}\\
            &((2,3),(4,3),(2,3)),\notag{}\\
            &((4,3),(1,1),(2,1)),\notag{}\\
            &((4,3),(2,1),(4,1)),\notag{}\\
            &\ldots,\label{eq:U2.4}\\
            &((4,3),(2,3),(4,3))\label{eq:U2.5}\\
            &\fcolorbox{rwth-blue}{white}{$((4,3),(4,3),(2,3))$}\}\notag{}\\
            k'=&7\notag{}
        \end{align}
    \end{minipage}\hfill
    \caption{Example of a reduction from \textsc{Exact Cover} to \textsc{3-Dimensional Matching}.}
    \label{fig:exact-cover-3Dmatch}
\end{figure}

\subsubsection{\textsc{Exact Cover} \SSP{} \textsc{Subset Sum}}
\EC
The problem \textsc{Subset Sum} is a special case of the KNAPSACK problem, where the weights of the items are equal to their values.
In Karp's original paper, the problem was called KNAPSACK.
\subsum
The polynomial reduction used in this SSP reduction was first discovered by Karp~\citeLiterature{karp1975}.
The idea of this reduction is basically to encode all sets as a string of 0 and 1, encoding which elements are in the set,
and to use this string as a number in base $m+1$ to represent the set. This way, each combination of elements forms a unique number
from which the elements in it can be directly read.\\
Now, the sum of all elements must be the string containing only ones, as we want all elements to be covered by the selected sets.
Thus, this directly translates to that the sum of all strings must be the string containing only ones, 
which translates to a set of sets that contains all elements exactly once.

\functionG
The reduction is exactly the one presented in~\citeLiterature{karp1975} and the reduction function 
$g$ going from $(U,\mathfrak{S})$ to $(A', M')$ is defined as follows:\\
We introduce a helper variable $d:=m+1$ and a helper matrix $E\in {\{0,1\}}^{m\times n}$ 
that contains the strings representing the sets as its rows with
\[
    e_{ij}:=\begin{cases}
        1 & \text{if } u_i \in S_j\\
        0 & \text{otherwise}  
    \end{cases}  
\]
With these, we can finally define the reduction function.
\begin{align*}
    n'&:=m\\
    M'&:=\frac{d^n-1}{d-1}\\
    a'_j&:=\sum_{i\in\{1,\ldots, n\}}e_{ij}d^{i-1}\\
\end{align*}
Where the target value $M'$ is equal to $\sum_{i\in\{1,\ldots, n\}}d^{i-1}$, thus, 
the string containing only ones in base $d$, and each $a'_j$
is the number in base $d$ that represents the elements in the set $S_j$.
The reduction runs in polynomial time, as the matrices have a size of at most $n\times m$, 
and calculating the sums is also polynomial.

\functionF
The functions $f_I$ are set to $f_{(U,\mathfrak{S})}(S_j)=a_j$ for all $j\in\{1,\ldots,m\}$.

\correctness
First, we take a look into the definition of $M'$. Using the geometric series formula, we see that 
\[\sum_{i=1}^{n}1\cdot d^{i-1}=\sum_{i=0}^{n-1}d^i = \frac{d^n-1}{d-1} = M'\]
Thus, $M'$ is exactly the sum of all elements in the set $U$.
As $d\geq2$, the $a'_i$ are constructed in a way that they directly encode the row $e_j$ of the matrix $E$.

Let us assume that $(A', M')$ is a YES-instance of \textsc{Subset Sum} with $A^*\subseteq A'$ being the solution.
Then, $\sum_{a'_i\in A^*}a'_i =M'$. 
Due to the construction of the $a'_j$, $\sum_{a'_j\in A^*}a'_j = M'$ if and only if $e'=\sum_{a'_j\in A^*} e_{ij} =(1,\ldots,1)$.
This means that the sets $S_j$ are disjoint, as otherwise at least one entry of $e'$ must be greater than 1, and,
that they cover each $u_i\in U$, as otherwise at least one entry of $e'$ must be 0.
Thus, $(U,\mathfrak{S})$ is a YES-instance of \textsc{Subset Sum}.\\
With the same argumentation, if $(U,\mathfrak{S})$ is a YES-instance of \textsc{Subset Sum}, then $(A', M')$ is a YES-instance of \textsc{Subset Sum}.\\

It is also clear by the argumentation, that the $a'_j\in A^*$ directly correspond to the $S_j$ that form a solution to the \textsc{Subset Sum} instance.
Thus, the reduction is a SSP reduction.
\begin{align*}
     & \{f_{(U,\mathfrak{S})}(S) \mid S\in\S(\textsc{Exact Cover})\}\\
    =& \{f_{(\mathfrak{S},U)}(T) \mid T\subseteq\mathfrak{S}, T_i, T_j\text{ are disjoint }\forall T_i, T_j\in T, \bigcup T_i = \bigcup \mathfrak{S} \} \\
    =& \{A^* \mid A^*\subseteq A', \sum_{a_i\in A^*}a_i = M'\}\\
    =& \{S'\mid S'\in \S(\textsc{Subset Sum})\}\\
    =& \{S'\cap f_{(U,\mathfrak{S})}(\mathfrak{S})\mid S'\in \S(\textsc{Subset Sum})\}
\end{align*}

\begin{figure}
    \centering
    \begin{minipage}[t]{0.48\textwidth}
        \ECexample
    \end{minipage}\hfill
    \begin{minipage}[t]{0.49\textwidth}
        \textsc{Subset Sum} instance:\\
        \vspace{-1em}
        \begin{align*}
            &\text{Help Variables:}\\
            d&=m+1=5\\
            E&=\begin{pmatrix}
                1 & 1 & 1 & 0 & 0\\
                1 & 0 & 1 & 1 & 0\\
                0 & 0 & 0 & 1 & 1\\
                1 & 0 & 0 & 1 & 0
            \end{pmatrix}\\
            &\text{Reduction:}\\
            n'&=m=4\\
            M'&=\frac{5^5-1}{5-1}=\frac{3124}{4}&=781\\
            \fcolorbox{rwth-blue}{white}{$a'_1$}&=5^0+5^1+5^2&=31\ \\
            a'_2&=5^0+5^2+5^3&=151\\
            \fcolorbox{rwth-blue}{white}{$a'_3$}&=5^3+5^4&=750\\
            a'_4&=5^0+5^3&=126\\
            &\text{Instance:}\\
            &(\{31,151,750,126\}, 781)
        \end{align*}
    \end{minipage}\hfill
    \caption{Example of the reduction from \textsc{Exact Cover} to \textsc{Subset Sum}.}
    \label{fig:EC-SubsetSum}
\end{figure}

\subsubsection{\textsc{Clique} \SSP{} \textsc{Independent Set}}\label{subsubsec:clique-is}
\clique
\IS

This reduction was first presented by Garey and Johnson in~\citeLiterature{garey1979computers}, section 3.1, 
and is probably the easiest reduction in this paper. 
The reduction from $(G,k)$ to $(G',k')$ just inverts the edges $e\in E$, as a clique and an independent set are complementary problems.
If some vertices form a clique in the original instance and all the edges are inverted, 
they are all pairwise not adjacent in the new instance, thus forming an independent set, and vice versa.

\begin{figure}
    \centering
    \begin{minipage}[t]{0.49\textwidth}
        \textsc{Clique} instance:\\
        \resizebox{\textwidth}{!}{
        \begin{tikzpicture}
            \node[dot=left:$v_1$, color=rwth-blue] (v_1) at (0,0) {};
            \node[dot=left:$v_2$, color=rwth-blue] (v_2) at (0,-2) {};
            \node[dot=left:$v_3$, color=rwth-blue] (v_3) at (2,-1) {};
            \node[dot=right:$v_4$] (v_4) [right=of v_3] {};
    
            \path[]
            (v_1) edge node {} (v_2)
            (v_2) edge node {} (v_3)
            (v_3) edge node {} (v_1)
            (v_3) edge node {} (v_4);
        \end{tikzpicture}
        }
        ~\\ $k=3$
    \end{minipage}\hfill
    \begin{minipage}[t]{0.49\textwidth}
        \textsc{Independent Set} instance:\\
        \resizebox{\textwidth}{!}{
        \begin{tikzpicture}
            \node[dot=left:$v_1$, color=rwth-blue] (v_1) at (0,0) {};
            \node[dot=left:$v_2$, color=rwth-blue] (v_2) at (0,-2) {};
            \node[dot=left:$v_3$, color=rwth-blue] (v_3) at (2,-1) {};
            \node[dot=right:$v_4$] (v_4) [right=of v_3] {};
        
            \path[]
            (v_1) edge node {} (v_4)
            (v_2) edge node {} (v_4);
        \end{tikzpicture}
        }
        ~\\$k'=3$
    \end{minipage}\hfill
    
    \caption{Graph $G$ with $k=3$ and the resulting graph $G'$ with $k'=3$.}
    \label{fig:clique-is}
\end{figure}
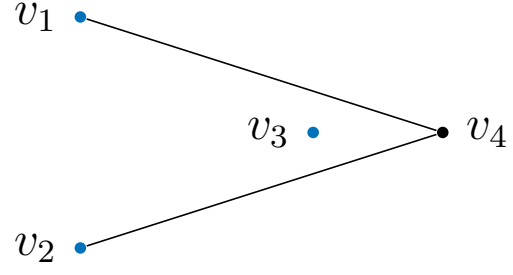

\functionG
The reduction $g$ from the \textsc{Clique} instance $(G,k)$ to the \textsc{Independent Set} instance $(G',k')$ works as follows:
We set $V':=V$ and invert the edges (if there is an edge $(v,w)\in G$ there is no edge $(v,w)\in G'$ and vice versa),
thus $E':=\{(v,w)\mid (v,w)\notin E\}$
and $k':=k$.
An example reduction is shown in \cref{fig:clique-is}.

\functionF
$f_{(G,k)}(v)=v\in V'$. 

\correctness
The correctness follows directly: If there is an independent set of size at least
$k$ in $G'$, the same set of vertices forms a clique in $G$, and if there is no independent set
of size at least $k$ in $G'$ there is no clique of that size in $G$, so we see the correctness of the
solution mapping directly.
Thus
\begin{align*}
    & \{f_{(G,k)}(S) \mid S\in\S(\textsc{Clique})\}\\
    =& \{f_{(G,k)}(V_1^*) \mid \forall v_i, v_j\in V_1^*: (v_i, v_j)\in E \} \\
    =& \{ V_2^* \mid V_2^*\subseteq V', \forall v_i, v_j\in V_2^*: (v_i, v_j)\notin E \}\\
    =& \{S'\mid S'\in \S(\textsc{Independent Set})\}\\
    =& \{S'\cap f_{(G,k)}(V)\mid S'\in \S(\textsc{Independent Set})\}
\end{align*}

\subsubsection{\textsc{3-SAT} \SSP{} \textsc{Exactly-One-3-SAT}}
\threesat
\exOne

In this reduction from an exercise by Arora and Barak~\citeLiterature{arora2009computational},
we need to transform a formula in which each clause needs to have \textit{at least one} literal set to true 
to a formula in which each clause has \textit{exactly one} literal set to true.
To do so, we introduce helper variables for each literal in each clause as well as new clauses, 
in a way that if some literal $\ell_j$ is set to true,
the corresponding helper variable $Z_{j,i}$ \textit{can} be set to true, but if $\ell_j$ is set to false, 
$Z_{j,i}$ \textit{must} also be false. Each clause is represented
by the helper variables $Z_{j,i}$ instead of the literals itself, plus clauses that ensure the correct 
behavior of the helper variables.

\functionG
The function $g$, mapping a \textsc{3-SAT} instance $(L,C)$ to a EXACTLY ONE 3-SAT instance $(L',C')$ 
is based on exercise 2.17 from \citeLiterature{arora2009computational} and is defined as follows:
Let the \textsc{3-SAT} instance be $\bigwedge_{i=1}^{m} C$, 
where $m$ is the number of clauses and $C_i=(\ell_{j_1}\vee \ell_{j_2}\vee \ell_{j_3})$.
Then, for each clause $C_i$, we introduce four new clauses and new variables:
\begin{align}
    C_i' =  & (Z_{j_1,i}     \vee Z_{j_2,i} \vee Z_{j_3,i})\tag{$C_i^1$}\\
    \wedge    & (\overline{\ell}'_{j_1} \vee Z_{j_1,i} \vee h_{j_1,i})\tag{$C_i^2$}\\
    \wedge    & (\overline{\ell}'_{j_2} \vee Z_{j_2,i} \vee h_{j_2,i})\tag{$C_i^3$}\\
    \wedge    & (\overline{\ell}'_{j_3} \vee Z_{j_3,i} \vee h_{j_3,i})\tag{$C_i^4$}
\end{align}
Thus, for each literal $\ell_j$ in $C_i$, we create two new variables $Z_{j,i}, h_{j,i}$ in a way that ensures that
if $\ell'_j$ is set to false, $Z_{j,i}$ must also be false, but if $\ell'_j$ is set to true, $Z_{j,i}$ 
can be either true or false, using the helper variable $h_{j,i}$ in $C_i^2, C_i^3, C_i^4$.
We set $g(L,C)=\bigwedge_{i=1}^{m} C_i'$.

\functionF
We define the functions $f_I:L\rightarrow L'$ to read a solution for our \textsc{3-SAT} instance $(L,C)$ from the
created \textsc{Exactly-One-3-SAT} instance $(L',C')$ as follows: $f_{(L,C)}(\ell_j)=\ell_j'$.

\correctness
Let us assume, $(L,C)$ over the literals $\ell_1,\ldots,\ell_n$ is a YES-instance of \textsc{3-SAT}. 
Then, there is an assignment $\alpha$ of the literals such that each clause is satisfied. 
If we set $\ell'_1,\ldots,\ell'_n$ to the same assignment, we obtain a partial assignment for $L'$, $L'(\alpha, Z, h)$.
The clauses $C_i^2, C_i^3, C_i^4$ can always be satisfied, 
as two of the literals in the clauses can be assigned without any constraints, and at least
one $Z_{j,i}$ can be set to true, as at least one $\ell_j$ is set to true, which means, $\overline{\ell}'_j$ is set to false, 
thus, $Z_{j,i}$ can be set to true. Thus, $Z, h$ can be assigned so that exactly one literal from each clause in $C'_i$ is true,
which leads to a satisfying assignment for $(L',C')$ through the function $f$.\\

Now, let us assume, $(L',C')$ is a YES-instance of \textsc{Exactly-One-3-SAT}. 
Then, in each clause in $C'$, exactly one literal is set to true. So, for one $C'_i$, one $Z_{j,i}$ is set to true,
which means that the corresponding $\ell'_j$ must be true, as otherwise, $C_i^j$ would not be satisfied. 
Thus, the assignment $\alpha'$ of the literals $\ell'_1,\ldots,\ell'_n$ corresponds to a satisfying 
assignment for $L$ with the same assignment of $\ell_1,\ldots,\ell_n$.\\
As for each clause 4 clauses are introduced, and the function $f$ is linear, 
the running time of the reduction is polynomial in the size of the input.

\begin{align*}
    & \{f_{(L,C)}(S) \mid S\in\S(\textsc{3-SAT})\}\\
    =& \{f_{(L,C)}(L_1^*) \mid L_1^*\subset L \text{ such that }\vert L_1^*\cap\{\ell_i,\overline{\ell}_i\}\vert=1 \text{ and }\vert L_1^*\cap C_j\vert\geq1\text{ for all }C_j\in C \} \\
    =& \{ L_2^* \mid L_2^*\subset L' \text{ such that } \vert L_2^*\cap\{\ell_i,\overline{\ell}_i\}\vert=1, L^*_2 \text{ only contains } \ell_i,\overline{\ell}_i\in L^*_1 \\
    &\text{ and } L^*_2\text{ is part of a satisfying assignment of the \textsc{Ex-One-3-SAT} instance} \}\\
    =& \{S'\cap f_{(L,C)}(L)\mid S'\in \S(\textsc{Ex-One-3-SAT})\}
\end{align*}

\begin{figure}
    \centering
    \threeSATexample
    \vspace*{2\baselineskip}

    \textsc{Exactly-One-3-SAT} instance:
    \begin{align*}
        C_1'' &= (\fcolorbox{rwth-blue}{white}{$Z_{1,1}$}\vee Z_{2,1}\vee Z_{3,1})\wedge (\overline{x}'_1\vee \fcolorbox{rwth-blue}{white}{$Z_{1,1}$}\vee h_{1,1})\wedge (\overline{x}'_2\vee Z_{2,1}\vee \fcolorbox{rwth-blue}{white}{$h_{2,1}$})\wedge (\fcolorbox{rwth-blue}{white}{$x'_3$}\vee Z_{3,1}\vee h_{3,1})\\
        C_2'' &= (Z_{1,2}\vee \fcolorbox{rwth-blue}{white}{$Z_{3,2}$}\vee Z_{4,2})\wedge (\fcolorbox{rwth-blue}{white}{$x'_1$}\vee Z_{1,2}\vee h_{1,2})\wedge (\overline{x}'_3\vee \fcolorbox{rwth-blue}{white}{$Z_{3,2}$}\vee h_{3,2})\wedge (\fcolorbox{rwth-blue}{white}{$\overline{x}'_4$}\vee Z_{4,2}\vee h_{4,2})
    \end{align*}
    Example solution:
    \begin{align*}
        x'_1&=1 && Z_{1,1}=1 && h_{1,1}=0\\
        x'_2&=1 && Z_{2,1}=0 && h_{2,1}=1\\
        x'_3&=1 && Z_{3,1}=0 && h_{3,1}=1\\
        x'_4&=0 && Z_{1,2}=0 && h_{1,2}=1\\
        &       && Z_{3,2}=1 && h_{3,2}=0\\
        &       && Z_{4,2}=0 && h_{4,2}=1
    \end{align*}
    \caption{Reduction from \textsc{3-SAT} to \textsc{Exactly-One-3-SAT}}
    \label{fig:3SAT-ex1-3SAT}
\end{figure}

\subsubsection{\textsc{3-SAT} \SSP{} \textsc{3-Dimensional Matching}}
\threesat{}
\threeDmatch{}

\begin{figure}
    \centering
    \begin{tikzpicture}
        \literalGadget{1}{2}{0}{4}{90}{1}
    \end{tikzpicture}
    \caption{The variable gadget for some variable $x_1$ in a formula with two clauses. The blue triples
    form the set $T_1^t$ and the white triples form the set $T_1^f$.}
    \label{fig:3SAT-3DM-gadget}
\end{figure}
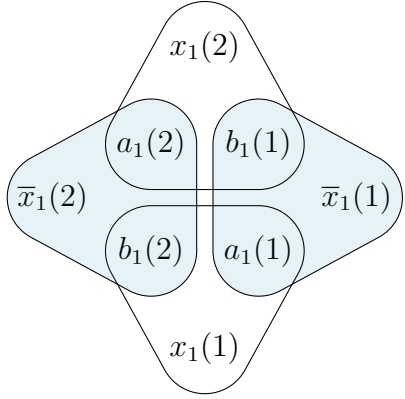

The reduction is inspired by~\citeLiterature{garey1979computers}, but changed slightly 
to match the definition of \textsc{3-Dimensional Matching} used in this paper.
Each variable is represented by a gadget containing triples that are connected in a way they form a set
of True-triples ($T_i^t$) and a set of False-triples ($T_i^f$), using literal-clause elements for each combination 
of the negative or positive literal and a clause, and helper elements.
Each literal-clause element forms a triple with two helper elements.
The blue triples are part of the True-triples and the white triples form the False-triples in,
as shown in~\cref{fig:3SAT-3DM-gadget} for a variable $x_1$ in a formula with two clauses.

\begin{figure}
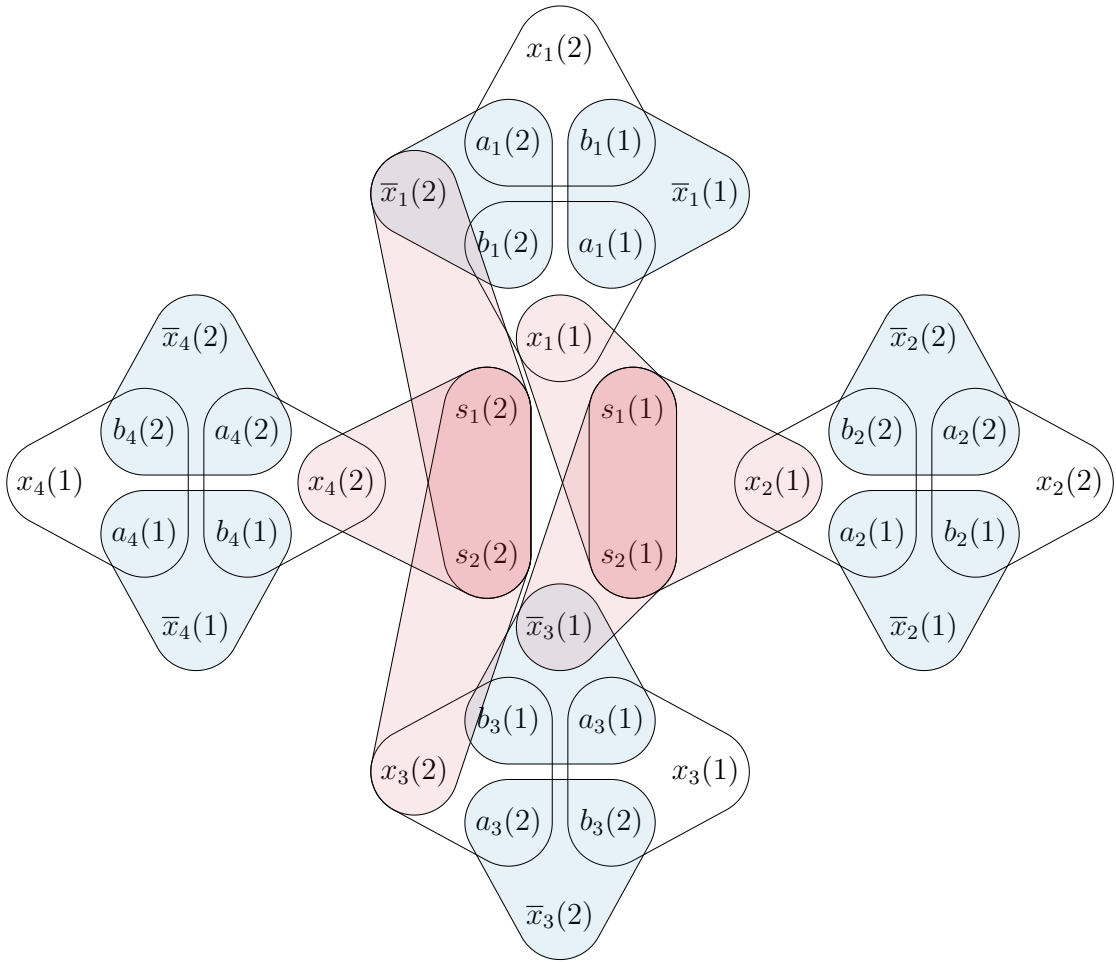

    \begin{minipage}[t]{0.29\textwidth}
        \threeSATexample
    \end{minipage}\hfill
    \begin{minipage}[t]{0.69\textwidth}
        \threeDimexample
    \end{minipage}\hfill
    \threeDimPic
    \centering
    \caption{An example reduction of \textsc{3-SAT} to \textsc{3-Dimensional Matching}.}
    \label{fig:3SAT-3DM-example}
\end{figure}

To simulate the clauses, for each clause, two elements are added, and these two elements then form triples with
the corresponding literal-clause elements in that clause, which are coloured red in~\cref{fig:3SAT-3DM-example}.

Now, we need to find the right $k$. For each variable gadget, we want to select $m:=\vert C\vert$ triples,
as this is exactly half of the triples, and since the triples are connected in a way that the neighbor of
a triple always corresponds to a variable-clause element that contains the negation of the literal in the triple,
this way we can ensure that either all variable-clause elements corresponding to the variable, or all corresponding
to the negated variable can be chosen, but never both. Thus, for all gadgets, this are $nm$ many. 
Additionally, for each clause, we need to select exactly one triple, corresponding to some literal that is 
set to true in the EXACTLY ONE 3-SAT instance, which are $m$ more. 
Thus, we set $k=m(n+1)$. An example for this reduction is depicted in~\cref{fig:3SAT-3DM-example}.

\functionG{}
Let us assume we have a \textsc{3-SAT} instance with the literals $L=\{x_1,\ldots,x_n,\overline{x}_1,\ldots,\overline{x}_n\}$ and the clauses $C=\{C_1,\ldots,C_m\}$.
We construct a \textsc{3-Dimensional Matching} instance with the sets $X, Y, Z$, the subset $U\subseteq(X\times Y\times Z)$ 
and the number $k$ as follows:
\begin{align*}
    X &= \{x_i(j), \overline{x_i}(j) \mid 1\leq i\leq n, 1\leq j\leq m\}\\
    Y &= \{a_i(j)\mid 1\leq i\leq n, 1\leq j\leq m \}\cup\{s_1(j)\mid 1\leq j\leq m\}\\
    Z &= \{b_i(j)\mid 1\leq i\leq n, 1\leq j\leq m \}\cup\{s_2(j)\mid 1\leq j\leq m\}\\
    k &= m(n+1)
\end{align*}
Thus, the set $X$ contains an element for each combination of a literal and an arbitrary clause,
$Y$ (and $Z$) contains the helper elements $a_i(j)$ ($b_i(j)$ respectively) and $s_1(j)$ ($s_2(j)$ respectively),
where the $a$ ($b$) elements will be part of the literal gadgets, and the $s$ elements will later form the clause triples.\\

Next, we need to construct $U$. First, the idea will be explained: $U$ is made up off the two sets $T$ and $C'$,
where the function of $T$ is to fix a valid assignment of the variables and the function of $C'$ is to ensure that
each clause is satisfied, as explained at the beginning of this reduction.
For each literal $\ell_i, 1\leq i\leq n$, we introduce the sets $T_i^t, T_i^f$ forming the literal gadget as follows:
\begin{align*}
    T_i^t &= \{(\overline{x}_i(j), a_i(j), b_i(j))\mid 1\leq j\leq m\}\tag{blue}\\
    T_i^f &= \{(x_i(j), a_i(j), b_i(j+1))\mid 1\leq j\leq m\}\text{ where } b_i(m+1)=b_i(1)\tag{white}\\
    \text{ and we set }\\
    T_i &=T_i^t\cup T_i^f\\
    T&=\bigcup_{i=1}^{n}T_i
\end{align*}
This gadgets form the circles described above, such that later, either all triples from $T_i^t$ or all triples from $T_i^f$ can be chosen.
We can see that the maximum number of triples that can be chosen from $T$ is bounded by $nm$, as otherwise,
two elements would overlap in the second or third dimension.

For each clause $C_j, 1\leq j\leq m$, we introduce the set $C'_j$ as follows:
\begin{align*}
    C'_j =& \{(\ell_i(j), s_1(j), s_2(j))\mid \ell_i\in C_j\}\\
    \text{ and we set}\\
    C'=&\bigcup_{j=1}^{m}C'_j
\end{align*}
The idea of $C'$ is that for each clause $C_j$, one triple can be chosen, which corresponds to a literal in the formula
that is set to true and thus ensures that each clause is satisfied.
The number of triples that can be chosen from $C'$ is bounded by $m$, due to the second and third dimension.
Finally, we set $U=T\cup C'$.

Thus, we see that in order to find a valid matching of size at least $k$, we need to select exactly $m$ triples 
from $C'$ and $nm$ triples from $T$, as we have established that $\vert Y\vert =\vert Z\vert= m(n+1)=k$.
This means for $T$ that for each $i$, either all triples from $T_i^t$ can be chosen or all triples from $T_i^f$, 
as that is the only way to select $nm$ triples without collisions in the second dimension.
From $C'$, for each clause, exactly one triple must be chosen.

\functionF
The functions $f_I: L\rightarrow U$ are defined as follows: 
    \[f_{(L,C)}(\ell_i)=
    \begin{cases}
        (\overline{x_i}(1), a_i(1), b_i(1))  &\text{if }\ell_i=x_i\\
        (x_i(1), a_i(1), b_i(2))             &\text{if }\ell_i=\overline{x}_i
    \end{cases}\]
In~\cref{fig:3SAT-3DM-example}, an example of the construction is shown. The graphic visualizes the \textsc{3-Dimensional Matching} instance,
with four gadgets corresponding to the set $T$, and the triples in the center corresponding to the clauses.
One can directly observe that for one $x_i$, the triples from $T_i^t$ exclude the triples from $T_i^f$ if they are selected, and vice versa.

\correctness
Let $L^*\subseteq L$ be a satisfying assignment for the \textsc{3-SAT} instance $(L,C)$.
For each $x_i\in L^*$ we select all triples $\{(\overline{x}_i(j), a_i(j), b_i(j))\mid 1\leq j\leq m\}$ from $T_i^t$
and for each $\overline{x}_i\in\phi$ all triples $\{(x_i(j), a_i(j+1), b_i(j))\mid 1\leq j\leq m\}$ from $T_i^f$. 
As for each $x_i$, either $x_i$ or $\overline{x}_i$, but never both are in $L^*$, 
$nm$ triples are selected from $T$. Thus, no more triples can be selected from $T$.

For each clause $C_j\in C$, we select one triple from $C'_j$, which is possible, because $L^*$ satisfies $(L,C)$,
thus some literal in $C_j$ is part of $L^*$, which means that in the corresponding triples chosen from $T_i$, 
the first dimension contains the negated literal. 
Thus, choosing the corresponding triple from $C'_j$ does not lead to a collision.

Thus, we obtain a matching $U^*\subseteq U$ of size $k$. For every possible solution of $(L,C)$, the corresponding
triples $f_{(L,C)}(\ell_i)$ are part of a valid matching, as each of the $T_i^t$ or $T_i^f$ respectively can be chosen without a collision.\\

Let $U^*\subseteq U$ be a matching of size $k$. Then, we know that $mn$ triples from $T$ and $m$ triples from $C'$ are chosen.
Because of the construction of $T$, for each $i$, either all triples from $T_i^t$ or all triples from $T_i^f$ are chosen.
If $T_i^t$ is chosen, $(\overline{\ell}_i(j), a_i(j), b_i(j))$ is also part of the matching, thus, according to the function $f_{(L,C)}$, 
$\ell_i$ is part of the satisfying assignment, and $\overline{l_i}$ otherwise.

From $C'$, each of the $m$ chosen triples must belong to a different $C'_j$, as otherwise, the second dimension of
two triples would match. Thus, for each clause $C_j$, at least one literal is satisfied, which directly completes the proof,
as the assignment of the literals provided by $f_I$ is a satisfying assignment for $(L,C)$.
Furthermore, all possible solutions of $(L,C)$ are also a solution of the constructed \textsc{3-Dimensional Matching} instance and vice versa,
since for the $i$ that are not chosen from $C'$, it can be arbitrarily decided whether to choose $T_i^t$ or $T_i^f$,
and the same goes for the literals that are not necessary for satisfying the clauses in $\phi$.
We have shown that the reduction is a complete SSP reduction.
\begin{align*}
    & \{f_{(L,C)}(S) \mid S\in\S(\textsc{3-SAT})\}\\
    =& \{f_{(L,C)}(L^*) \mid L^*\subset L \text{ is a solution to }(L,C) \} \\
    =& \{ W^* \mid W^*\subseteq U, \vert W^*\vert = n, t\in W^*\text{ contains } \ell_i(1) \text{ if } \ell_i\in L^* \}\\
    =& \{ W \mid W\subseteq U, W\text{ is solution to } g(L,C) \}\\
    & \cap \{ W^* \mid W^*\subseteq U, \vert W^*\vert = n, t\in W^*\text{ contains } \ell_i(1) \}\\
    =& \{S'\cap f_{(L,C)}(L)\mid S'\in \S(\textsc{3-Dim Matching})\}
\end{align*}

\subsubsection{\textsc{Directed Hamiltonian Path} \SSP{} \textsc{Undirected Hamiltonian Path}}
\dirHamPath
\undirHamPath

This reduction works analogous to the reduction from \textsc{Directed Hamiltonian Cycle} to \textsc{Undirected Hamiltonian Cycle} in~\citeLiterature{grune2023large}.
It forces the path in the \textsc{Undirected Hamiltonian Path} instance $(G',s',t')$ to keep the direction 
of the edges in the \textsc{Directed Hamiltonian Path} instance $(G,s,t)$, 
through adding in- and out-vertices for each vertex.
The edges are then added from the out-vertex of $v$ to the in-vertex of $w$ for each $(v,w)\in A$.

\begin{figure}
    \centering
    \begin{minipage}[t]{0.49\textwidth}
        \textsc{Directed Hamiltonian Path} \\instance:\\
        \resizebox{\textwidth}{!}{
        \begin{tikzpicture}
            \node[color=rwth-magenta] (v1) {$v_1$};
            \node[color=rwth-magenta] (v2) [right=2of v1] {$v_2$};
            \node[] (v3) [below left=3 and 1of v1] {$v_3$};
            \node[] (v4) [right=2of v3] {$v_4$};
            \node[] (v5) [right=2of v4] {$v_5$};
    
            \path[->]
                (v1) edge [color=rwth-blue] node {} (v3)
                (v2) edge node {} (v1)
                (v3) edge [color=rwth-blue] node {} (v4)
                (v4) edge [bend left, color=rwth-blue] node {} (v5)
                (v4) edge node {} (v1)
                (v5) edge [bend left] node {} (v4)
                (v5) edge [color=rwth-blue] node {} (v2);
        \end{tikzpicture}
        }~\\\\\\\\
        $s=v_1, t=v_2$
    \end{minipage}\hfill
    \begin{minipage}[t]{0.49\textwidth}
        \textsc{Undirected Hamiltonian Path} instance:\\
        \resizebox{\textwidth}{!}{
        \begin{tikzpicture}
            \node[] (v1) {$v_1$};
            \node[color=rwth-magenta] (v1in) [above=0.3 of v1] {$v'_{1in}$};
            \node[] (v1out) [below=0.3of v1] {$v'_{1out}$};
    
            \node[] (v2) [right=2of v1] {$v_2$};
            \node[] (v2in) [above=0.3 of v2] {$v'_{2in}$};
            \node[color=rwth-magenta] (v2out) [below=0.3of v2] {$v'_{2out}$};
    
            \node[] (v3) [below left=3 and 1 of v1] {$v_3$};
            \node[] (v3in) [above=0.3 of v3] {$v'_{3in}$};
            \node[] (v3out) [below=0.3of v3] {$v'_{3out}$};
    
            \node[] (v4) [right=2of v3] {$v_4$};
            \node[] (v4in) [above=0.3 of v4] {$v'_{4in}$};
            \node[] (v4out) [below=0.3of v4] {$v'_{4out}$};
    
            \node[] (v5) [right=2of v4] {$v_5$};
            \node[] (v5in) [above=0.3 of v5] {$v'_{5in}$};
            \node[] (v5out) [below=0.3of v5] {$v'_{5out}$};
    
            \foreach \vertex in {1,...,5}{
                \path[color=rwth-blue]
                    (v\vertex) edge node {} (v\vertex in)
                    (v\vertex) edge node {} (v\vertex out);
            }
            \path[]
                (v1out) edge [color=rwth-blue] node {} (v3in)
                (v2out) edge node {} (v1in)
                (v3out) edge [color=rwth-blue] node {} (v4in)
                (v4out) edge [color=rwth-blue] node {} (v5in)
                (v4out) edge node {} (v1in)
                (v5out) edge node {} (v4in)
                (v5out) edge [color=rwth-blue] node {} (v2in);
        \end{tikzpicture}
        }\\~
        $s'=v'_{1in}, t'=v'_{2out}$
    \end{minipage}\hfill
    \caption{Example of the reduction from \textsc{Directed Hamiltonian Path} to \textsc{Undirected Hamiltonian Path}.}
    \label{fig:dirHamPath-undirHamPath}
\end{figure}
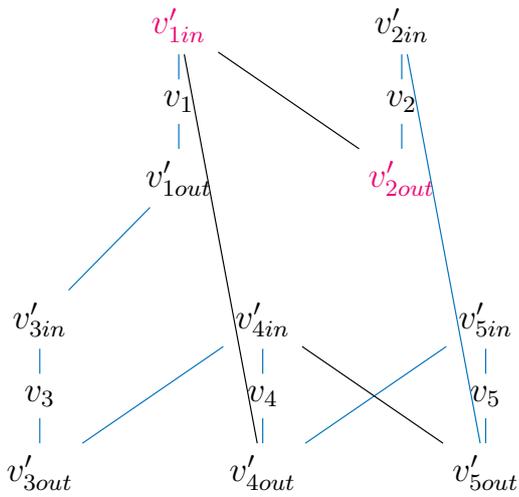

\functionG
We define the reduction function $g$ from an \textsc{Directed Hamiltonian Path} instance $(G,s,t)$ to an 
\textsc{Undirected Hamiltonian Path} instance $(G',s',t')$ as follows:
We first transform each vertex $v\in V$ to three vertices $v'_{in}, v', v'_{out}\in V'$.
Then, for each $v\in V$, we add the edges $(v'_{in},v')$ and $(v',v'_{out})$ to $E'$.
Finally, we add the edges $(v'_{out},w'_{in})$ for all $(v,w)\in A$ to $E'$ and set $s'=s'_{in}$ and $t'=t'_{out}$.
In~\cref{fig:dirHamPath-undirHamPath}, 
an example with 5 vertices is shown that illustrates the reduction.

\functionF
We then define $f_I: A\rightarrow E'$ as $f_{(G,s,t)}((v,w))=(v_{out}, w_{in})$.

\correctness
Let $(G,s,t)$ be a YES-instance of \textsc{Directed Hamiltonian Path}, thus, $G$ has a Hamiltonian path $s, v_1,\ldots, v_l, t$ from $s$ to $t$.\\
Then, $s'_{in}, s', s'_{out}, v'_{1,in}, v_1', v'_{1,out}, \ldots, v'_{l, in}, v_l', v'_{., out}, t'_{in}, t', t'_{out}$ 
is a Hamiltonian path in $G'$ from $s'_{in}$ to $t'_{out}$, as each vertex is included exactly once. 
Also, if $(v,w)\in A$ is part of a solution of $(G,s,t)$, then $(v'_{out},w'_{in})\in E'$ is part of the corresponding solution in $(G',s',t')$, 
thus, the solution mapping also works correctly.\\

Let $(G',s'_{in},t'_{out})$ be a YES-instance of \textsc{Undirected Hamiltonian Path}.
The path must start with the edge $(s'_{in},s')$, as otherwise, $s'$ can never be visited, since then it would 
be entered through the edge $(s'_{out}, s')$, and that would result in a dead end. 
The same argumentation holds for every other vertex-gadget: once $v'_{i,in}$ or $v'_{i, out}$ is entered, all three of the vertices belonging to $v_i$ must be followed.
Thus, as each edge that does not connect vertices belonging to the same $v\in V$ is of the form $(v'_{out},w'_{in})$
such that $(v,w)\in A$, 
the whole path must be of the form $s'_{in}, s', s'_{out}, v'_{1, in}, v_1', v'_{1, out}, \ldots, v'_{l, in}, v_l', v'_{l, out}, t'_{in}, t', t'_{out}$
and therefore correspond to a directed path $s, v_1,\ldots, v_l, t$ in $G$, that contains exactly the $(v,w)$ such that
$(v'_{out},w'_{in})$ is part of the solution.\\
The reduction runs in polynomial time, as $3\vert V\vert$ vertices and $2\vert V\vert+\vert A\vert$ edges are generated.
Thus, we have a correct SSP reduction.

\begin{align*}
    & \{f_{(G,s,t)}(S) \mid S\in\S(\textsc{Dir Hamiltonian Path})\}\\
    =& \{f_{(G,s,t)}(C) \mid C\subseteq A, C\text{ forms a Hamiltonian path from } s\text{ to }t \}\\
    =& \{ C^* \mid C^*\subseteq E', C^*=\{(v_{out}, w_{in}) \mid (v,w)\in A\} \text{ and } C^*\text{ is part of a}\\
    &\text{\textsc{Hamiltonian path} in } G' \}\\
    =& \{S'\cap f_{(G,s,t)}(A)\mid S'\in \S(\textsc{Undir Hamiltonian Path})\}
\end{align*}

\subsubsection{\textsc{Undirected Hamiltonian Path} \SSP{} \textsc{Traveling Salesperson Problem}}
\undirHamPath
\TSP
The reduction works analogous to the reduction from \textsc{Undirected Hamiltonian Cycle} to \textsc{Traveling Salesperson Problem} in~\citeLiterature{grune2023large}.
The graph is transformed into a complete graph, but only the edges from the original graph or the edge connecting the start and end of the path can be used
without exceeding the weight limit. Thus, a path in the original graph will be a TSP tour in the new graph and vice versa.

\begin{figure}
    \centering
    \begin{minipage}[t]{0.49\textwidth}
        \textsc{Undirected Hamiltonian Path} instance:\\
        \begin{tikzpicture}
            \node[] (v1) {$v_1$};
            \node[] (v2) [right=of v1] {$v_2$};
            \node[] (v3) [below=of v1] {$v_3$};
            \node[] (v4) [right=of v3] {$v_4$};
    
            \path[rwth-blue]
                (v1) edge node {} (v2)
                (v1) edge node {} (v3)
                (v3) edge node {} (v4);

            \path[]
                (v2) edge node {} (v3);
        \end{tikzpicture}
        ~\\ $s=v_2, t=v_4$
    \end{minipage}\hfill
    \begin{minipage}[t]{0.49\textwidth}
        \textsc{Traveling Salesperson Problem} instance:\\
        \begin{tikzpicture}
            \node[] (v1) {$v_1$};
            \node[] (v2) [right=of v1] {$v_2$};
            \node[] (v3) [below=of v1] {$v_3$};
            \node[] (v4) [right=of v3] {$v_4$};
    
            \path[]
                (v1) edge [color=rwth-blue] node[pos=0.5, above] {0} (v2)
                (v1) edge [color=rwth-blue] node[pos=0.5, left] {0} (v3)
                (v1) edge node[pos=0.5, below] {1} (v4)
                (v2) edge node[pos=0.5, above] {0} (v3)
                (v2) edge [color=rwth-blue] node[pos=0.5, right] {0} (v4)
                (v3) edge [color=rwth-blue] node[pos=0.5, below] {0} (v4);
        \end{tikzpicture}
        ~\\ $k:=0$
    \end{minipage}\hfill
    \caption{An example of the reduction from \textsc{Undirected Hamiltonian Path} to \textsc{Traveling Salesperson Problem}. The weigths are 0 if the edge is in the original graph, or connects $s$ and $t$.}
    \label{fig:undirHamPath-TSP}
\end{figure}
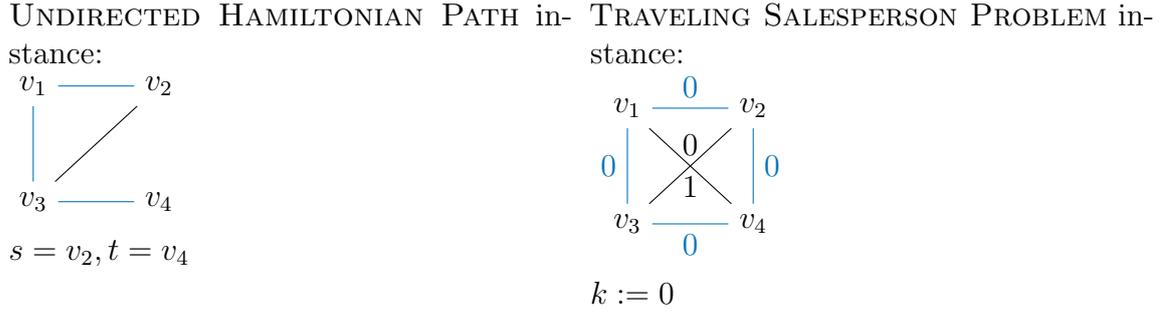

\functionG
We define the reduction function $g$ as follows:
We set $V':=V$ and $E'$ contains all possible edges between the vertices. 
We define the weight function $w':E\rightarrow\mathbb{Z}$ as follows:
\[w(e')=
\begin{cases}
    0, & \text{if }e'\in E\\ 
    0, & \text{if }e'=(s,t)\\
    1, &\text{else}\end{cases}\]
At last, we set $k'=0$, so that only edges from $E$ or the edge connecting the start and end of the path can be used.
\cref{fig:undirHamPath-TSP} shows an example of the reduction.
\functionF
We set $f_{(G,s,t)}(e)=e \in E$.

\correctness
Let $(G,s,t)$ be a YES-instance of \textsc{Undirected Hamiltonian Path}, thus, 
$G$ has a Hamiltonian path $s, v_1,\ldots, v_l, t$ from $s$ to $t$.\\
Then, $s, v_1,\ldots, v_l, t, s$ is a TSP tour in $G'$, as all used edges have a weight of $0$, 
specifically, the tour includes all edges that are part of the solution of $(G,s,t)$.\\

Let $(G',k')$ be a YES-instance of the \textsc{Traveling Salesperson Problem}. 
As only edges from $E$ and the edge $(s,t)$ can be part ot the tour,
and all vertices are covered, the tour without the edge $(s,t)$ is a Hamiltonian path in $G$.
The reduction runs in polynomial time, as at most $\vert V\vert^2$ edges are generated 
and the weight function can be computed in polynomial time.
As the mapping of the solutions is correct, we can conclude that

\begin{align*}
    & \{f_{(G,s,t)}(S) \mid S\in\S(\textsc{Undir Hamiltonian Path})\}\\
    =& \{f_{(G,s,t)}(E_1^*) \mid E_1^*\subseteq E,  E_1^* \text{ is a Hamiltonian path in } G \} \\
    =& \{ E_2^* \mid E_2^*\subseteq E', (s,t)\notin E_2^*, E_2^*\cup\{(s,t)\}\text{ is a TSP path in } G' \}\\
    =& \{S'\cap f_{(G,s,t)}(E)\mid S'\in \S(\textsc{Traveling Salesperson})\}
\end{align*}

\subsubsection{\textsc{Undirected Hamiltonian Path} \SSP{} \textsc{Undirected Hamiltonian Cycle}}
\undirHamPath
\undirHam

The reduction is analogous to the SSP reduction from the directed variants of the two problems from~\citeLiterature{grune2023large}.
In the reduction, an additional vertex $w$ is added, as well as edges connecting the endpoints of the path to $w$.
This way, a Hamiltonian path in the original graph directly forms a Hamiltonian cycle in the new graph.
The reduction was inspired by Arora and Barak~\citeLiterature{arora2009computational}.

\functionG
We define $g$ mapping an \textsc{Undirected Hamiltonian Cycle} instance $(G, s, t)$ to a \textsc{TSP} instance
$(G')$ as follows:
We set $V'=V\cup\{w\}$ and $E'=E\cup\{(s,w), (w,t)\}$. 

\functionF
We let $f_{(G,s,t)}((v,w))=(v',w')\in E'$.

\correctness
The reductions runs in polynomial time, as only one vertex and two edges are added.
If $G$ has a Hamiltonian path from $s$ to $t$, then $G'$ clearly has a Hamiltonian cycle, 
formed by the path and the additionally added edges $(s,w), (w,t)$.

The other way around, if $G'$ has a Hamiltonian cycle, the edges $(s,w)$ and $(w,t)$ have to be part of it,
as otherwise, the vertex $w$ cannot be a part of the cycle. 
Thus, there must be a Hamiltonian path from $s$ to $t$ in $G$, as that is $G'$ without $w$ and the edges incident to $w$.
Clearly, the reduction is a SSP reduction, i.e. the following holds:
\begin{align*}
    & \{f_{(G,s,t)}(S) \mid S\in\S(\textsc{Undir Hamiltonian Path})\}\\
    =& \{f_{(G,s,t)}(E_1^*) \mid E_1^* \subseteq E, E_1^*\text{ is a Hamiltonian path in } G \} \\
    =& \{ E_2^* \mid E_2^*\subseteq E', \{(s,w), (w,t)\}\notin E_2^*, E_2^*\cup\{(s,w),(w,t)\}\text{ is a Hamiltonian cycle in } G'\}\\
    =& \{S'\cap f_{(G,s,t)}(E)\mid S'\in \S(\textsc{Undir Hamiltonian Cycle})\}
\end{align*}

\begin{figure}
    \centering
    \begin{minipage}[t]{0.49\textwidth}
        \textsc{Undirected Hamiltonian Path} instance:\\
        \begin{tikzpicture}
            \node[] (v1) {$v_1$};
            \node[] (v2) [right=of v1] {$v_2$};
            \node[] (v3) [below=of v1] {$v_3$};
            \node[] (v4) [right=of v3] {$v_4$};
    
            \path[rwth-blue]
                (v1) edge node {} (v2)
                (v1) edge node {} (v3)
                (v3) edge node {} (v4);
                
            \path[]
                (v2) edge node {} (v3);
        \end{tikzpicture}
        ~\\ $s=v_2, t=v_4$
    \end{minipage}\hfill
    \begin{minipage}[t]{0.49\textwidth}
        \textsc{Undirected Hamiltonian Cycle} instance:\\
        \begin{tikzpicture}
            \node[] (v1) {$v_1$};
            \node[] (v2) [right=of v1] {$v_2$};
            \node[] (v3) [below=of v1] {$v_3$};
            \node[] (v4) [right=of v3] {$v_4$};
            \node[] (w)  [above right=0.5 and 0.5 of v4.center] {$w$};
    
            \path[rwth-blue]
                (v1) edge node {} (v2)
                (v1) edge node {} (v3)
                (v3) edge node {} (v4)
                (v4) edge node {} (w)
                (w) edge node {} (v2);
            
            \path[]
                (v2) edge node {} (v3);
        \end{tikzpicture}
    \end{minipage}
    \caption{The reduction from \textsc{Undirected Hamiltonian Path} to \textsc{Undirected Hamiltonian Cycle}. 
    The vertex $w$ is added to ensure that a cycle in $G'$ always indicates a path in $G$.}
    \label{fig:undirHamPath-undirHamCyc}
\end{figure}
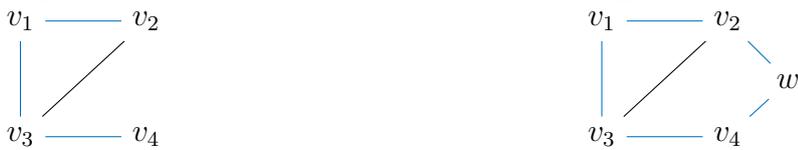

\subsubsection{\textsc{3-Dimensional Matching} \SSP{} \textsc{3-Dimensional Exact Matching}}
\threeDmatch
\threeDExMatch

This reduction maps instances of \textsc{3-Dimensional Matching}, where a $k$ is specifically given as part of the input,
to instances of \textsc{3-Dimensional Exact Matching}, where the matching needs to include all elements of the sets $X,Y,Z$.
The reduction has two steps; the padding step, that pads all sets to the same size, and the counting step,
which simulates the number $k$ in the new instance.
This reduction is not inspired by Karp, Garey and Johnson or Arora and Barak, but is needed to connect the two
versions of \textsc{3-Dimensional Matching}.\\

\begin{figure}
    \centering
    \begin{subfigure}{0.47\textwidth}
        \begin{tikzpicture}
            \node[] (X) {$X$};
            \node[] (Y) [right=of X] {$Y$};
            \node[] (Z) [right=of Y] {$Z$};
    
            \node[dot] (x1) [below=of X] {};
            \node[dot] (x2) [below=of x1] {};
            \node[dot] (x3) [below=of x2] {};
            \node[dot, color=rwth-green] (x4) [below=of x3] {};
            \node[dot, color=rwth-green] (x5) [below=of x4] {};
    
            \node[] (1) [left=of x1] {$1$};
            \node[] (2) [left=of x2] {$2$};
            \node[] (3) [left=of x3] {$3$};
            \node[] (4) [left=of x4] {$4$};
            \node[] (5) [left=of x5] {$5$};

            \node[dot] (y1) [below=of Y] {};
            \node[dot] (y2) [below=of y1] {};
            \node[dot] (y3) [below=of y2] {};
            \node[dot] (y4) [below=of y3] {};
            \node[dot, color=rwth-green] (y5) [below=of y4] {};
    
            \node[dot] (z1) [below=of Z] {};
            \node[dot] (z2) [below=of z1] {};
            \node[dot] (z3) [below=of z2] {};
            \node[dot] (z4) [below=of z3] {};
            \node[dot] (z5) [below=of z4] {};

            \foreach \y in {1,2,3,4}{
                \path[color=rwth-green]
                (x4) edge node {} (y\y);
                \foreach \z in {1,2,3,4,5}{
                    \path[color=rwth-green]
                        (y\y) edge node {} (z\z);
                    \path[color=rwth-grass]
                        (y5) edge node {} (z\z);
                }
            }
            \path[color=rwth-grass]
                (x5) edge node {} (y5);
        \end{tikzpicture}
        ~\\ $k=2$\\
        New triples added to $U$:\\
        $(x_4, y_1,z_1), (x_4, y_1, z_2),\ldots,(x_4, y_1, z_5), \\
        (x_4, y_2, z_1),\ldots,(x_4, y_4, z_5)$ and\\ 
        $(x_5, y_5,z_1),(x_5,y_5,z_2),\ldots,(x_5,y_5,z_5)$.
        \caption{The black dots represent the original elements of the sets $X,Y,Z$, 
        and the green dots represent the new elements added.}
        \label{fig:3DM-3DEM-garbage:a}
    \end{subfigure}\hfill
    \begin{subfigure}{0.47\textwidth}
        \begin{tikzpicture}
            \node[] (X) {$X$};
            \node[] (Y) [right=of X] {$Y$};
            \node[] (Z) [right=of Y] {$Z$};
    
            \node[dot] (x1) [below=of X] {};
            \node[dot] (x2) [below=of x1] {};
            \node[dot] (x3) [below=of x2] {};
            \node[dot, color=rwth-green] (x4) [below=of x3] {};
            \node[dot, color=rwth-green] (x5) [below=of x4] {};
            \node[dot, color=rwth-red] (x6) [below=of x5] {};
    
            \node[] (1) [left=of x1] {$1$};
            \node[] (2) [left=of x2] {$2$};
            \node[] (3) [left=of x3] {$3$};
            \node[] (4) [left=of x4] {$4$};
            \node[] (5) [left=of x5] {$5$};
            \node[] (6) [left=of x6] {$6$};

            \node[dot] (y1) [below=of Y] {};
            \node[dot] (y2) [below=of y1] {};
            \node[dot] (y3) [below=of y2] {};
            \node[dot] (y4) [below=of y3] {};
            \node[dot, color=rwth-green] (y5) [below=of y4] {};
            \node[dot, color=rwth-red] (y6) [below=of y5] {};
    
            \node[dot] (z1) [below=of Z] {};
            \node[dot] (z2) [below=of z1] {};
            \node[dot] (z3) [below=of z2] {};
            \node[dot] (z4) [below=of z3] {};
            \node[dot] (z5) [below=of z4] {};
            \node[dot, color=rwth-red] (z6) [below=of z5] {};

            \foreach \y in {1,2,3,4}{
                \path[color=rwth-red]
                    (x6) edge node {} (y\y);
                \foreach \z in {1,...,5}{
                    \path[color=rwth-red]
                        (y\y) edge node {} (z\z);
                }
            }
            \path[color=rwth-carmine]
                (y6) edge node {} (z6);
            \foreach \x in {1,2,3}{
                \path[color=rwth-carmine]
                    (x\x) edge node {} (y6);
            }
        \end{tikzpicture}
        ~\\ $k=2$
        \vspace*{2\baselineskip}
        \caption{The second step of the reduction: the red dots represent the $\vert X\vert-k=3-2=1$ new elements added to the sets $X,Y,Z$.}
        \label{fig:3DM-3DEM-garbage:b}
    \end{subfigure}\hfill
    \caption{An example of the first and second step of the reduction from \textsc{3-Dimensional Matching} to \textsc{3-Dimensional Exact Matching}.
    The red and green edges each represent the triples added in the step.}
    \label{fig:3DM-3DEM-garbage}
\end{figure}
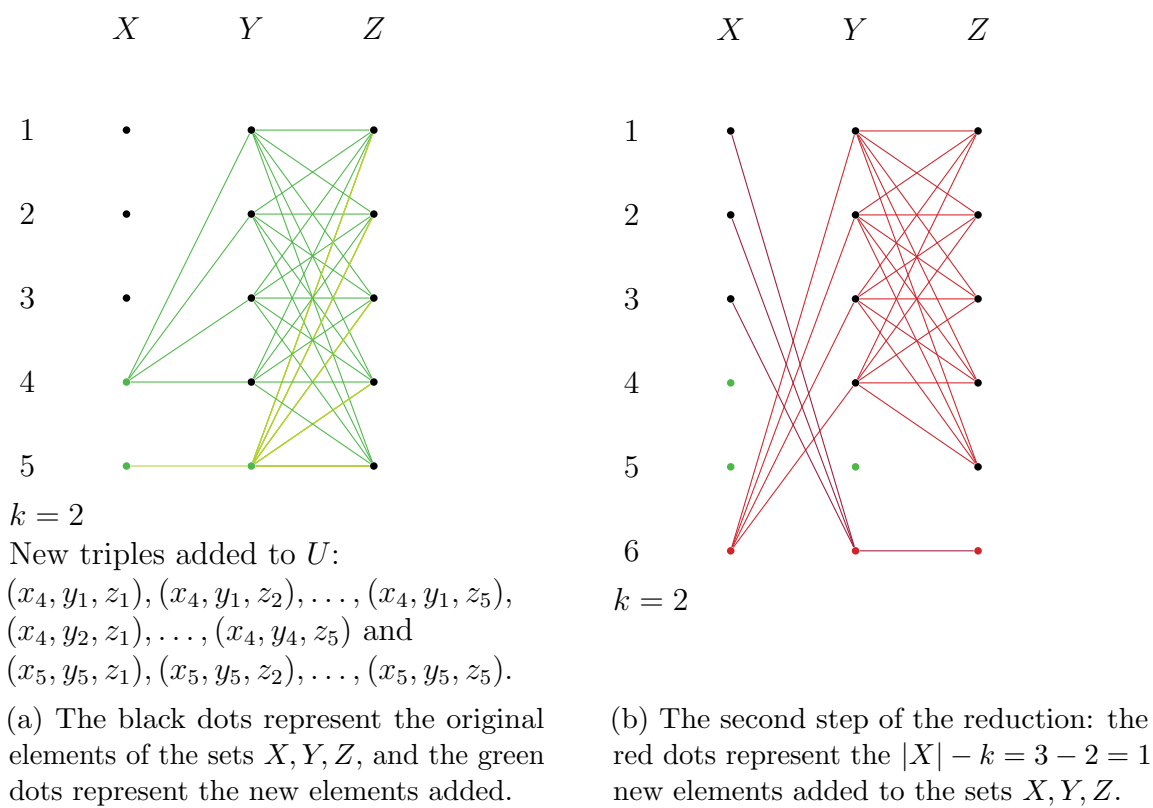

W.l.o.g. we assume that $\vert X\vert\leq\vert Y\vert\leq\vert Z\vert=:p$. In the padding step, the first two sets
might need to be padded with new elements to achieve the same size.
The new elements are then paired to triples with all elements of the other sets,
so that any element that is not part of the matching yet can be included. 
The idea is shown in~\cref{fig:3DM-3DEM-garbage:a}. 
But then, the number of triples that can be chosen without collisions is still limited to $k+(p-\vert X\vert)$.
This is why we simulate the number $k$ in the counting step, 
because we need a matching of the same size as our sets.\\
In the counting step, to all sets, $\vert X\vert-k$ new elements are added, 
that are again combined to triples with all elements of the other sets,
to ensure the up to $\vert X\vert -k$ elements that are not part of a triple in the matching yet can be included.
This step is illustrated in~\cref{fig:3DM-3DEM-garbage:b}. Each set has the size $p+\vert X\vert-k$,
and the number of triples in the matching is $k+(p-\vert X\vert)+2(\vert X\vert-k)=p+\vert X\vert-k$.\\
The solution mapping function simply matches each triple to itself, as the matchings are expanded, thus,
if a triple is part of the matching in the \textsc{3-Dimensional Matching} instance, it is also part of the new matching
in the\textsc{3-Dimensional Exact Matching} instance.

\functionG
Formally, the reduction function $g$ mapping an instance $(U,(X,Y,Z),k)$ to an instance $(U',(X',Y',Z'))$ is defined as follows:
    \begin{enumerate}
        \item[1.]
        W.l.o.g. we assume that $\vert X\vert\leq\vert Y\vert\leq\vert Z\vert$. 
        First, we set $X':=X, Y':=Y, Z':=Z$ and $U':=U$.
        In this step, the sets $X',Y',Z'$ are extended to the size of $Y'$, which we denote as $p$. 
        First, $\vert Y\vert-\vert X\vert$ new elements are added to $X'$, and these new elements each form triples
        with all possible combinations of elements from $Y'$ and $Z'$, thus 
        \[\{(x_a, y_b, z_c)\in U'\mid \vert X\vert < a\leq\vert Y\vert, 1\leq b\leq\vert Y\vert, 1\leq c\leq p\}\]
        Then, to both $X'$ and $Y'$, $p-\vert Y\vert$ new elements are added, and these new elements each form triples
        \[\{(x_a, y_a, z_b)\in U'\mid \vert Y\vert < a\leq p, 1\leq b\leq p\}\]
        Then, each set has the size $p$.
        This step is shown in~\cref{fig:3DM-3DEM-garbage} as an example.
        \item[2.] 
        Now, we need to somehow simulate the number $k$ in our new instance. 
        Therefore, we introduce $\vert X\vert-k$ new elements to the sets $X',Y',Z'$, and form triples
        \[\{
            (x_a, y_b, z_c)\in U'\mid p < a\leq p+\vert X\vert-k, 1\leq b,c\leq p
        \}\]
        as well as triples
        \[\{
            (x_a, y_b, z_b)\in U'\mid 1\leq a\leq p, p< b\leq p+\vert X\vert-k
        \}\]
        The idea is that with the $x$-triples, the elements of $Z$ and $Y$ that are not part of the matching yet can be included, 
        as all combinations of any $y$ and $z$ from the original instance are combined with any $x_a$,
        and with the $y-z$-triples, the elements of $X$ can be included in the matching in the same way.
    \end{enumerate}
    In the end, each set has the size $p+\vert X\vert-k$.

    \functionF
    The solution mapping function is the identity function, thus $f_{(U,(X,Y,Z),k)}((x,y,z))=(x,y,z)\in U'$ for all $(x,y,z)\in U$.

\correctness
First, we can see that the running time of the reduction is polynomial, as there are $O(\vert Y\vert^2)$ new tuples
added.
Let $(U,(X,Y,Z),k)$ be a YES-instance of \textsc{3-Dimensional Matching}. 
Then, the reduction function $g$ maps this instance to $(U',(X',Y',Z'))$ as described above.
Let $M\subseteq U$ be the matching of the \textsc{3-Dimensional Matching} instance of size at least $k$. 
We start by setting $M^*:=M$.
After the first step, $\vert Z\vert-\vert X\vert$ triples from the triples generated in that step
can be added to $M^*$, as the added triples contain all possible combinations of elements from $Y$ and $Z$, thus,
not-yet-matched elements can be chosen.\\
In other words:
\begin{align*}
    M^*:=M&\cup\{(x_a, y_b, z_c)\in U'\mid \vert X\vert < a\leq\vert Y\vert, y_b, z_c\text{ are not part of the matching yet}\}\\
    &\cup\{(x_a, y_a, z_b)\in U'\mid \vert Y\vert < a\leq p, z_b\text{ is not part of the matching yet}\}
\end{align*}
Thus, in each dimension, there are $\vert X\vert-k$ elements left over after this step, 
and the size of $M^*$ is $k+p-\vert X\vert$.\\
After the second step, these left over elements can be added to the matching, as the added triples are constructed
in a way that allows any combinations of elements from $X,Y,Z$ to be matched, thus\\
\begin{align*}
    M^*:=M^*&\cup\{(x_a, y_b, z_c)\in U'\mid p < a\leq p+\vert X\vert-k, y_b, z_c\text{ are not part of the matching yet}\}\\
    &\cup\{(x_a, y_b, z_b)\in U'\mid p< b\leq p+\vert X\vert-k, x_a\text{ is not part of the matching yet}\}
\end{align*}
This are $2(\vert X\vert-k)$ triples that can be added to the matching (2 from each of the new rows), 
thus, the matching $M^*$ has size $k+p-\vert X\vert+2(\vert X\vert -k)=p+\vert X\vert-k$, 
which is exactly the size of $X',Y',Z'$, and the triples from the function $f$ are part of $M^*$.\\

The other way around, let $(U',(X',Y',Z'))$ be a YES-instance of \textsc{3-Dimensional Exact Matching}. 
From the way the instance was produced, we can conclude that
\begin{enumerate}
    \item[a)] $X',Y',Z'$ have the size $q=p+\vert X\vert-k$,
    \item[b)] the matching $M^*$ has the size $q$, 
    \item[c)] there are exactly $2(\vert X\vert-k)$ triples in $M^*$ that contain an element from the new rows 
    added in step 2, as there are $\vert X\vert-k$ new rows and in each of them, the $y_i$- and $z_i$-element always
    appear together in a triple together with an $x_a$ from the original set $X$, and $x_i$ on its own,
    paired with some $y_b\in Y, z_c\in Z$. Thus, a triple never contains an $x$ from the new set 
    as well as an $y,z$-pair from the new set. Furthermore, all of these elements need to be covered by the matching,
    thus, for each row there are $2$ triples in $M^*$,
    adding up to a total of $2(\vert X\vert-k)$ triples in $M^*$ that contain an element from the new rows,
    \item[d)] there are exactly $p-\vert X\vert$ triples in $M^*$ that contain an element from the new row
    added in step 1, since there are $p-\vert X\vert$ new rows and in each row, all new elements only appear
    together in a triple, combined with some elements from the original sets, 
    and as all of them need to be covered by the matching, for each row, one triple is part of $M^*$.
\end{enumerate}

Thus, if we remove the extra elements and triples containing them, we are left with a matching of size
\[q-2(\vert X\vert-k)-(p-\vert X\vert)=q-p-\vert X\vert+2k = p-p+\vert X\vert-\vert X\vert +2k-k=k\]
in which all triples are of the form $(x,y,z)\in X\times Y\times Z$, thus, 
$(U,(X,Y,Z),k)$ is a YES-instance of \textsc{3-Dimensional Matching}, retrieved from the function $f$.
Thus,
\begin{align*}
    & \{f_{(U,(X,Y,Z),k)}(S) \mid S\in\S(\textsc{3-Dim Matching})\}\\
    =& \{f_{(U,(X,Y,Z),k)}(W) \mid W\subseteq U, \vert W\vert\geq k \text{ such that for any }(x_i, y_i, z_i),(x_j,y_j,z_j)\in W:\\
    & x_i\neq x_j,y_i\neq y_j, z_i\neq z_j \} \\
    =& \{ W^* \mid  W^*\subseteq U' \text{ such that } W^*\subseteq U \text{ and }\\
    & W^* \text{ is part of a solution to the \textsc{3-Dim Exact Matching} instance} \}\\
    =& \{S'\cap f_{(U,(X,Y,Z),k)}(U)\mid S'\in \S(\textsc{3-Dim Exact Matching})\}
\end{align*}

\subsubsection{\textsc{3-Dimensional Exact Matching} \SSP{} \textsc{Partition}}\label{subsubsec:3DEM-Partition}
\threeDExMatch{}
\partition
Garey and Johnson present the reduction on which this reduction is based in~\citeLiterature{garey1979computers}.
They call the problem from which is reduced \textsc{3-Dimensional Matching}, but as in this paper, 
that name has already been used for a slightly different variant of the problem, 
we call this version \textsc{3-Dimensional Exact Matching} to avoid confusion.\\

The reduction maps each triple from the \textsc{3-Dimensional Exact Matching} instance $(U(X,Y,Z))$ 
to a binary number in the \textsc{Partition} instance $(A',w)$ representing the elements in the triple. 
For this, an order of the elements is fixed, $x_1,\ldots, x_q, y_1,\ldots, y_q, z_1,\ldots, z_q$,
and each triple $u_i\in U$ is mapped to an element $a_i\in A'$ with $w'(a_i)$ being the binary number 
such that \[w'(a_i)[j]=\begin{cases}
    1, & \text{if }x_j\in u_i\\
    1, & \text{if }y_{j-q}\in u_i\\
    1, & \text{if }z_{j-2q}\in u_i\\
    0, & \text{else}
\end{cases}\]

We know that in order to have a matching,
all elements need to be covered, thus, the sum of the corresponding numbers is fixed, we call it $B$.
Hence, when we put all numbers corresponding to the triples from the matching into a set, and all other 
numbers into the other, the first set has the total sum of $B$, and the other set, denoted as garbage set, 
has the total sum of $\sum_{i=1}^k w'(a)-B$, with $k=\vert U\vert$.
Using this trick, two more elements can be added to the set, with weights even out the difference between the two sets,
namely $a_{b_1}=\sum_{i=1}^k w'(a)-B$ and $a_{b_2}=\sum_{i=1}^k w'(a)+B$. This way, both sets have exactly the sum $2\sum_{i=1}^k w'(a)$,
as we can combine $a_{b_1}$ with the set corresponding to the matching, and $a_{b_2}$ with the garbage set.

\functionG
In the reduction from the \textsc{3-Dimensional Exact Matching} instance $(U,(X,Y,Z))$ to the
\textsc{Partition} instance $(A',w')$, the mapping of each combination of the elements in $X, Y, Z$ to a specific number is fixed as follows:
A string of bits is constructed, that is divided into $3q$ zones of $p=\lceil log_2(k+1)\rceil$ bits each, 
where $k$ is the number of triples in the subset $U$. 
Each of the zones corresponds to one element in $X, Y, Z$. $p$ is chosen to be the size of the zones 
to avoid overflow into the next zone when adding up the different numbers corresponding to triples.
As there can be at most $l$ triples containing some element, the sum of all the numbers corresponding to the
elements will still not overflow into the next zone.

For each triple $u_i=(x_{b(i)}, y_{c(i)}, z_{d(i)})\in U$, where $b(i), c(i), d(i)$ are the indices of the elements in $X, Y, Z$ of $u_i$,
we add $a_i$ to $A'$, with \[w'(a_i)=2^{p(3q-b(i))}+2^{p(2q-c(i))}+2^{p(q-d(i))}\] thus the number, where in the zones
corresponding to $x_{b(i)}, y_{c(i)}, z_{d(i)}$ the rightmost bit is set to one, and all other bits are set to zero.
After this step, $A'$ contains $q$ elements, and for each element, its weight directly encodes the elements in the corresponding triple.
We now set $B=\sum_{j=0}^{3q-1}2^{pj}$ and add two more elements: $a_{3q+1}$ with $w'(a_{3q+1})=\sum_{i=1}^k w'(a)+B$, 
thus the element padding up the garbage set, and $a_{3q+2}$ with $w'(a_{3q+2})=2\sum_{i=1}^k w'(a)-B$,
thus the element padding up the matching set.

\functionF
Each triple $u_i$ is mapped to its corresponding element $a_i\in A'$: $f_{(U,(X,Y,Z))}(u_i)=a_i$.

\correctness
Let $(U, (X, Y, Z))$ be a yes-instance of \textsc{3-Dimensional Exact Matching}, with $U^*\subseteq U$ being a solution.
Then, $A^*=\{ a_i\mid u_i\in U^* \}\cup \{ a_{3q+2} \}$ is a solution of the \textsc{Partition} instance.
This is because we know that the sum of the numbers corresponding to the triples in $U^*$ is $B$, since $B$ is
the binary number that has ones at the rightmost bit of each zone. Thus, 
\[\sum_{a\in A^*} w'(a)=B+2\sum_{i=1}^k w'(a)-B=2\sum_{i=1}^k w'(a)\]
Also, the sum of the garbage set $A'\setminus A^*$ is 
\[\sum_{a\notin A^*} w'(a)=\sum_{i=1}^k w'(a)-B+\sum_{i=1}^k w'(a)+B=2\sum_{i=1}^k w'(a)\] 
thus the sum of the garbage set is equal to the sum of the matching set.
Thus, the elements $a_i$ with $u_i\in U^*$ are part of the solution, thus if $u_i\in U^*$ then $f_{(U,(X,Y,Z))}(u_i)\in A^*$.\\

For the other direction, let $(A',w')$ be a yes-instance of \textsc{Partition}, with $A^*\subseteq A'$ being a solution set.
Since summing up all entries in one zone can never be more than $k$, thus maximum $2^p-1$, there are no overflows into neighbor zones.
The total weight of all elements in $A'$ is $\sum_{a\in A'} w'(a) = 4\sum_{i=1}^k w'(a)$.
As by definition $a_{3q+2}$ is part of the solution, $a_{3q+1}$ is not, as their weights combined have a total
weight of $3\sum_{i=1}^k w'(a)$, and all remaining elements have a total weight of only $\sum_{i=1}^k w'(a)$.
Thus, the remaining $a\in A^*$ have a total weight of 
\begin{align*}
    \sum_{a\in A^*} w'(a) &= 2\sum_{i=1}^k w'(a)-w'(a_{3q+1})\\
    &= 2\sum_{i=1}^k w'(a) - (2\sum_{i=1}^k w'(a)-B)\\
    &= B
\end{align*}
Which means that the binary representation of the sum of the $w'_(a_1),\ldots, w'(a_k)$ is the binary number with
ones at the rightmost bit of each zone, and zeros everywhere else. Thus, the elements $u_i$ with $a_i\in A^*$ 
must form the solution for the \textsc{3-Dimensional Exact Matching} instance, and
\begin{align*}
    & \{f_{(U,(X,Y,Z))}(S) \mid S\in\S(\textsc{3-Dim Exact Matching})\}\\
    =& \{f_{(U, (X,Y,Z))}(W) \mid W\subseteq U, \vert W\vert=q \text{ such that for any }(x_i, y_i, z_i),(x_j,y_j,z_j)\in W:\\
    & x_i\neq x_j,y_i\neq y_j, z_i\neq z_j \} \\
    =& \{ A^* \mid A^*\subseteq A', \{a_{3q+2}\}\notin A^*, A^*\cup\{a_{3q+2}\}\text{ is a solution of \textsc{Partition}} \}\\
    =& \{S'\cap f_{(U,(X,Y,Z))}(U)\mid S'\in \S(\textsc{Partition})\}
\end{align*}

The construction runs in polynomial time, as we create $3q$ elements and numbers with at most $3qp$ bits each.

\newcommand{\drawOneElement}[3]{

    \coordinate (pos) at #2;

    \foreach \x in {#1} {
        \coordinate (pos) at ($(pos) + (0.25, 0)$);
        \ifnum\x=1
            \ifnum#3=1
                \node[draw=rwth-blue] at (pos) {$1$};
            \else
                \node[] at (pos) {$1$};
            \fi
        \else
            \node[] at (pos) {$0$};
        \fi
    }
}

\def\drawNumber#1#2#3#4#5#6{

    \node[] at (-0.75, #4-0.5) {#5};
    \ifthenelse{#1=1}{
        \drawOneElement{0,0,1}{(0, #4-0.5)}{#6};
    }{
        \drawOneElement{0,0,0}{(0, #4-0.5)}{#6};
    }
    \ifthenelse{#1=2}{
        \drawOneElement{0,0,1}{(1, #4-0.5)}{#6};
    }{
        \drawOneElement{0,0,0}{(1, #4-0.5)}{#6};
    }
    \ifthenelse{#1=3}{
        \drawOneElement{0,0,1}{(2, #4-0.5)}{#6};
    }{
        \drawOneElement{0,0,0}{(2, #4-0.5)}{#6};
    }
    \ifthenelse{#2=1}{
        \drawOneElement{0,0,1}{(3, #4-0.5)}{#6};
    }{
        \drawOneElement{0,0,0}{(3, #4-0.5)}{#6};
    }
    \ifthenelse{#2=2}{
        \drawOneElement{0,0,1}{(4, #4-0.5)}{#6};
    }{
        \drawOneElement{0,0,0}{(4, #4-0.5)}{#6};
    }
    \ifthenelse{#2=3}{
        \drawOneElement{0,0,1}{(5, #4-0.5)}{#6};
    }{
        \drawOneElement{0,0,0}{(5, #4-0.5)}{#6};
    }
    \ifthenelse{#1=1}{
        \drawOneElement{0,0,1}{(6, #4-0.5)}{#6};
    }{
        \drawOneElement{0,0,0}{(6, #4-0.5)}{#6};
    }
    \ifthenelse{#1=2}{
        \drawOneElement{0,0,1}{(7, #4-0.5)}{#6};
    }{
        \drawOneElement{0,0,0}{(7, #4-0.5)}{#6};
    }
    \ifthenelse{#1=3}{
        \drawOneElement{0,0,1}{(8, #4-0.5)}{#6};
    }{
        \drawOneElement{0,0,0}{(8, #4-0.5)}{#6};
    }
}

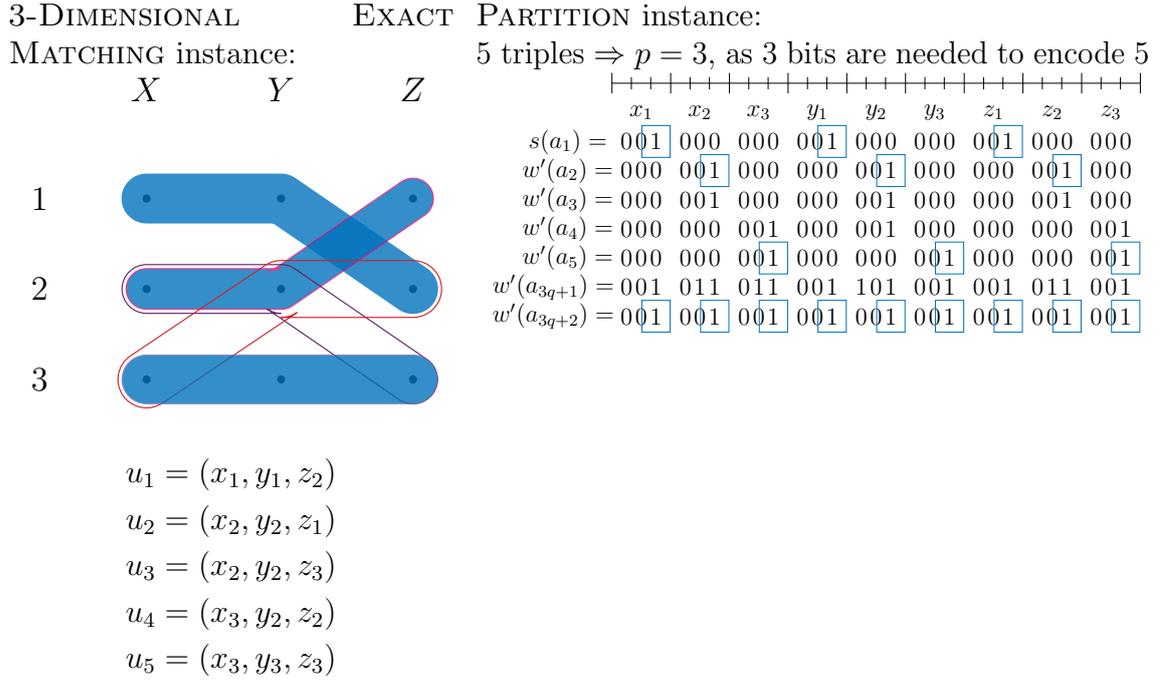
\begin{figure}
    \centering
    \begin{minipage}[t]{0.39\textwidth}
        \textsc{3-Dimensional Exact Matching} instance:\\
        \resizebox{\textwidth}{!}{
        \begin{tikzpicture}
            \node[] (X) {$X$};
            \node[] (Y) [right=of X] {$Y$};
            \node[] (Z) [right=of Y] {$Z$};
    
            \node[dot] (x1) [below=of X] {};
            \node[dot] (x2) [below=of x1] {};
            \node[dot] (x3) [below=of x2] {};
    
            \node[] (1) [left=of x1] {$1$};
            \node[] (2) [left=of x2] {$2$};
            \node[] (3) [left=of x3] {$3$};

            \node[dot] (y1) [below=of Y] {};
            \node[dot] (y2) [below=of y1] {};
            \node[dot] (y3) [below=of y2] {};
    
            \node[dot] (z1) [below=of Z] {};
            \node[dot] (z2) [below=of z1] {};
            \node[dot] (z3) [below=of z2] {};

            \draw[draw=rwth-blue, fill=rwth-blue, opacity=0.8] \hedgem{x1}{y1}{z2, y1}{3mm};
            \draw[draw=rwth-magenta, fill=rwth-blue, opacity=0.8] \hedgem{x2}{y2}{z1, y2}{2.5mm};
            \draw[draw=rwth-purple] \hedgem{x2}{y2}{z3, y2}{3mm};
            \draw[draw=rwth-red] \hedgem{x3}{y2}{z2, y2}{3.5mm};
            \draw[draw=rwth-violet, fill=rwth-blue, opacity=0.8] \hedgem{x3}{y3}{z3}{3mm};
            
        \end{tikzpicture}
        }
        \begin{align*}
            u_1&=(x_1, y_1, z_2)\\
            u_2&=(x_2, y_2, z_1)\\
            u_3&=(x_2, y_2, z_3)\\
            u_4&=(x_3, y_2, z_2)\\
            u_5&=(x_3, y_3, z_3)
        \end{align*}
       
    \end{minipage}\hfill
    \begin{minipage}[t]{0.59\textwidth}
        \textsc{Partition} instance:\\
        $5$ triples $\Rightarrow p=3$, as $3$ bits are needed to encode $5$
        \resizebox{\textwidth}{!}{
        \begin{tikzpicture}
            \draw(0,0)--(9,0);
            \foreach \x in {0,...,9} {
                \draw (\x,5pt) -- (\x,-5pt);
                \foreach \y in {0.33,0.66} {
                    \ifthenelse{\x=9}{}{\draw (\x+\y,3pt) -- (\x+\y,-3pt);}
                    
                }
            }
            \foreach \x/\xtext in {0.5/$x_1$, 1.5/$x_2$, 2.5/$x_3$, 3.5/$y_1$, 4.5/$y_2$, 5.5/$y_3$, 6.5/$z_1$, 7.5/$z_2$, 8.5/$z_3$} {
                \node[] at (\x,-0.5) {\xtext};
            }
            \drawNumber{1}{1}{2}{-0.5}{$s(a_1)=$}{1}
            \drawNumber{2}{2}{1}{-1}{$w'(a_2)=$}{1}
            \drawNumber{2}{2}{3}{-1.5}{$w'(a_3)=$}{0}
            \drawNumber{3}{2}{2}{-2}{$w'(a_4)=$}{0}
            \drawNumber{3}{3}{3}{-2.5}{$w'(a_5)=$}{1}
            \node[] at (-1, -4) {$w'(a_{3q+2})=$};
            \foreach \x in {0,...,8} {
                \drawOneElement{0,0,1}{(\x, -4)}{1};
            }
            \node[] at (-1, -3.5) {$w'(a_{3q+1})=$};
            \drawOneElement{0,0,1}{(0,-3.5)}{0};
            \drawOneElement{0,1,1}{(1,-3.5)}{0};
            \drawOneElement{0,1,1}{(2,-3.5)}{0};

            \drawOneElement{0,0,1}{(3,-3.5)}{0};
            \drawOneElement{1,0,1}{(4,-3.5)}{0};
            \drawOneElement{0,0,1}{(5,-3.5)}{0};

            \drawOneElement{0,0,1}{(6,-3.5)}{0};
            \drawOneElement{0,1,1}{(7,-3.5)}{0};
            \drawOneElement{0,0,1}{(8,-3.5)}{0};

        \end{tikzpicture}
        }
    \end{minipage}
    \caption{An example of the reduction from \textsc{3-Dimensional Exact Matching} to \textsc{Partition}.
    $a_{3q+2}$ is part of the solution set for \textsc{Partition}, and $a_{3q+1}$ is not.
    }
    \label{fig:3DEM-Partition}
\end{figure}

\subsubsection{\textsc{Exactly-One-3-SAT} \SSP{} \textsc{Combinatorial Auction}}
\exOne
\combiAuction

In this reduction, a \textsc{Exactly-One-3-SAT} instance $(L,C)$ is reduced to a \textsc{Combinatorial Auction} 
instance $(n,k,P)$ in a quite intuitive way: We create the sets so that for each variable, there are two sets,
one for the positive and one for the negative literal. Of course, only one of these sets should be able to be 
selected, which is why the index of the variable is also included. 
Also, each clause should only be satisfied by one literal,
thus, the clauses which contain the literal are also included in the set. 
The price of each set corresponds to the number of clauses that would be satisfied by choosing that literal plus one,
to make sure for each variable, at least one set, positive or negative, must be chosen.
Then, we set $k$ to the number of variables plus the number of clauses,
because in total, we want to satisfy all clauses and select one set for each variable, thus we need a total price of $k$.
Thus:
\functionG
Each literal $x_i$ is transformed to a set \\$S_{2i-1} = \{ x_i, i, C_{a_1}, \ldots, C_{a_b} \}$, 
and $\overline{x}_i$ to $S_{2i}=\{\overline{x}_i, i,C_{a_1},\ldots,C_{a_b}\}$ respectively,
where $C_a$ are all clauses in which the variable appears. 
This way we can ensure that not both $x_i$ and $\overline{x}_i$ can be selected, 
and for each clause only one literal that is part of the clause. 
We define the pair $p_i=( S_i, x_i)$ with $x_i$ being one plus the number of clauses that contain the literal
belonging to the set $S_i$, to ensure that for each variable, at least $x_i$ or $\overline{x}_i$ must be selected. 
Finally, we set $k:=\frac{\vert L\vert}{2}\vert C\vert$. An example is shown in~\cref{fig:ex1-3SAT-combAuc}.

\functionF
We set $f_{(L,C)}(\ell_i)=\begin{cases}
    S_{2i-1} & \text{if } x_i=\ell_i\\
    S_{2i} & \text{if } x_i=\overline{\ell_i}
\end{cases}$, \\because each pair directly translates to the corresponding literal being chosen.

\correctness
The reduction runs in polynomial time, as the number of sets is linear in the number of variables and clauses
and each set contains at most $2+m$ elements, where $m$ is the number of clauses,
and $f_I$ can be computed in linear time.\\
Let $(L,C)$ be the \textsc{Exactly-One-3-SAT} instance and let $L^*$ be the solution of the instance.
Then, the pairs $p_{2i-1}$ for each $x_i\in L^*$ and $p_{2i}$ for each $\overline{x}_i\in L^*$ 
form a solution for the \textsc{Combinatorial Auction} instance, because the sets $S$ are disjoint, 
as per definition for each literal, only either $x_i$ or $\overline{x}_i$ can be in $L^*$, and
$\vert L^*\cap C_j\vert=1$ for each clause $C_j\in C$, thus, each $C_j$ only appears in one set $S_i$.
Thus, the sets are disjoint. As we have set $x_i$ to the number of clauses containing the literal in the
set $S_i$ plus one, the sum over all $x_i$ in the solution must be $k$, as $L^*$ satisfies all clauses in $C$.
Thus, the solution is given by the mapping function $f$.\\

Let $S^*\subseteq\{S_1,\ldots, S_m\}$ be a solution to the \textsc{Combinatorial Auction} instance $(n,k,P)$.
Then, $\sum_{S_i\in S^*}x_i\geq k$. 
We observe that the maximum weight that can be achieved is $k$, because for one variable, either the set corresponding
to $x_i$ or the set corresponding to $\overline{x}_i$ can be chosen, but not both, and the price of each set
is higher than one if and only if a $C_j$ is covered by the set, and there are only $\vert C\vert$ different $C_j$, 
thus, the maximum weight is $\frac{\vert L\vert}{2}\vert C\vert=k$.
In order to have a weight of $k$, each $C_j$ must be covered at least once, and as it only can be 
covered at most once in order to be a valid solution, one set containing $C_j$ must be in $S^*$ for each $C_j$.
Additionally, for each variable $x_i$, either the set corresponding to $x_i$ or the set corresponding to $\overline{x}_i$
must be in $S^*$ to achieve a weight of $k$.
As all sets are disjoint, the assignment of the literals given by $f_I$ is valid, as since the index of the 
variable is included in the set corresponding to the literal, and thus only one of the sets can be in $S^*$.
Thus, when we choose the literals corresponding to the sets according to our function $f_{(L,C)}$, 
we get a solution for the \textsc{Exactly-One-3-SAT} instance, as each clause is covered exactly once and for 
each variable, only $x_i$ or $\overline{x}_i$ can be part of it, but not both.\\

\begin{align*}
    & \{f_{(L,C)}(S) \mid S\in\S(\textsc{Exactly-One 3-SAT})\}\\
    =& \{f_{(L,C)}(L^*) \mid L^*\subset L \text{ such that } \vert L^*\cap\{\ell_i,\overline{\ell}_i\}\vert=1 \text{ and }\vert L^*\cap C_j\vert=1\text{ for all }C_j\in C \} \\
    =& \{ S^* \mid S^*\subseteq S\text{ with } \sum_{S_i\in S^*} x_{i}\geq k \}\\
    =& \{ S' \mid S'\in \S(\textsc{Combinatorial Auction})\}\\
    =& \{S'\cap f_{(L,C)}(L)\mid S'\in \S(\textsc{Combinatorial Auction})\}
\end{align*}

\begin{figure}
    \centering
    \begin{minipage}[t]{0.49\textwidth}
        \textsc{Exactly-One-3-SAT} instance:
        \begin{align*}
            \varphi&=(x_1\vee x_2\vee \overline{x_3})\wedge(\overline{x_1}\vee x_3\vee x_4)
        \end{align*}
        Example solution:
        \begin{align*}
            \fcolorbox{rwth-blue}{white}{$x_1$}&=1\\
            x_2&=0\\
            \fcolorbox{rwth-blue}{white}{$x_3$}&=1\\
            x_4&=0
        \end{align*}
    \end{minipage}
    \begin{minipage}[t]{0.49\textwidth}
        \textsc{Combinatorial Auction} instance:
        \begin{align*}
            S_1&=\{x_1,1,C_1\}              && x'_1=2 && \fcolorbox{rwth-blue}{white}{$p_1=(S_1, x'_1)$}\\	
            S_2&=\{\overline{x}_1,1, C_2\}  && x'_2=2 && \fcolorbox{white}{white}{$p_2=(S_2, x'_2)$}\\
            S_3&=\{x_2,2,C_1\}              && x'_3=2 && \fcolorbox{white}{white}{$p_3=(S_3, x'_3)$}\\
            S_4&=\{\overline{x}_2,2\}       && x'_4=1 && \fcolorbox{rwth-blue}{white}{$p_4=(S_4, x'_4)$}\\
            S_5&=\{x_3,3,C_2\}              && x'_5=2 && \fcolorbox{rwth-blue}{white}{$p_5=(S_5, x'_5)$}\\	
            S_6&=\{\overline{x}_3,3,C_1\}   && x'_6=2 && \fcolorbox{white}{white}{$p_6=(S_6, x'_6)$}\\
            S_7&=\{x_4, 4, C_2\}            && x'_7=2 && \fcolorbox{white}{white}{$p_7=(S_7, x'_7)$}\\
            S_8&=\{\overline{x}_4, 4\}      && x'_8=1 && \fcolorbox{rwth-blue}{white}{$p_8=(S_8, x'_8)$}\\
            k&=6
        \end{align*}
    \end{minipage}
    \caption{Example for the reduction of \textsc{Exactly-One-3-SAT} to \textsc{Combinatorial Auction}}
    \label{fig:ex1-3SAT-combAuc}
\end{figure}

\subsubsection{\textsc{Exactly-One-3-SAT} \SSP{} \textsc{Subset Sum}}
\exOne
\subsum

This reduction is similar to the reduction from 3-SAT to \textsc{Subset Sum} in~\citeLiterature{grune2023large}
and is inspired by Arora and Barak~\citeLiterature{arora2009computational}.
The idea is, again, that we need to transfer the meaning of the assignment of some literal into a number that 
directly encodes which clauses are satisfied by this literal. To do so, we need to encode each clause and
each variable in a unique way, ensure that for each variable $i$, either $x_i$ or $\overline{x}_i$ is set to true, 
but never both, and ensure that each clause is satisfied by exactly one literal in it. 
This is done by introducing strings of bits for each literal that encode to which variable it belongs 
(to ensure the sum later adds up to one for each variable) and in which clauses it appears (to count for each clause how many literals in it are satisfied).
Furthermore, the base must be chosen wisely, as otherwise, there could be overflows into the next bit, which would lead to an incorrect encoding.

\functionG
We define the function $g: \{0,1\}^*\rightarrow \{0,1\}^*$, 
mapping a instance $(L,C)$ with $n:=\vert L\vert, m:=\vert C\vert$ to a set of numbers $a_1,\ldots, a_{n'}$ 
and a target value $M'$ as follows:
We create a table $T$ with columns $x_1,\ldots, x_{n/2}, C_1, \ldots, C_m$, and $n$ rows, where each row represents 
one literal, and each column referring to one variable or one clause.
Each row now has a one in the column of the variable it belongs to,
and a one in each column of a clause it appears in, thus, 
for $j>n/2$:
\[t_{i,j}=\begin{cases}
    1 & \text{if }\ell_i\in C_{j-n/2}\\
    0 & \text{otherwise}
\end{cases}\] and if  $j\leq n/2$, 
\[t_{i,j}=\begin{cases}
    1 & \text{if }i=j\\
    0 & \text{otherwise}
\end{cases}\]
Each row is viewed as a encoding of a number $a(\ell_i)$ to the base $n+1$.
Thus, $n':=n$.
Then, we set the target value $M':=1,\ldots,1$ to the base $n+1$.~\cref{fig:ex1-3SAT-SubsetSum} shows an example of this encoding.

\functionF
We set $f_{(L,C)}(\ell_i)=a(\ell_i)$, where $a(\ell_i)$ is the number encoded in the row of the literal $\ell_i$ in the table.

\correctness
First of all, the reduction runs in polynomial time, as we introduce $n$ numbers, each calculated in polynomial time.\\
Let us assume, $(L,C)$ is a YES-instance of 3-SAT. 
Then, there is set of literals $L^*= \{ \ell_{a_1}, \ldots, \ell_{a_{n/2}} \} \subset L$ of literals that are set to true such that
$\vert L^*\cap\{\ell_i,\overline{\ell}_i\}\vert=1$ for all $i\in\{1,\ldots,n\}$, and $\vert L^*\cap C_j\vert=1$ for 
all $C_j\in C, j\in \{1,\ldots,\vert C\vert\}$.
The corresponding numbers $a(\ell_{a_1}),\ldots,a(\ell_{a_{n/2}})$ per definition must add up to $M'$, 
as each clause is satisfied by exactly one literal in it
and for each variable $x_i$, exactly either $x_i$ or $\overline{x}_i$ is part of the solution set.
Thus, the elements given by $f_I$ exactly form the solution.\\

For the other direction, we first need to consider how the base was chosen.
As there are $n$ numbers $a(\ell_{a_1}),\ldots,a(\ell_{a_{n/2}})$, all the numbers added up on one bit can be at 
most $n$, thus, it is not possible to overflow into the next bit, as the base is $n+1$. 
Thus, as the sum of all numbers is $M'$, which means that each bit must be one, this must mean that for each bit, exactly one
of the numbers in the solution has a one in that bit. As each number has a one in the column of the variable it belongs to,
this means that for each variable, either $x_i$ or $\overline{x}_i$ is part of the solution for \textsc{Exactly-One-3-SAT}, but never both.
Additionally, as each number has a one in the column of the clause it belongs to, 
this also means that each clause is satisfied by exactly one literal in it.
Thus, if $a(\ell_{a_1}),\ldots,a(\ell_{a_{n/2}})$ is a solution for the \textsc{Subset Sum} instance,
$\ell_{a_1},\ldots, \ell_{a_{n/2}}$ is a solution for the \textsc{Exactly-One-3-SAT} instance.
Thus,
\begin{align*}
    & \{f_{(L,C)}(S) \mid S\in\S(\textsc{Exact-One 3-SAT})\}\\
    =& \{f_{(L,C)}(L^*) \mid L^*\subset L \text{ such that } \vert L^*\cap\{\ell_i,\overline{\ell}_i\}\vert=1 
    \text{ and }\vert L^*\cap C_j\vert=1\text{ for all }C_j\in C \} \\
    =& \{ A^* \mid A^*=\{a(\ell_{a_1}),\ldots,a(\ell_{a_{n/2}})\}, L^*=\{\ell_{a_1},\ldots, \ell_{a_{n/2}}\} \}\\
    =& \{ A^* \mid \sum_{a_i\in A^*}a_i=M' \}\\
    =& \{ S' \mid S'\in\S(\textsc{Subset Sum})\}\\
    =& \{S'\cap f_{(L,C)}(L)\mid S'\in \S(\textsc{Subset Sum})\}
\end{align*}

\begin{figure}
    \centering
    \begin{minipage}[t]{0.39\textwidth}
        \textsc{Exactly-One-3-SAT} instance:
        \begin{align*}
            \varphi&=(x_1\vee x_2\vee \overline{x_3})\wedge(\overline{x_1}\vee x_3\vee x_4)
        \end{align*}
        Example solution:
        \begin{align*}
            \fcolorbox{rwth-blue}{white}{$x_1$}&=1\\
            x_2&=0\\
            \fcolorbox{rwth-blue}{white}{$x_3$}&=1\\
            x_4&=0
        \end{align*}
    \end{minipage}\hfill
    \begin{minipage}[t]{0.54\textwidth}
        \textsc{Subset Sum} instance:\\
        \resizebox{\textwidth}{!}{
        \begin{tabular}{r|cccc|cc}
            & $x_1$ & $x_2$ & $x_3$ & $x_4$ & $C_1$ & $C_2$\\
            \hline
            \fcolorbox{rwth-blue}{white}{$a(x_1)$}            & 1 & 0 & 0 & 0 & 1 & 0\\
            $a(\overline{x}_1)$                               & 1 & 0 & 0 & 0 & 0 & 1\\
            $a(x_2)$                                          & 0 & 1 & 0 & 0 & 1 & 0\\
            \fcolorbox{rwth-blue}{white}{$a(\overline{x}_2)$} & 0 & 1 & 0 & 0 & 0 & 0\\
            \fcolorbox{rwth-blue}{white}{$a(x_3)$}            & 0 & 0 & 1 & 0 & 0 & 1\\
            $a(\overline{x}_3)$                               & 0 & 0 & 1 & 0 & 1 & 0\\
            $a(x_4)$                                          & 0 & 0 & 0 & 1 & 0 & 1\\
            \fcolorbox{rwth-blue}{white}{$a(\overline{x}_4)$} & 0 & 0 & 0 & 1 & 0 & 0\\
            \hline
            $M'$                                               & 1 & 1 & 1 & 1 & 1 & 1\\
        \end{tabular}
        }
        Each row is a number to the base $n+1=5$.\\
        Thus, e.g. $a(x_1)=5^5+5^1=3125+5=3130$, and the target sum $M'=\sum_{i=0}^{5}5^i=3906$.
    \end{minipage}
    \caption{The reduction from \textsc{Exactly-One-3-SAT} to \textsc{Subset Sum}. The table shows the encoding of the literals and clauses in the \textsc{Subset Sum} instance as a number in base $5$.}
    \label{fig:ex1-3SAT-SubsetSum}
\end{figure}

\subsubsection{\textsc{3-SAT} \SSP{} \textsc{Exact Cover}}
\threesat
\EC
This reduction does not originate in one of the three papers that are covered in this thesis,
as~\citeLiterature{karp1975} uses a reduction to \textsc{Graph Coloring} as an intermediate step, but \textsc{Graph Coloring}
was excluded here, as explained in~\cref{subsec:excludedProblems}, so we had to come up with another reduction.
It was inspired by the lecture notes from the University of Pennsylvania~\citeWebsites{pennsylvania}, 
where a similar reduction is used to show the NP-completeness of \textsc{Exact Cover}.

\functionG
We want to reduce a \textsc{3-SAT} instance $(L,C)$ to an \textsc{Exact Cover} instance $(U, \mathfrak{S})$, 
which we do by creating sets $x_i^+$ and $x_i^-$ for each variable $x_i$. The sets encode whether the variable
is set to true or false. Additionally we include sets $C_j$ for each clause $C_j$.
These sets contain elements $Z_{j,k}$ for each $k$-th element of the $j$-th clause, as well as elements $C'_j$
for each clause and $x'_i$ for each variable. 
The clause set $C_j$ contains the element $C'_j$ and each combination of $Z_{j,k}$,
thus representing by which assignments the clause can be satisfied.
The sets $x_i^+$ and $x_i^-$ are constructed such that
they each contain the elements $Z_{j,k}$ that correspond to the negation of the literal.
This way, when lets say $x_i^-$ is chosen, the $Z_{j,k}$ elements with the $k$-th variable of $C_j$ being $x_i$ are already covered, 
and the remaining $Z_{j,k}$ elements, thus the ones with the $k$-th variable of $C_j$ being $\overline{x}_i$, can be chosen with the clause sets.
Thus, for each clause, exactly one of the clause sets is chosen, for each variable, exactly one of the variable sets is chosen,
and each $Z_{j,k}$ element must be part of these choices, as they can only be covered this way.
\Cref{fig:3SAT-EC} shows the idea of the reduction.

Formally, we define the reduction function $g$ as follows:
Let $(L,C)$ be a \textsc{3-SAT} instance with variables $x_1, \ldots, x_n$ and clauses $C_1, \ldots, C_m$.
We now set the universe $U$ of the \textsc{Exact Cover} instance to be
\[
    U:=\{x'_i\mid 1\leq i\leq n\} \cup \{C'_j\mid1\leq j\leq m\}\cup\{Z_{j,k}\mid 1\leq j\leq m, 1\leq k\leq 3\}.
\]
This are the elements corresponding to the variables, the clauses, and the $k$-th element of the $j$-th clause each.

Then, we define the sets as follows:
\begin{align*}
    x_i^+&:=\{x'_i\}\cup\{ Z_{jk}\mid \text{ the } k\text{-th element of } C_j \text{ is } \overline{x}_i\}\\
    x_i^-&:=\{x'_i\}\cup\{ Z_{jk}\mid \text{ the } k\text{-th element of } C_j \text{ is } x_i\}\\
    \text{ for each } C_j:\\
    &\{C'_j, Z_{j,1}\}, \{C'_j, Z_{j,2}\}, \{C'_j, Z_{j,3}\},\\
    &\{C'_j, Z_{j,1}, Z_{j,2}\}, \{C'_j, Z_{j,1}, Z_{j,3}\}, \{C'_j, Z_{j,2}, Z_{j,3}\},\\
    &\{C'_j, Z_{j,1}, Z_{j,2}, Z_{j,3}\}
\end{align*}
Thus we get the true and false sets for each variable, and a clause set for each possible combination of satisfied
literals for each clause.

\functionF
The functions $f_I$ are defined as $f_{(L,C)}(\ell_i)=\begin{cases}
    x_i^+ & \text{if } \ell_i=x_i\\
    x_i^- & \text{if } \ell_i=\overline{x}_i
\end{cases}$.

\correctness
The reduction runs in polynomial time, as there are 7 sets for each clause and 2 sets for each variable,
as well as $4m+n$ elements in the universe introduced, which is polynomial in the size of the \textsc{3-SAT} instance.

Let us assume $(L,C)$ is a satisfiable \textsc{3-SAT} instance.
Then, we can choose the following sets to form an exact cover:
First, we choose $x_i^+$ for each variable $x_i$ that is true in the satisfying assignment, 
and $x_i^-$ for each variable $x_i$ that is false, as given by the function $f_I$. 
Thus, all $f_{(L,C)}(\ell_i)$ are chosen for all $\ell_i\in L$.
As each variable is set to either true or false, either $x_i^+$ or $x_i^-$ is chosen for each variable,
resulting in the elements $x'_i$ being covered exactly once.
Then, for each clause $C_j$, we choose the set that contains the $Z_{j,k}$ elements such that the $k$-th 
literal of $C_j$ is satisfied by the satisfying assignment of $(L,C)$.
Thus, each $C'_j$ is covered exactly once. 
Next, we need to show that the $Z_{j,k}$ elements are covered exactly once as well.
We know that each $Z_{j,k}$ appears in exactly one $x_i^+$ or $x_i^-$ set and in three clause sets.
If $x_i^+$ is chosen, the $Z_{j,k}$ elements corresponding to the negation of $x_i$ are already covered by $x_i^+$,
and thus must not appear in the clause sets, and vice versa for $x_i^-$.
However, these are exactly the literals that are not satisfied, thus, they do not appear in the chosen clause sets.
The remaining $Z_{j,k}$ elements are covered by the chosen clause sets, as they are the ones that correspond to the satisfied literals.
Thus, we have an exact cover in the created instance.\\

Let us assume $(U', \mathfrak{S}')$ is a YES-instance of \textsc{Exact Cover}.
Then, each element of $U$ is covered exactly once by the sets in a solution of the instance.
Specifically, this means that for each $i$, exactly the set $x_i^+$ or $x_i^-$ is chosen,
thus, each variable is set to either true or false, according to $f_I$. 
For each $j$ exactly one of the clause sets is chosen. With each $x$ set,
the elements $Z_{j,k}$ in that set are covered, thus, the clause sets containing these $Z_{j,k}$ can not be part of the solution.
Thus, as each $C'_j$ can only be covered by a clause set, and they always contain at least one $Z_{j,k}$,
the clause set that is chosen must contain all $Z_{j,k}$ elements that are not already covered by the $x$ sets.
Thus, setting the $x_i$ variables to true or false according to the chosen $x_i^+$ or $x_i^-$ sets results
in a satisfying assignment for the \textsc{3-SAT} instance, because each clause is satisfied by at least one literal,
as it otherwise would not be covered by a clause set.\\
Thus, the reduction is a correct SSP reduction from \textsc{3-SAT} to \textsc{Exact Cover}.
\begin{align*}
    & \{f_{(L,C)}(S) \mid S\in\S(\textsc{3-SAT})\}\\
    =& \{f_{(L,C)}(L^*) \mid L^*\subset L \text{ such that }\vert L^*\cap\{\ell_i,\overline{\ell}_i\}\vert=1 
    \text{ and }\vert L^*\cap C_j\vert\geq1\text{ for all }C_j\in C \} \\
    =& \{ X \mid X\subseteq\{ x_i^+, x_i^-\mid \ell_i\in L^*\}, \exists S'\text{ such that } X\subseteq S', S'\in \S(\textsc{Exact Cover}) \}\\
    =& \{S'\cap f_{(L,C)}(L)\mid S'\in \S(\textsc{Exact Cover})\}
\end{align*}

\begin{figure}
    \centering
    \begin{minipage}[t]{0.25\textwidth}
        \threeSATexample
    \end{minipage}
    \begin{minipage}[t]{0.74\textwidth}
        \textsc{Exact Cover} instance:\\
        \begin{align*}
        U=& \{x'_1, x'_2, x'_3, x'_4, C'_1, C'_2, Z_{1,1}, Z_{1,2}, Z_{1,3}, Z_{2,1}, Z_{2,2}, Z_{2,3}\}
        \end{align*}
        \begin{align*}
        \fcolorbox{rwth-blue}{white}{$x_1^+$}=&\{x'_1, Z_{2,1}\} && C_1: && \{C_1, Z_{1,1}\}, \{C_1, Z_{1,2}\}, \{C_1, Z_{1,3}\},\\
        x_1^-=&\{x'_1, Z_{1,1}\} && && \fcolorbox{rwth-blue}{white}{$\{C_1, Z_{1,1}, Z_{1,2}\}$}, \{C_1, Z_{1,1}, Z_{1,3}\},\\
        \fcolorbox{rwth-blue}{white}{$x_2^+$}=&\{x'_2\} && && \{C_1, Z_{1,2}, Z_{1,3}\}, \{C_1, Z_{1,1}, Z_{1,2}, Z_{1,3}\}\\
        x_2^-=&\{x'_2, Z_{1,2}\} && C_2: && \{C_2, Z_{2,1}\}, \fcolorbox{rwth-blue}{white}{$\{C_2, Z_{2,2}\}$}, \{C_2, Z_{2,3}\},\\
        \fcolorbox{rwth-blue}{white}{$x_3^+$}=&\{x'_3, Z_{1,3}\} && && \{C_2, Z_{2,1}, Z_{2,2}\}, \{C_2, Z_{2,1}, Z_{2,3}\},\\
        x_3^-=&\{x'_3, Z_{2,2}\} && && \{C_2, Z_{2,2}, Z_{2,3}\}, \{C_2, Z_{2,1}, Z_{2,2}, Z_{2,3}\}\\
        x_4^+=&\{x'_4\}\\
        \fcolorbox{rwth-blue}{white}{$x_4^-$}=&\{x'_4, Z_{2,3}\}
        \end{align*}
    \end{minipage}
    \caption{Example of the reduction from \textsc{3-SAT} to \textsc{Exact Cover}. 
    The sets highlighted in blue form the solution corresponding to the satisfying assignment given for the \textsc{3-SAT} instance.}
    \label{fig:3SAT-EC}
\end{figure}

This concludes our examination of reductions that could be transformed into SSP reductions.

\subsection{Reductions That Are No SSP-Reductions}\label{subsec:noreductions}
In this section, all reductions from Section 4 in~\citeLiterature{karp1975}, 
that do not fulfill the SSP property will be listed and examined.

\subsubsection{\textsc{Clique} $\leq_p$ \textsc{Vertex Cover}}\label{subsubsec:clique-vc}
\clique
\VC
This reduction by Karp~\citeLiterature{karp1975} is not an SSP reduction and can also not easily be transformed into one.

In the reduction from the \textsc{Clique} instance $(G,k)$ to the \textsc{Vertex Cover} instance $(G',k')$,
we set $V'=V$, and invert the edges, i.e., $E'=\{\{u,v\}\mid \{u,v\}\notin E\}$.
Then, we set $k'=n-k$.
The solution sets are exactly the opposite, as the clique of size $k$ in $G$, which we call $V^*$, 
is an independent set in $G'$ as we have seen in~\cref{subsubsec:clique-is}. 
This is equivalent to all vertices in $V^*$ being pairwise disjoint in $G'$,
thus, each edge $e\in E'$ has at least one endpoint in $V'\setminus V^*$, making $V'\setminus V^*$ a vertex cover of $G'$.
Conversely, if $V^*$ is a vertex cover in $G'$, then all vertices $v\in V'\setminus V^*$ are pairwise disjoint,
resulting in an independent set, thus $V\setminus V^*$ is a clique in $G$.
Thus, the solution set of the \textsc{Clique} instance contains exactly the vertices not in the solution set 
of the \textsc{Vertex Cover} instance, which means that the function $f_{(G,k)}$ cannot be derived, 
as we do not know before which vertices are in the solution set of the 
\textsc{Vertex Cover} instance and which are not.

\subsubsection{\textsc{Exact Cover} $\leq_p$ \textsc{Hitting Set}}
In his paper, Karp defines the \textsc{Hitting Set} problem as follows:
\begin{problem}
    \problemtitle{\textsc{Hitting Set} as in Karp's Paper}
    \probleminstance{Set of sets $S_j\subseteq\{1,\ldots,n\}$ for $j\in\{1,\ldots,m\}$.}
    \problemuniverse{$\U:=\{1,\ldots,n\}$.}
    \problemfeassol{-}
    \problemsol{The set of all $H\subseteq\{1,\ldots,n\}$ such that $H\cap S_j\neq\emptyset$ for all $j\in\{1,\ldots,m\}$.}
\end{problem}
\begin{problem}
    \problemtitle{\textsc{Hitting Set} - usual definition}
    \probleminstance{Set of sets $S_j\subseteq\{1,\ldots,n\}$ for $j\in\{1,\ldots,m\}$, number $k\in\NN$.}
    \problemuniverse{$\U:=\{1,\ldots,n\}$.}
    \problemfeassol{The set of all $H\subseteq\{1,\ldots,n\}$ such that $H\cap S_j\neq\emptyset$ for all $j\in\{1,\ldots,m\}$.}
    \problemsol{Set of all feasible solutions with $\vert H\vert\leq k$.}
\end{problem}
However, all other literature known to me defines the problem it is written here.
Even Karp himself defines \textsc{Hitting Set} in the commonly known way in a later paper~\citeLiterature{chandrasekaran2011algorithms}.
The two definitions are not equivalent, which can be seen with a simple example:
Lets say we have the sets $S_1=\{1\}, S_2=\{2\}$ over the universe $\{1,2\}$, then one can easily use the universe as
a solution when using Karp's definition. 
If we take the same set, but add an k, lets say $k=1$, we cannot find a solution that contains at most $k$ 
elements and covers all $S_j$, thus, there is no solution to the usual definition of \textsc{Hitting Set} for the same instance.
As Karp's definition is not used in any other literature known to us, 
it is unnecessary to consider the reduction from \textsc{Exact Cover} to \textsc{Hitting Set} in Karp's paper.

\subsubsection{\textsc{Subset Sum} $\leq_p$ \textsc{Sequencing}}
\sequencing

Karp again uses a formulation of the \textsc{Sequencing} problem that differs from the typical definition in literature.
He defines the reduction as follows:\\
\begin{align*}
    p'&:=n, &T'_i=D'_i:=a_i\\
    D'_i&:=M, &k':=\sum_{i=1}^n a_i - M\\
\end{align*}
The problem that occurs either way is that each job is part of the solution, 
and the question is only in which order the jobs are executed, which makes it impossible to define the
functions $f_I:(a_i, i\in\{1,\ldots,n\})\rightarrow(i'\in\{1,\ldots,n\})$, as only the order can be considered.
Thus, each solution only contains one element, namely the order of the jobs,
and thus, all elements belonging to one solution of the \textsc{Subset Sum} instance must be mapped
to the same element in the Universe of the \textsc{Sequencing} instance, which is not possible, 
because the functions $f_I$ must be injective.

\subsubsection*{\textsc{SAT} $\leq_p$ \textsc{0-1-Integer Programming}}
\addcontentsline{toc}{subsubsection}{\textsc{SAT} $\leq_p$ \textsc{0-1-Integer Programming}}
\sat
The definition of the problem \textsc{0-1-Integer Programming} differs slightly from the definition Karp used in his paper~\citeLiterature{karp1975},
as this is the more common definition in literature:
\IP

In this reduction, a \textsc{SAT} instance $(L,C)$ is mapped into a matrix and a vector $(C',b')$. 
The rows of the matrix $C'$ correspond to the clauses, with the columns standing for a variable each. 
Each entry directly encodes whether the variable of that column is included ($-1$)
in the clause corresponding to the row, or its negation ($1$), or neither ($0$). 
The vector $b'$ basically counts the number of negated variables in each clause.\\
Thus, the solution vector $X'$ corresponds to an assignment of the variables from the \textsc{SAT} instance, 
and the product of the matrix $C'$ with $X'$ must be less or equal to $b'$ in each entry.
If a non-negated variable is included, the entry will be negative, and if a negated variable is included, 
the entry will be zero, but the entry from $b'$ will be greater or equal to zero, as then, 
at least one negated literal is in the corresponding clause, thus, the clause is satisfied by the assignment 
given by $X'$, if $X'$ is a solution to the \textsc{0-1-Integer Programming} instance.

\functionG
The reduction is based on Karp's reduction, using some adaption to ensure all solutions can be mapped to a solution of the \textsc{0-1-Integer Programming} instance.
We set the function $g$ as follows:
\begin{align*}
    c'_{ij}&=\left\{\begin{array}{rl}
        -1 & \text{if }x_j\in C_i \\
        1 & \text{if }\overline{x_j}\in C_i\\  
        0 & \text{otherwise}
    \end{array}\right.\\
    b'_i&= -1+\vert\{ \overline{x_j}\mid \overline{x_j}\in C_i, 1\leq j\leq n\}\vert
\end{align*}
for $1\leq i \leq p,\: p$ denotes the number of clauses in the \textsc{SAT} instance $\phi$.
The reduction $g$ is polynomial, as the size of the matrix $C'$ is $p\times n$ and the size of the vector $b'$ is $p$.
\functionF
The problem that occurs now is that we need to define $f_{\phi}(\ell_i)$, but both literals $x_i$ and $\overline{x}_i$
share one entry in the solution vector, which means we would have to say that $x_i$ 
is chosen if the corresponding entry is one and $\overline{x}_i$ is chosen if it is zero,
which is not possible because a boolean variable $x'_i$ is not included in the solution if it is set to zero,
which means that the mapping of the negated variables does not work.
An example is given in~\cref{fig:sat-int}.

\begin{figure}
    \centering
    \begin{minipage}[t]{0.39\textwidth}
        \SATexample
    \end{minipage}\hfill
    \begin{minipage}[t]{0.59\textwidth}
        \textsc{0-1-Integer Programming} instance:\\
        \begin{align*}
            C=&\begin{blockarray}{ccccc}
                & x_1 & x_2 & x_3 & x_4 \\
                \begin{block}{c(cccc)}
                C_1 &-1 &-1 & 1 & 0\\
                C_2 & 1 &-1 & 0 & 0\\
                C_3 & 1 & 0 & 0 & 1\\
                \end{block}
            \end{blockarray}\\
            b=&\begin{blockarray}{ccc}
                C_1 & C_2 & C_3\\
                \begin{block}{(ccc)}
                    0 & 0 & 1\\
                \end{block}
            \end{blockarray}
        \end{align*}
    \end{minipage}\hfill
    \caption{Example of the polynomial reduction by Karp from \textsc{SAT} to \textsc{0-1-Integer Programming} 
    which does not fulfill the SSP property.}
    \label{fig:sat-int}
\end{figure}

\subsection{Problems that are not Covered in the SSP Context}
\label{subsec:excludedProblems}
Some problems are not considered in the context of SSP reductions.
As we have learned, our main motivation for finding SSP reductions is to find new SSP-NP-complete problems,
because then the min-max variants of these problems are $\Sigma_2^p$-complete, and there are not too many
problems known to be $\Sigma_2^p$-complete. Additionally, the min-max-min variant is proven to be $\Sigma_3^p$-complete.
For some problems, it does not really make sense to define a min-max variant, because the question of how
the variant would be defined exactly is not clear and it would be quite unnatural to define it.
This is the case for the problem \textsc{Graph Coloring}.

\subsubsection{\textsc{Graph Coloring}}
\graphColoring
In Karp's paper, the \textsc{Graph Coloring} problem is found under the name \textsc{Chromatic Number}, 
but in modern literature, it is mostly referred to as \textsc{Graph Coloring}.\\

Lets take a look at the \textsc{Graph Coloring} problem. 
This problem asks whether there is a function that assigns $k$ different
colors to the vertices of a graph such that no two adjacent vertices have the same color.

Next, let us try to define the network interdiction variant of \textsc{Graph Coloring}.
According to~\citeLiterature{grune2023large}, the question of the network interdiction 
variant is given a threshold $t$, some cost function $c$, 
that assigns weights to the elements of the universe, in this case all combinations of vertices and colors,
whether there is some subset of the universe such that the total cost of that subset does not exceed $t$, 
and each solution contains an element from that subset, formally: 
\[ \exists B\subseteq\U(x)\text{ such that } c(B)\leq t\text{ and }\forall S\in \S(x): B\cap S\neq\emptyset \]
This is an interesting constraint, as it now means that some vertices can not be colored with some specific colors.
However, each solution can simply be permuted and is still valid,
as an example, if we have an instance with $k=2$ and a solution where e.g. $v_1$ is colored with color 1 
and $v_2$ with color 2, but $(v_1, 1)$ is prohibited, simply switching each vertex colored 
in color 1 to color 2 and vice versa still yields a valid solution.
But we can also not ban a vertex to be colored by any color, as that would interfere with the definition of the problem.

Thus, we see that defining the network interdiction variant of \textsc{Graph Coloring} is not possible in a meaningful way.
This holds for partition problems in general, as they can always be permuted and therefore provide multiple
possible equivalent solutions. In the problem \textsc{Partition}~\cref{subsubsec:3DEM-Partition}
we were able to solve this problem by defining that the element $a_n$ must be part of the solution,
and as there are only two partitions and thus possible identical solutions, this works, but for $k!$ many
possible solutions, this is not feasible.

\chapter{Documentation of the Website}\label{chap:documentation}
The website that is documented in this chapter is a compendium website that currently focuses on SSP-NP-complete
problems and SSP reductions. It can be found under the URL \url{https://reductions.network}~\citeWebsites{ReductionsNetwork}.
The source code of the project can be found in the GitLab repository under \url{https://git.rwth-aachen.de/femkepfa/ba-compendiumwebsite},
and the documentation refers to the commit \code{9c3fb54f}, which is the version of the project that is submitted as part of this thesis.
In this chapter, first, the general structure and architecture
of the website is given in~\cref{sec:concept,sec:structure}, then, some information on accessibility and how it was tried to incorporate are given,
followed by descriptions of the libraries used in the project in~\cref{sec:libraries}.

\section{Concept of the Website}\label{sec:concept}
The website serves as a visual collection for problems and reductions between them. 
At the moment, all known SSP-NP-complete problems are included, as well as SSP-NP reductions, but the intent is to
expand the collection to include more problems and reductions of different complexity.
The most important feature hence is the display of the network, as well as the ability to interact with it,
thus getting access to more information as well as being able to add entries to or edit entries in the collection.
As everyone interested should be able to contribute, but it needs to be ensured that the information they submit is correct,
and that the server is not flooded with spam or malware, two different roles are implemented: Visitors and registered users.
Visitors can submit their contributions, which are not directly added to the database, as they need to be approved
by a registered user. Registered users are basically the admins of the page: they can add, edit, and delete entries,
as well as approve or reject contributions from visitors and register new users.\\

\begin{figure}
    \centering
    \includegraphics[width=\textwidth]{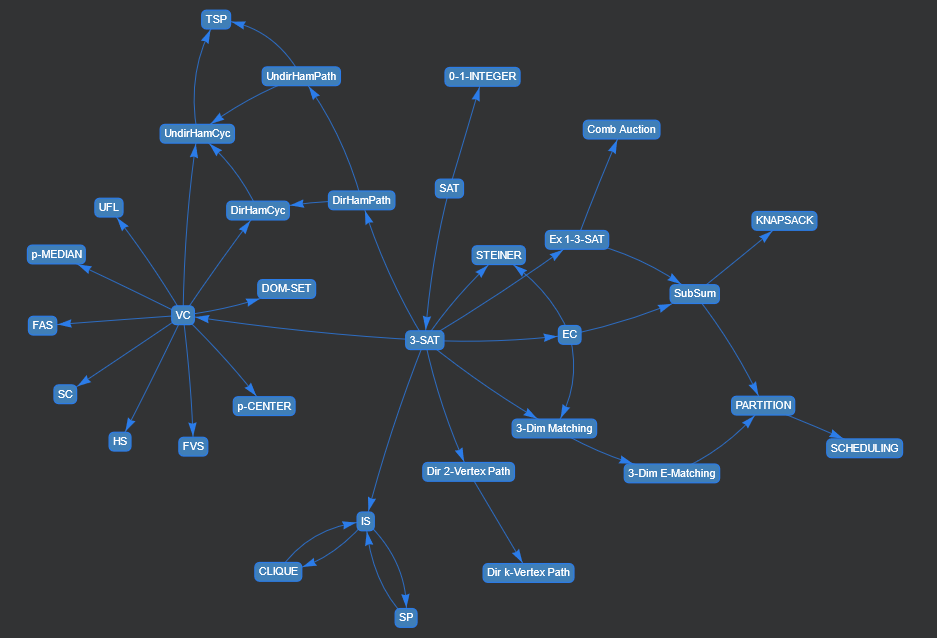}
    \caption{The graph showing problems and reductions between them on the compendium website.}
    \label{fig:websiteGraph}
\end{figure}

The problems and reductions are displayed as nodes and edges in a graph, where the nodes represent
the problems and the edges, going from one node to another, represent the reductions between the
problems. The nodes are labeled with the abbreviation of the problem they represent,
as the complete name looks cluttered.~\Cref{fig:websiteGraph} shows how this looks like.

\begin{figure}
    \centering
    \includegraphics[width=\textwidth]{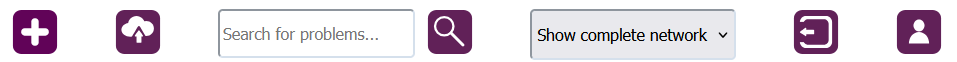}
    \caption{The icons for the navigation tools in the header of the website for a user that is logged in.
    The plus opens a form to add a new entry, the upload symbol opens a form in which data can be
    submitted in the JSON format, the magnifying glass submits the search form, the rectangle with 
    an outgoing arrow logs the user out, and the person icon opens the page to register a new user. }
    \label{fig:header}
\end{figure}

Furthermore, the website needs some navigation tools, as the network can become quite large and crowded.
For this, a search function was implemented, which the user can use to search for a specific node, which is then
focused on and its information is displayed. Additionally, there is a filter function, which allows the user
to focus on the entries relevant for them, for example if they want to investigate only reductions that are SSP-NP
reductions, they can use the filter to gray out all entries to which the filter does not apply.
The icons for the navigation tools are shown in~\cref{fig:header}.\\

\begin{figure}
    \centering
    \includegraphics[width=\textwidth]{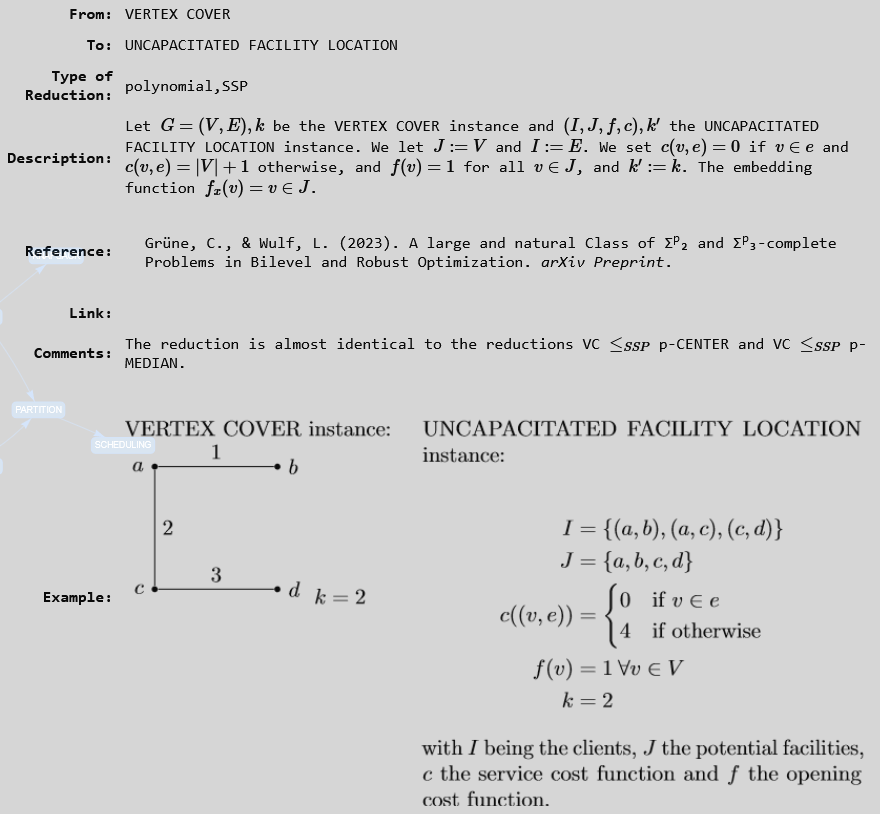}
    \caption{The additional information panel that opens when a user clicks on a node or edge in the graph.}
    \label{fig:addInfoDiv}
\end{figure}

To access further information on a problem or reduction, the user can click on the corresponding node or edge,
which then opens a panel with detailed information on the element, including references for further reading,
as can be seen in~\cref{fig:addInfoDiv} for the reduction from \textsc{Vertex Cover} to \textsc{Uncapacitated Facility Location}.
For reductions, the definitions of the problems involved are displayed as well, and for most cases also a visual
example of the reduction, as this helps immensely to understand the idea behind the reduction.\\

\section{Structure and Architecture}\label{sec:structure}
The website is created using node.js, which is a runtime environment for JavaScript for the creation
of servers, apps, etc.~\citeWebsites{nodejs}. Node.js runs JavaScript code outside of a browser,
using an asynchronous design, which makes it very efficient for I/O-heavy tasks, as the thread is not blocked,
but the return value is awaited, when an I/O operation is executed.
One advantage of node.js is that one can use JavaScript in the backend, and writing the whole project
using the same language makes maintenance as well as readability better and easier,
as the website also uses JavaScript in the frontend. 
Also, node.js provides many libraries and functionalities,
either build-in or usable through npm, the node package manager, which prevents reinventing the wheel,
as many features have already been implemented by other programmers and can be used. 

The project is developed without a framework for the client-side code such as React or Angular. 
As the project is not very complex, only containing three pages in total, a framework was not necessarily
needed.

The main page, index.ejs, is the page that displays the network and everything related to it.
Additionally, there is a page for the login, login.ejs, and a page to register new users, register.ejs.

\begin{figure}[]
    \centering
    \includegraphics[width=\textwidth]{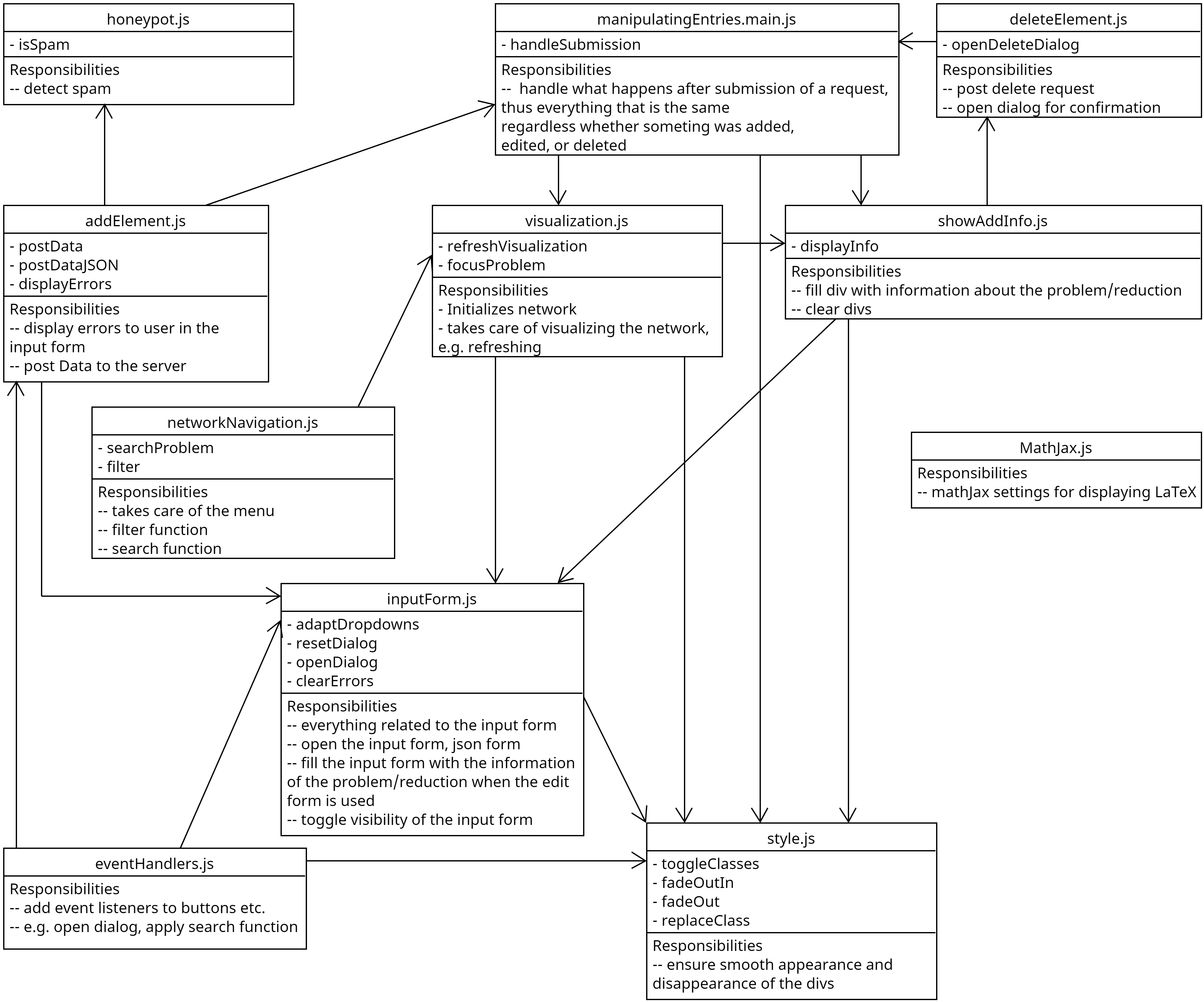}
    \caption{Dependency graph of the frontend files}
    \label{fig:class_diagram}
\end{figure}

In the frontend, HTML, CSS, and JavaScript are used exclusively. 
In~\cref{fig:class_diagram}, a diagram of the files in the frontend is shown. 
The arrows indicate the dependencies between the files, thus e.g. \code{eventHandler.js} uses functions from \code{visualization.js}.

The file \code{eventHandlers.js} is responsible for adding event handlers to the buttons and forms on the website.
One example is the event handler for the toggle buttons, which change the visibility of the divs that are displayed,
or the button which opens the input form for adding a new problem or reduction.

Some of the functions used in \code{eventHandlers.js} are from the file \code{inputForm.js}.
This file contains functions that control the input form, such as clearing the inputs, filling in the already
existing values when a problem or reduction is to be edited, or setting the input fields of the correct form
to required.

In \code{addElement.js}, the \code{post} functions are found, as well as the functions for displaying errors
during the post process to the users so that they can fix them and submit again. The \code{post} functions
are explained in~\cref{subsec:postData}. The file handles the posting of data by visitors, registered users,
and the posting of JSON data to the server.

For this, the \code{isSpam} function from the file \code{honeypot.js} is needed, to ensure that spam by bots
can be detected and is not posted to the server. This is explained in~\cref{subsec:honeypot}.

After the data is posted, the file \code{manipulatingEntries.main.js} is used to handle what happens next.
This means in the case of registered users to display the new problem or reduction in the network,
and otherwise to give the user a response so that they know their data was submitted successfully.

When a element is to be deleted, this happens in \code{deleteElement.js}. This file contains the function
to delete an entry, which includes an alert to confirm the deletion, and calling the backend to delete the entry.
After deletion, the network is updated to reflect the changes.

The file \code{visualization.js} takes care of initializing the network using vis.js~\cref{subsec:visjs},
updating the network, and interacting with it. The file includes a method to focus on a specific node.

When a node or edge is clicked, information on that element is displayed on the right side of the screen.
This is handled in \code{showAddInfo.js} and includes fetching the information from the backend and 
displaying it in the correct format, calling MathJax to render LaTeX code, and displaying the example image 
for reductions.

In \code{style.js}, the smooth fading in and out of the divs that are displayed on top of the network is realized.
This is done by using different classes with CSS styling options and transitions.

\code{NetworkNavigation.js} handles the navigation functions that can be used through the buttons in the header.
Specifically, the search and filter functions and everything related to them are implemented here. 
They will be explained in~\cref{subsec:searchFilter}.

The last file in the frontend is \code{mathjax.js}, in which the configuration for MathJax is set up.
This includes what delimiters are used for the math mode.

\begin{figure}
    \centering
    \includegraphics[width=\textwidth]{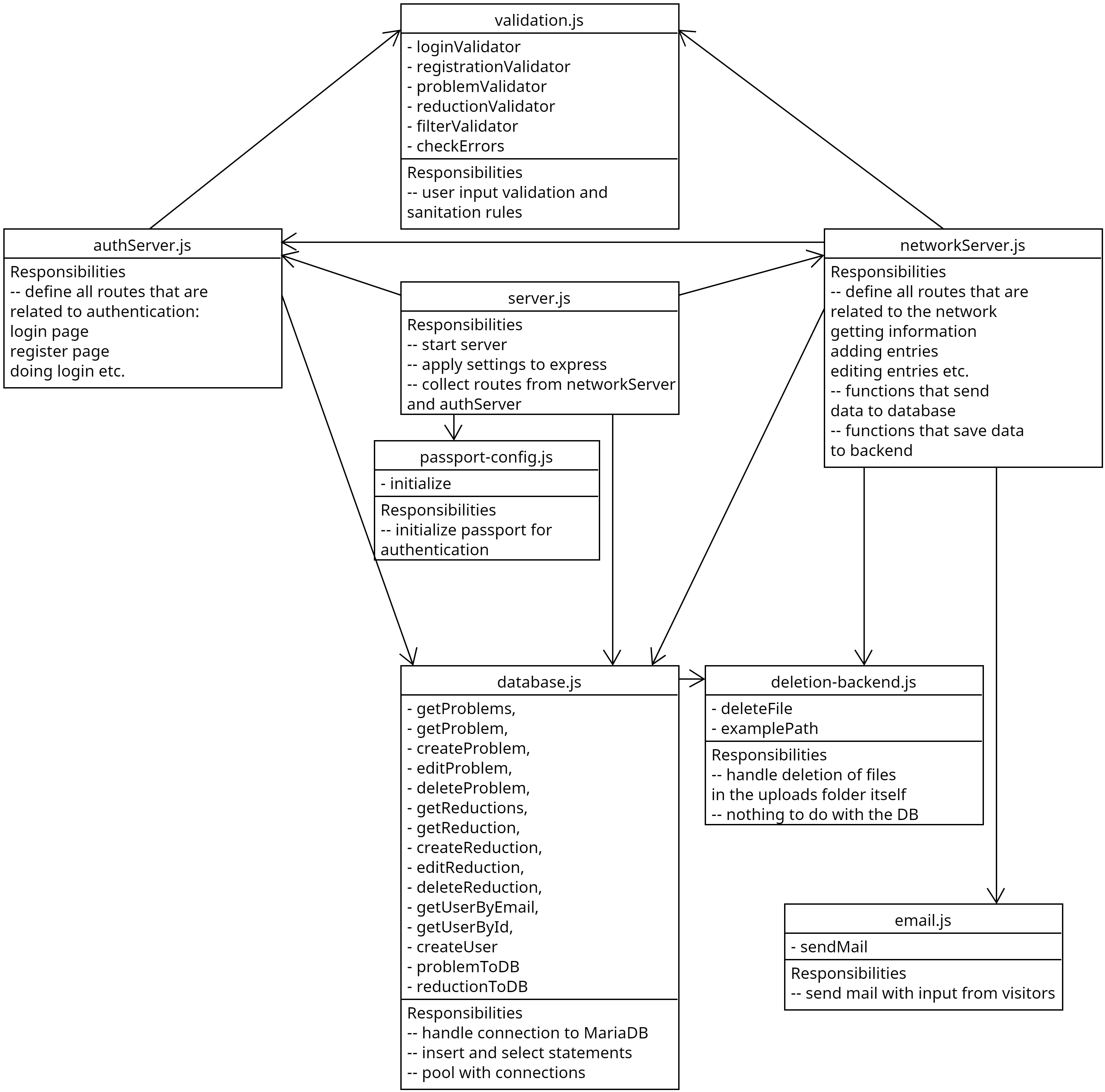}
    \caption{Dependency graph of the backend files}
    \label{fig:backend_class_diagram}
\end{figure}

In the backend, the project is structured as shown in~\cref{fig:backend_class_diagram}.
There are three server files, that contain the routes for connecting the frontend with the backend:
\code{server.js}, \code{authServer.js}, and \code{networkServer.js}.
In \code{server.js}, the connection to the database is established, the server is set up,
and the routes from \code{authServer.js} and \code{networkServer.js} are imported.
\code{Authserver.js} includes all routes related to the login and registration of users, 
and \code{networkServer.js} contains all routes related to the network, e.g. fetching the data for the network,
posting new data, etc.

The file \code{validation.js} contains the validation functions for the input data, which are used in the routes.
\code{Passport-config.js} initializes the passport module, which is used for the authentication of users.

In \code{database.js}, the connection to the database is established, and the queries for the database are defined,
as well as the functions \code{reductionToDB} and \code{problemToDB}, on which~\cref{subsec:reductionToDB} focuses.
In these functions, \code{deletion-backend.js} is used to delete files from the upload folder.

When visitors submit an entry, an email is sent containing the input in JSON format.
This is done in \code{email.js}, which uses the nodemailer module to send the email.

\section{Accessibility}\label{sec:accessibility}
Accessibility for as many people as possible is always an important task.
In this project, the following aspects were considered:
\begin{itemize}
    \item \textbf{Colorblindness:} The colors used in the network visualization are chosen in a way that they are distinguishable for people with colorblindness.
    The graph is blue, which is a color that most people can see, and the only other color used is purple, which is
    used for the navigation tools and thus does not necessarily need to be distinguished from the blue graph.
    Besides that, a dark grey and white are used.
    \item \textbf{Contrast:} The contrast between text and background is high enough to be readable for people with visual impairments.
    \item \textbf{Screen reader:} We tried to make the website compatible with screen readers, so that visually impaired people can use it.
    All input fields have labels, and it was made sure to always used proper methods to e.g. hide elements that should not be visible,
    so that they are invisible for screen readers as well. 
    However, as no one involved in this project has a screen reader nor the knowledge to use one,
    it was not possible to test whether one can actually use and navigate the website with a screen reader.
    As the main point of the website is to visually display relations between problems and reductions,
    it is hard to make it fully accessible for people with visual impairments, so it was decided to not prioritize
    this aspect.
    \item \textbf{Simple design:} The design of the website is simple, so that one does not get distracted by unnecessary elements,
    and it is clear what the navigation elements do, also by using icons. 
    All navigation elements are colored to distinguish them from the rest of the website.
    \item \textbf{Focus:} The element the user interacts with is always focused on, e.g. when a form is opened, 
    the user will want to enter something into the first input field of the form, so that is focused.
\end{itemize}

To gain information on what to consider regarding accessibility, 
mainly, the Accessibility Developer Guide by the Foundation "Access for all"~\citeWebsites{accDevGuide}
and mozilla's developer guide on accessibility~\citeWebsites{mdnAccessibility} were used.

\section{External Libraries}\label{sec:libraries}

\subsection{Vis.js}\label{subsec:visjs}
Vis.js is the most important library used in this project. 
It is a browser based visualization library that can handle large amounts of dynamic data~\citeWebsites{visjs}, which is important for 
this use case, as its objective is to create a compendium website with potentially thousands of entries.
The library was originally developed by Almende B.V.~\citeWebsites{almende} and is now maintained by the community.
In this project, the network library~\citeWebsites{visnetwork} is used to create the network visualizations on the website.\\

Important factors for choosing the right visualization library were
\begin{enumerate}
    \item Scalability, to ensure that the library can handle a large amount of data,
    \item Manipulation, such that the data and appearance can be changed dynamically,
    \item Interaction, so that the user can e.g. click on an element to get more information displayed
    \item Open source and free to use
\end{enumerate}
Vis.js was chosen because it fulfills all the needs mentioned above and is widely used and therefore, 
a lot of documentation, help, and examples are available.
Also, it is easily integrable in a website, as it is a JavaScript library, whereas some alternatives are based on other languages,
such as Pyvis~\citeWebsites{pyvis}, which is a Python library, or 
NetworkD3~\citeWebsites{networkD3}, which, even though it is a JavaScript library,
comes from an R background. Again, sticking to JavaScript makes it easier to maintain the code.\\

In the documentation of vis.js Network, a basic network is presented, which forms the basis for the network used in this project.
The website uses the standalone build of vis.js, which means it includes all necessary dependencies for 
the network visualization and is ready to use out of the box.
However, the standalone build is not combinable with other vis.js modules~\citeWebsites{visjs}.\\

Vis.js has many features implemented, such as physics simulation, which is used in this project to create a dynamic network,
or the ability to manipulate the edges and nodes dynamically, e.g. changing colors, adding or removing nodes and edges.
In the project, the function to initialize the network, thus setting the options and data, 
the \code{get()} function to get nodes and edges from the network, the \code{update()} and \code{remove()}
functions, which update the elements in the network but without reloading it, 
resulting in a smooth display of new elements, as well as the function for removal of elements are used. 
Furthermore, the \code{focus()} method is used, which
centers the network on a specific node, needed for the search function implemented.\\

Some functionalities needed to be added. Firstly, a zoom limit was added, as the network zoomed so far out
that the elements are not visible anymore quite quickly, and with the zoom limit, the zooming always
stays within a reasonable range.\\
Another feature that would be useful is to be able to define whether scrolling zooms the network
or scrolls the page, as this is not trivial to implement, but very useful. It would be beneficial to 
have the option to adjust vis.js such that pinching, and on computers, control in combination 
with scrolling, zooms the network, and normal scrolling scrolls the page up and down.\\
Also, a function to change the color of the nodes and edges as part of a filter function was added, 
which definitely would be a good feature for the library.
Additionally, it would be nice to also include an option that disables elements, thus in the case
of the filter function, one could disable clicking on nodes or edges to which the filter does not
apply, and at the moment, this needs to be implemented by the user, by handling the click event.

\subsection{MariaDB}\label{subsec:mariadb}
To store the data, thus all the information on the problems and reduction, a MariaDB database is used.

MariaDB is an open source relational database management system using SQL.
In this project, the package mariadb is used to connect to the database and execute queries.
It provides a method to generate a pool of connections, such that multiple queries can be executed in parallel,
and there is no need to initiate new connections for each query. Thus, MariaDB is used for storing the data,
and the library mariadb is used to access and manipulate the data.

\subsection{Express}\label{subsec:express}
Express is a web framework for node.js~\citeWebsites{express}. Its purpose is to handle the routes, 
create APIs, and provide many middleware modules. 
In this project, express is used to serve files, manage sessions using express-session, validate data with express-validator,
limit the access to avoid flooding with express-rate-limit,
and create and handle routes as well as APIs.

\subsubsection*{Express-session}
\addcontentsline{toc}{subsubsection}{Express-session}
Express-session is a middleware for express that manages the sessions of the users. This is done through storing
a cookie in the client's browser, which is sent with each request, so that express-session can identify the user
and validate that they actually are the same user. The session automatically expires after some time, in this case after two hours.
Express is needed to check if a user is logged in, when requests are sent that require authentication, such as
manipulating the database. The routes that connect to the database are protected so that only logged in users can access them.

\subsubsection*{Express-validator}
\addcontentsline{toc}{subsubsection}{Express-validator}
This library is used for validation and sanitization of the data that is sent to the server to prevent attacks
such as SQL injection or cross-site scripting \citeWebsites{express-validator}. 
It can validate and sanitize as well the request body, but also the query, cookies, and more.
In this project, for most inputs, the length is checked, such that the input cannot be too long and flood the server.
As the inputs, e.g. the description of the reductions, should be short anyway, this also does not affect the user experience.
Additionally, the query for the filter function is validated and sanitized, as the content of the query is used to query the database.
Furthermore, for some inputs, the format is checked, and some inputs are sanitized to prevent attacks.
This means that special characters are escaped, e.g. to prevent HTML injection. 
However, since LaTeX code is allowed and needed for example in the definitions of the problems or 
the descriptions of the reductions, these inputs are not sanitized, as some special characters needed in the
LaTeX code would then be escaped.

\subsubsection*{Express-rate-limit}
\addcontentsline{toc}{subsubsection}{Express-rate-limit}
As the name suggests, the middleware express-rate-limit~\citeWebsites{express-rate-limit} 
is used to limit the number of requests to the server, to prevent denial of service (DoS) attacks. 
This is done by setting a limit on the number of requests that can be sent in a time window.
In this case, the rate for a specific user is limited, as well as for an IP address.
As the website is not expected to have many users trying to edit data at the same time, the rate is rather low,
with five requests per 15 minutes for login attempts as well as 15 requests per 15 minutes for the 
submission of data by visitors, which can be done by anyone and is therefore vulnerable to misuse of bots or other attacks.

\subsection{Passport}\label{subsec:passport}
Passport is an important library, as it is used for the authentication of the users.
This middleware provides many different means of authentication, some of them are via Google,
GitHub, or local authentication, which is used here~\citeWebsites{passport}. The local authentication strategy is used
as to not rely on services like Google, and to keep the control over the login process.
Passport is used in combination with bcrypt, which handles the hashing of the passwords.
The authentication process establishes a session, for which express-session is used, so that a user does not
have to sign in every time they want to access a route.\\
The library is very easy to use and popular.

\subsection{Bcrypt}\label{subsec:bcrypt}
Closely related to passport, bcrypt is used for the hashing of the passwords, using salt, 
so that they can be stored securely in the database~\citeWebsites{bcrypt}. 
Salt is some random extra data that is added to the input before encryption and thereby
provides more security, as it prevents dictionary attacks,  since the same passwords with 
different salt result in a different entry in the table storing the credentials.
Bcrypt handles the hashing as well as the comparison of the submitted passwords
and the stored password, when a user wants to log in. Of course, it still depends on the users to choose
secure passwords, as for insecure passwords that can be guessed easily, attackers can still gain
access to the account. To motivate users to choose secure passwords, the password is required to have a minimum
length as well as to contain at least one number, one uppercase letter, and one special character.
This is done in the frontend, which can of course be bypassed, but the users can be trusted 
to have no malicious intent, and only registered users can register new users.

\subsection{Nodemailer}\label{subsec:nodemailer}
Visitors should also be able to submit new reductions or problems.
However, these submissions should not be added to the database and displayed on the website immediately,
as they could contain malicious content and need to be reviewed for correctness.
This is where Nodemailer comes into play. Nodemailer is a module for node.js that allows to send and receive emails via SMTP
directly from the website~\citeWebsites{nodemailer}.
The library is easy to use, as one simply needs to provide the host, port, and authentication details of 
the email account with which one wishes to send emails, and then can send emails via the sendMail function.
In this project, whenever a visitor contributes a new submission,
an email is sent automatically to the website administrator, containing a JSON object with the details
of the submission as plain text. This way, the administrator gets notified about the new submission and can review it,
to then add it to the database by copying the JSON object into an input form on the website, that 
only accepts JSON objects as an input.

\subsection{Multer}\label{subsec:multer}
Multer handles multipart data~\citeWebsites{multer}, and is used for file uploads in this case. 
It handles where uploaded files are stored, as well as size limits and validation of file types.
Multer is included in the post routes, either with a file included, or without.

\subsection{Libraries for Formatting}\label{subsec:citation-js}
Citation-js is used for displaying the citations using BibTeX correctly.
It runs in the frontend, and can handle different citation formats. It was only tested it with BibTeX,
but if users add other standard citation formats, such as BibJSON or BibLaTeX, it should still render correctly~\citeWebsites{citation-js}.\\

Mathjax renders mathematical formulas in LaTeX, such that they are displayed correctly~\citeWebsites{mathjax}.
It also runs in the frontend, and, importantly, is accessible to screen readers.
As it runs in the frontend, it is not the fastest rendering solution, 
but as the LaTeX code that needs to be rendered only contains a few hundred characters at most,
and during development there did not appear to be any noticeable delays or performance issues, 
Mathjax is still a good choice, as it also cuts the workload of the server.

\section{Code Documentation}\label{sec:code}
In this section, we will dive deeper into the code of the website.
It is written in plain HTML, CSS, and JavaScript, using node.js and express.js as backend and middleware.
This approach was chosen, as we were not experienced with web development before starting this project and
wanted to understand the basics of web development before getting into more advanced frameworks like Angular or React.

We refer to logged in users as registered users, and to users that are not logged in as visitors.
The most important functions will be described in the following. 
These are the functions for the submission of data, both for registered users and visitors, 
the honeypot mechanism that prevents spam to be submitted through the form, the search- and filter functions,
as well as the function that handles the visualization of the network.
Additionally, one post route will be given as an example, to show which measures for protecting the server are included.

\begin{figure}
    \centering
    \includegraphics[width=\textwidth]{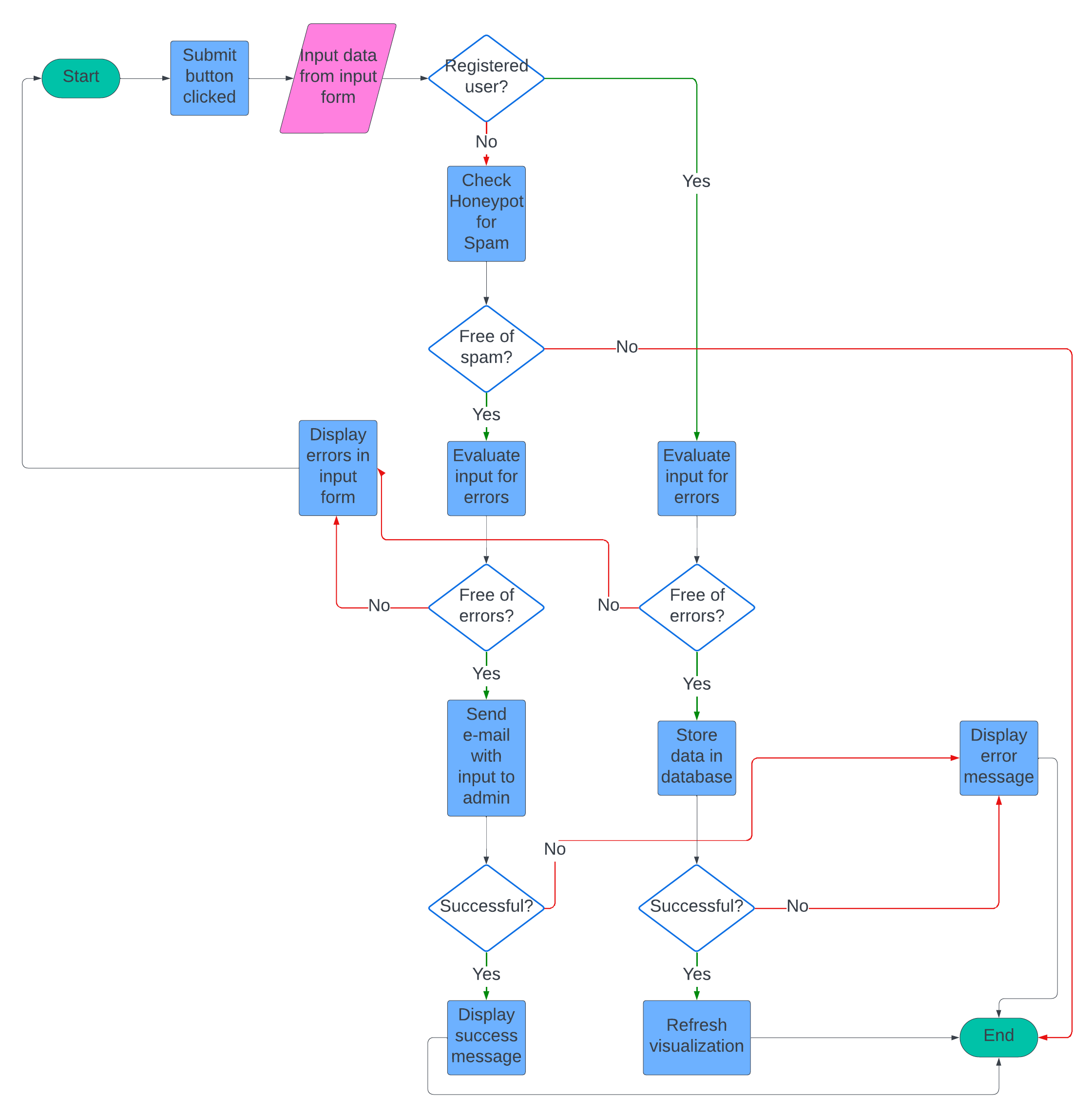}
    \caption{Flow Chart showing the data submission process}
    \label{fig:dataSubmission}
\end{figure}

\subsection{\code{postData}}\label{subsec:postData}
The functions \code{problemToDB, reductionToDB}
(\cref{subsec:reductionToDB}), and \code{isSpam} (\cref{subsec:honeypot}) are part of the
process of adding new data to the database. A flow chart of this process can be seen in~\cref{fig:dataSubmission}.
On clicking the submit button, the function \code{postData} is called
with an event and a boolean whether a visitor or a registered user called the function.
Note that this does not suffice to ensure only users can access the routes
that directly add data to the database, as this function runs in the frontend and can easily
be called with manipulated data.

The data is collected from the form using the \code{FormData} object, and the form contains 
the input fields.
The object compiles key-value pairs that can be sent to the backend~\citeWebsites{formData}.
The data from the input form can now be accessed, and the type, which is whether it is a
request to add a new entry or to edit an existing entry, as well as the selection, which
is either problem or reduction, thus, the value chosen by the user in the dropdown menu,
are stored. They are needed to later fetch the correct route.

In the case a visitor made the submission, the honeypot is checked, and then, if there is no spam,
the route \code{post('/addProblemUnauth')} (or the route for adding reductions respectively) is called.
If a registered user made the submission, the route \code{post('/addProblem')} 
(again, the route for reductions respectively) is called immediately.
In the post routes, the input is evaluated for potential errors, and if any occur, 
these are displayed and the user is asked to fix them. Otherwise,
the data is stored in the database or an email is sent containing the input data for submissions of visitors,
which is handled in the backend by the functions \code{reductionToDB} and \code{sendMail}.
If everything goes well, visitors get a success message displayed, and registered users
get the refreshed visualization, which includes their submission, which is done in \code{handleSubmission}.

\begin{lstlisting}[language=JavaScript, caption={function \code{postData} from \code{addElement.js}}, label=lst:postData]
function postData(e, authorized) {
    e.preventDefault();
    // close form and return if honeypot is filled
    if(!isSpam()) {
        closeDialog();
        resetDialog();
        return;
    }
    // otherwise, handle the submission
    const formData=new FormData(form); // FormData stores all the inputs in the form
    
    const type=Object.fromEntries(formData.entries()).actionType; // edit or add
    let selection=Object.fromEntries(formData.entries()).selection; // problem or reduction
    if(authorized) {
        fetch(baseUrl+"add"+selection+"?type="+type, {
        method: 'POST',
        body: formData,
        }).then(async res => {
        const data=await res.json();
        if(res.status === 400) {
            displayErrors(data.errors);
        } else {
            handleSubmission(data, selection);
            closeDialog();
        }
        });
    } else { // unauthorized
        fetch(baseUrl+"add"+selection+"UnAuth?type="+type, {
        method: 'POST',
        body: formData,
        }).then(async res => {
        const data=await res.json();
        if(res.status === 400) {
            displayErrors(data.errors);
        } else {
            handleSubmission(data, 'submission');
            closeDialog();
        }
        });
    }
}

async function handleSubmission(data, type=false) {
    if(data.success) {
        refreshVisualization(); // If submission was successful, update content
        if(type) {
            let id=data.id;
            if(type === 'problem') {
                displayInfo({
                edges: [],
                nodes: [id]});
            } else if(type === 'reduction') {
                displayInfo({
                edges: [id],
                nodes: []});
            } else if(type === 'submission') {
                alert(successMessage);
            } else {
                console.error('Invalid type');
            }
        } else {
            console.error(data.message);
        }
    } else {
    // If submission fails, display errors in website 
    console.error('Error:', data.error);
    alert('An error occurred while adding new entries. Please try again.');
    }
}
\end{lstlisting}
\suppressfloats[t]

\subsection{\code{post('/addReduction')}}
The post route for adding reductions to the database is a good example for a typical post route in this project.
In~\cref{lst:networkServer}, both the route itself and the multer middleware for storing the uploaded file on the server
are presented.
When a user calls the route \code{'/addReduction'}, first of all, it is checked if the user is 
logged in, using the \code{checkAuthenticated} method, which is provided by passport. 
This is important to ensure that the route can only be executed if the user is actually logged in. 
Then, all post routes have a limiter, that forces an upper bound to the number of times which a 
specific IP address can try to post something with that route within a given time frame. 
The values are chosen such that an actual user should never fail their post process,
as one cannot post all the time, since it does take some time to input the right values properly, 
but that bots trying to spam the server are rejected after a few times. 
In the next step, the upload of the example file is handled,
which will be explained in more detail in~\cref{subsec:reductionToDB}. 
Then, \code{reductionValidator} validates and sanitizes the user input. It is important that 
this is done \textit{after} the upload, because multer, which handles the upload, 
parses the body of the request, and without that, the body cannot be accessed.
The validator ensures for example that the required inputs are not empty, not too long, and of the correct format.
However, the input is not sanitized, as that interferes with the usage of LaTeX code on the website.
After all of these steps are done, the actual function that the route should execute is handled.
In this case, this means that if any errors occurred in the validation step, these errors are sent back
to the frontend and displayed to the user, so that they can fix their input. 
If no errors occurred, the data is sent to the backend and the function \code{reductionToDB} is called, and if 
everything \code{reductionToDB} was successful, an ok response is sent back to the frontend to complete the process.

\begin{lstlisting}[language=JavaScript, caption={post route from \code{networkServer.js}}, label=lst:networkServer]
networkRouter.post('/addreduction', checkAuthenticated, 
postLimiter, upload.single('reductionExample'), reductionValidator, 
async (req, res) => {
    // handle validation results
    const errors=validationResult(req);
    if(!errors.isEmpty()) {
        console.log("errors:", errors.array());
        return res.status(400).json({ errors: errors.array() });
    }
    // if the validation is ok, handle the data (send to db and send ok response)
    try {
        const { type }=req.query;
        let id=await db.reductionToDB(req.body, req.file, type);
        res.status(200).send({ success: true, id: id });
    } catch(error) {
        const { type }=req.query;
        console.error(`error ${type}ing data: `, error);
        res.status(500).json({ error: 'Internal Server Error' });
    }
});
\end{lstlisting}
\suppressfloats[t]

\subsection{\code{problemToDB, reductionToDB}}\label{subsec:reductionToDB}
These functions are used to insert data into the database, either into the problem table, 
or into the reduction table.
The two functions are build analogously, with the only difference being into which table the data is stored,
and that for reductions, it is necessary to handle the upload of the example. 
As \code{reductionToDB} is the slightly more complicated function, it is used as an example here.
\Cref{lst:reductionToDB} shows the relevant code for the function. 

First, we need to handle the completeness property, which is an array, as it is composed of checked
checkboxes in the input form and stored as an array in the \code{formData} object.
In oder to be stored in the database, it needs to be converted to a string.
If only one value is given, it already is a string, and otherwise, the entries are joined together by a comma.
This way, the string can simply be added into the database.

Then, the file name must be stored. 
The file itself is stored using the multer middleware, and multer stores the file 
in the folder \code{'/public/uploads'}. The file name is composed of a timestamp
and a random number, to ensure that each file name is unique, and again to ensure that 
nothing malicious can be injected this way.\\
In the function \code{reductionToDB} itself, the file name is stored in the database, 
as the function is only called if the file upload through multer was successful.
If the route was called to edit a reduction, it is also checked if there already was a file uploaded,
which is then deleted, so that only data is stored on the server that is still relevant and not outdated.
The data is added to the database using the function \code{createReduction} (or \code{createProblem, editReduction}, etc.).
The function \code{createReduction} uses prepared statements, which is another safety measure to 
prevent SQL injections.
Prepared statements treat the input not as SQL code, but first insert placeholders to pre-compile the SQL
statement and parse the actual inputs later as strings, such that even if the input were malicious 
SQL code, it would not alter the SQL statement. 
This approach is used everywhere where queries to the database are executed.

\begin{lstlisting}[language=JavaScript, caption={functions \code{reductionToDB} and \code{createReduction} from \code{backend/database.js}},label=lst:reductionToDB]
async function reductionToDB(body, file, type) {
    let id;
    let typeString=body.reductionType; // make string if there is only one value
    if(Array.isArray(body.reductionType)) {
        types=body.reductionType.join(", ");
    }
    const fileName = file ? file.filename : null;
    if(type === "add") {
        id=await createReduction(body.reductionFrom, 
            body.reductionTo, typeString, 
            body.reductionDescription, body.reductionRef, 
            body.reductionLink, body.reductionComments, fileName);
    } else if(type === "edit") {
        let oldFileName=await getReduction(body.reductionID)
        .then((result) => {return result.exampleImage});

        id=await editReduction(body.reductionID, 
            body.reductionFrom, body.reductionTo, typeString, 
            body.reductionDescription, body.reductionRef, 
            body.reductionLink, body.reductionComments, fileName)
        .then((result) => {
            if (fileName) {
                try {
                    deleteFile(oldFileName);
                } catch (error) {
                    console.error("error deleting file", error);
                }
            }
        })
        .catch((error) => {
            console.error("database deletion error:", error);
            res.status(500).json({success: false, error: 'Database Error'});
        });
    }
    return id;
}

async function createReduction(from, to, type, description, reference, referencelink, comments, filePath){
    const result = await query(`
        INSERT IGNORE INTO reductions 
        (\`from\`, \`to\`, type, description, reference, referenceLink, comments, exampleImage)
        VALUES (?,?,?,?,?,?,?,?);`,
        [from, to, type, description, reference, referencelink, 
        comments, filePath],
        true);
    return(result);
}

\end{lstlisting}
\suppressfloats[t]

\subsection{\code{isSpam}}\label{subsec:honeypot}
The function \code{isSpam} evaluates the honeypot fields that are part of the input form.
To detect bots an prevent them from spamming the server, a simple honeypot was implemented~\cref{fig:honeypot}.
It consists of a button and a text input, which are invisible and thus are usually not filled out by human users.
In the function \code{postData}, the function \code{isSpam} is called. The spam check should also be in 
the route itself, because a check in the frontend can always be bypassed, but this was not implemented.
The function \code{isSpam} simply checks if any of the fields are filled out, and returns \code{true} if 
spam was detected.
\begin{figure}
    \begin{lstlisting}[language=HTML5]
    <input type="radio" id="radioBtnProb" class="randomClass" name="radioBtnProb" value="direct" style="margin-left:-100px">
    <input id="randomID" class="randomClass" type="text" name="randomID">
    \end{lstlisting}
    \begin{lstlisting}[language=CSS]
    .randomClass {
        display: none !important;
    }
    \end{lstlisting}
    \begin{lstlisting}[language=JavaScript]
    function isSpam() {
        if (honeypot.value.length != 0 || honeypotProblem.checked || honeypotReduction.checked) {
            return false;
        }
        return true;
    }
    \end{lstlisting}
    \caption{HTML and CSS code for the honeypot, and the function which checks if the honeypot fields are filled out.}
    \label{fig:honeypot}
\end{figure}

Next, some important functions for navigating the network are to be explained.

\subsection{Search and Filter Functions}\label{subsec:searchFilter}
As the network can become large and hard to navigate quickly, a search and a filter function were implemented.
The search function compares the input string with the names and abbreviations of the nodes in the network,
using case-non-sensitivity and Levenshtein distance to enable finding the problem even if the spelling differs a bit.
If a match is found, the corresponding node is focused in the center of the screen, and the information is displayed.
The problem is found by getting all names and abbreviations from the network. If there is a direct match found,
this is returned. Otherwise, it is checked if there is a problem with a Levenshtein distance of three or lower.
If a match is found, the node is focused, using the focus method that zooms in on the node and 
displays the information of the problem.

\FloatBarrier
\begin{lstlisting}[language=JavaScript, caption={Functions \code{searchProblem} and \code{focusProblem}}, label=lst:searchProblem]
function searchProblem(problemName) {
    let problemNonCaseSensitive = problemName.toLowerCase();
    let matchedNode = nodesDataset.get({
        filter: function (node) {
            let nameNonCaseSensitive = node.label.toLowerCase();
            let abbNonCaseSensitive = node.title.toLowerCase();
            let res = nameNonCaseSensitive === problemNonCaseSensitive || abbNonCaseSensitive === problemNonCaseSensitive;
            return res;
        }
      })[0];
      if (!matchedNode) {
        let bestDistance = 1000;
        // If no exact match is found, try to find the best match
        nodesDataset.forEach(node => {
            node.label = node.label.toLowerCase();
            node.title = node.title.toLowerCase();
            let distance = Math.min(levenshteinDistance(problemNonCaseSensitive, node.label), levenshteinDistance(problemNonCaseSensitive, node.title));

            if (distance < bestDistance) {
                bestDistance = distance;
                matchedNode = node;
            }
        });
        
        if (bestDistance > 3) {
            matchedNode = null;
        }
    }
    if (matchedNode) {
        return matchedNode.id;
      } else {
        alert('Problem not found. Please try another formulation, short form or full name.');
        return null;
      }
}

function focusProblem(id) {
    // Focus on the matched node
    network.focus(id, {
        scale: 1.0,
        animation: {
        duration: 1000,
        easingFunction: 'easeInOutQuad'
        }
    });
    displayInfo({
        edges: [],
        nodes: [id]
    });
}
\end{lstlisting}
\suppressfloats[t]

The filter function takes its input from the dropdown menu, in which all possible choices are presented.
It uses the function \code{getCompleteness}, which fetches the route \code{'/filter'}.
The route \code{'/filter'} fetches all nodes and edges~\cref{lst:filter}, 
and then only returns the nodes which include the given filter in their completeness property, 
and reductions, that are of a type that is relevant for that chosen complexity class.
In the case of the class SSP-NP, SSP reductions need to be shown, and for the class NP, polynomial reductions need to be displayed. 
Then, all nodes are colored in a light gray, and only the elements that match
the filter are colored in the blue that is used in the network again. 
Sadly, it was not feasible to implement the method in a way that disables clicking on the grayed 
out elements, as that would involve changing the source code of vis.js. 
Thus, the filter function is only a visual filter, but nevertheless, 
it helps to only focus on the problems and reductions that are relevant for the user.

\begin{lstlisting}[language=JavaScript, caption={The important functions for the network filter}, label=lst:filter]
async function filter(e) {
    e.preventDefault();
    let filter = filterInput.value;
    
    let elements = await getCompleteness(filter);
    if (!elements) { // getCompleteness returns null if there was an error
        resetColors(color=rwthBlue);
    }
    resetColors();
    colorElements(elements, rwthBlue, 1); // blue is the standard color for the network
}

networkRouter.get('/filter', navigationLimiter, filterValidator, async (req, res) => {
    try{
        const filter=req.query.completeness.toLowerCase();
        let filterReductions=filter;
        const problems=await db.getProblems();
        const reductions=await db.getReductions();
        if(filter === "np" || filter === "p") {
            filterReductions="polynomial";
        } else if(filter === "ssp-np") {
            filterReductions="ssp";
        } // insert more constraints here for other classes

        // find the problems that have the filter as completeness class
        filteredProblems=problems.filter(problem =>
            problem.completeness.map(completeness => completeness.toLowerCase())
            .includes(filter));
        filteredReductions = reductions.filter(reduction => 
            reduction.type.map(type => type.toLowerCase())
            .includes(filterReductions));
        res.status(200).json({nodes: filteredProblems, edges: filteredReductions});

    } catch(error) {
        console.error("error fetching data: ", error);
        res.status(500).json({error: 'Internal Server Error'});
    }

});
\end{lstlisting}
\suppressfloats[t]

\subsection{Network visualization}\label{subsec:networkVis}
In this section, the way in which the network is initialized and displayed is mentioned.
As we know, vis.js was used for the visualization of the graph.
The graph contains the problems and reductions, which are retrieved from the database using \code{fetchData}.
The function \code{fetchData} creates two \code{vis.DataSet}, which is a set of keys and values. 
One \code{DataSet} for the nodes is created, with the parameters being the id, 
name, abbreviation, and completeness, which come from the problems table,
and one \code{DataSet} for the edges, which correspond to the reductions and contain the ids of the problem from
which and to which are reduced, the id of the reduction, and the type of the reduction.
For the nodes, these values are needed as the abbreviations are displayed on the nodes, 
to avoid the network looking very cluttered, and the full names are displayed on hover.

The options in~\cref{lst:network} define the layout of the network, and using the options, many other things can be
customized. In this project, the colors are defined to match the RWTH color scheme, which simultaneously also 
provides distinguishable colors for better accessibility, and the boxes are
rectangular, as that gives the network a cleaner look. Also, we want directed edges, to show the direction
of the reductions. On click, the network calls the function \code{displayInfo}, which takes the parameters
that are sent with the click event, and takes which node or edge was clicked, to then display the corresponding
information on the right side of the screen.

Also, a zoom limit is added to the network, as vis.js seems to be quite sensitive to zooming, and the zoom
often turned out to be far too much. \code{AdaptDropdowns} is called on initialization to add all entries
that were found in the database to the dropdown menu that is part of the input form for adding new entries,
because it should always contain exactly the problems that are in the database.

\begin{lstlisting}[language=JavaScript, caption={Network visualization using vis.js}, label=lst:network]
function startNetwork() {
    fetchData()
    .then(({nodes, edges}) => {
        // create a network
        nodesDataset = nodes;
        edgesDataset = edges;

        var container = document.getElementById("network");
        var networkData = {
        nodes: nodesDataset,
        edges: edgesDataset
        };
        var options = {
            nodes: {
                shape: 'box',
                color: {background: rwthBlue, highlight: rwthBlue }, 
                font: {color: '#ffffff'}
            },
            edges: { arrows: 'to' },
        };
        network = new vis.Network(container, networkData, options);

        network.on('click', displayInfo);

        // add a zoom limit to preven zooming too far
        let lowerzoomlimit = {
            scale: 0.3,
        }
        let upperzoomlimit = {
            scale: 2,
        }
        network.on("zoom",function(){ //while zooming 
            if(network.getScale() <= lowerzoomlimit.scale ) {
                network.moveTo(lowerzoomlimit); // stop zooming out
            } else if(network.getScale() >= upperzoomlimit.scale) {
                network.moveTo(upperzoomlimit); // stop zooming in
            }
        });

        adaptDropdowns(nodesDataset, []);
    });
}
\end{lstlisting}
\suppressfloats[t]

\subsection{Database Structure}\label{subsec:databaseStructure}
\begin{figure}
    \centering
    \includegraphics[width=\textwidth]{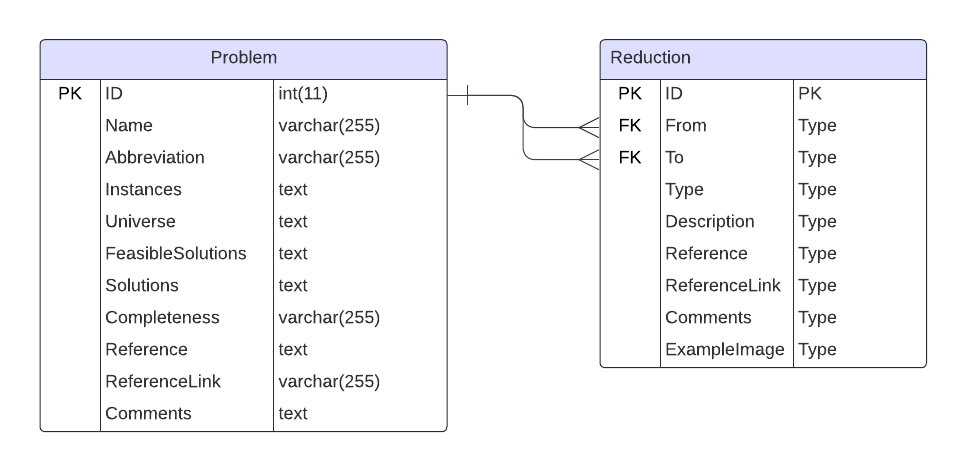}
    \caption{Entity-Relationship diagram of the tables storing the problems and the reductions in the MariaDB database.
    FK stands for foreign key, thus \code{From} and \code{To} reference the id of the problems table, and PK is the primary key.}
    \label{tab:tableStructure}
\end{figure}

Next, the database structure is described and it is explained why the database was set up this way.
There are three tables in the database: 
\begin{itemize}
    \item The table credentials, which is needed for the login and authentication,
    \item the table problems, which stores the data related to the problems, and
    \item the table reductions, in which the data related to the reductions is stored.
\end{itemize}
The table credentials stores the id, email, and (salted and hashed) password of the users.
As only the problems and reductions are relevant for the visualization, the structure of these two tables is described in detail.
The Entity-Relationship diagram displaying the relation of the two tables
is shown in~\cref{tab:tableStructure}. It can be seen that a \code{from} or \code{to} field references exactly one
problem, and a problem can be referenced by arbitrarily many reductions.
Each problem and reduction has a unique \code{id},
which is the primary key and is used to reference it, e.g., 
reductions reference the ids of the problems involved in the reduction.
The \code{abbreviation} of a problem is later used as a label for the node, 
as always displaying the complete name takes up too much space.
Next, the problem definitions follow with the fields \code{instances}, \code{universe}, \code{feasiblesolutions}, and \code{solutions}.
In the reduction table, the columns \code{from} and \code{to} reference the ids of the problems involved in the reduction.
\code{Description} contains detailed description of the reduction function, it might also include correctness.
The field \code{exampleImage} contains the file name of the corresponding image, the image
itself is not stored in the database, but on the server directly.
Both tables have the columns \code{reference} and \code{referenceLink}, 
which are used to store the reference from which the definition of the problem
or the reduction is found, and where a more detailed explanation is given. If one wants to 
add any more information, thoughts etc., this can be done in the \code{comments} field.

The types are chosen to not be excessively large, as e.g. names are generally not very long, 
and descriptions should be short and not take up too much space.
The fields \code{from} and \code{to} in the table reductions have foreign key constraints to the id of the problems table.

The division into two tables is very natural, as the problems and reductions are two different entities,
and the reductions reference the problems. It is of course also important to reference the ids of the problems,
as they are the primary key and thus unique for each problem and naturally used to reference them.

For displaying the graph, the ids, names, and abbreviations of the problems are needed,
and for the reductions, it is only needed to know the id and which problems are involved in the reduction.
The functions \code{getProblems} and \code{getReductions} do also return the completeness value, but this is 
only because that function is also used for the filter function, where that information is needed.
The ids are used to reference the problems and reductions, and the \code{from} and \code{to} fields
in the reduction table reference the ids of the problems involved. 
The names and abbreviations are needed as label and hover text for the nodes.

\chapter{Conclusion and Future Work}\label{chap:conclusion}
In this thesis, many reductions were examined regarding the question whether they fulfill the SSP property.
For most of the reductions, this turned out to be true. 
All reductions that were not yet examined in~\citeLiterature{grune2023large} were examined, excluding reductions 
containing the problems \textsc{Chromatic Number} or \textsc{Max Cut} from Karp's paper on 21 NP-complete problems~\citeLiterature{karp1975},
and most of them could be modified to fulfill the SSP property.
Also, five out of six reductions covered in~\citeLiterature{garey1979computers} turned out to be easily transformable
into polynomial SSP reductions, and most of the reductions given or hinted at in~\citeLiterature{arora2009computational}
were examined successfully and turned out to be transformable into reductions preserving the SSP property as well. 
Through this, in total 19 reductions were newly discovered and eight new problems were shown to be SSP-NP-complete.

Additionally, a compendium website was built that serves as a collection of SSP-NP-complete problems and 
reductions fulfilling the SSP property, that can easily be extended to include other complexity classes and reduction types as well.
This helps collecting and sorting knowledge regarding any kind of reductions, as it is designed to easily understand
the relations between problems, be able to take submissions from theoretical computer scientists from all over the world
and thereby works as a knowledge base for reductions.

In the future, it would be interesting to examine more polynomial reductions for the SSP property,
as well as finding a scheme for reductions that cannot be transformed
into SSP reductions. With regards to the compendium website, it would be nice to add a functionality
such that enables visitors to still be able to view their submission after sending it.
This way, they can fix problems, e.g. with the visualization of formulas, as it happens often that users
use LaTeX macros defined by themselves, and these macros are not available on the server, which leads to problems.
These problems can only be detected if the user can see  how the submission will look like later, 
and could thereby also protect the page administrator from a lot of work trying to find out what the user meant. 
Additionally, problems from more complexity classes should be added,
which also requires an option to switch between the different graphs.
If the websites becomes more popular, it would be important to add more roles, so that there are
visitors, normal users, and administrators. This way, visitors still can submit new problems,
but only admins can review and add them, and only admins can create new users. Normal users
can directly input data to the database, but can not handle submissions of visitors or add new users.
This way, people who are trustworthy and have a good understanding of the topic can be given
the user role, so that not everything has to be reviewed by an admin.

\bibliographystyleLiterature{alpha}
\bibliographyLiterature{literature}
\bibliographystyleWebsites{alpha}
\bibliographyWebsites{websites}
\end{document}